\documentclass[amsmath,amssymb,twocolumn,superscriptaddress]{revtex4}
\usepackage{doi}
\usepackage{hyperref}
\hypersetup{colorlinks=true,linkcolor=blue,
  citecolor=cyan,}
\usepackage{dcolumn}
\usepackage{bm}
\usepackage{xcolor}
\usepackage{amsmath}
\usepackage{amssymb}
\usepackage[T1]{fontenc}
\usepackage{lmodern}
\usepackage{textcomp}
\usepackage{newunicodechar}

\providecommand{\dif}{\mathrm{d}} \def\d{\dif}
\usepackage{graphicx}
\usepackage{dcolumn}
\usepackage{bm}
\usepackage{color}
\usepackage{enumitem}
\usepackage{amsmath}
\usepackage{amssymb}
\def\EE{{\cal E}}
\def\LL{{\cal L}}
\newcommand{\beq}{\begin{equation}}
\newcommand{\eeq}{\end{equation}}
\newcommand{\bea}{\begin{eqnarray}}
\newcommand{\eea}{\end{eqnarray}}
\newcommand{\non}{\nonumber}
\newunicodechar{−}{\ensuremath{-}}

\begin{document}

\title{Relativistic accretion process onto rotating black holes in Einstein-Euler-Heisenberg nonlinear electrodynamic gravity}

\author{Orhan~Donmez}
\email{orhan.donmez@aum.edu.kw}
\affiliation{College of Engineering and Technology, American University of the Middle East, Egaila 54200, Kuwait}

\author{G. Mustafa}
\email{gmustafa3828@gmail.com (Corresp. Author)}
\affiliation{Department of Physics, Zhejiang Normal University, Jinhua 321004, People's Republic of China}
\affiliation{Research Center of Astrophysics and Cosmology, Khazar University, Baku, AZ1096, 41 Mehseti Street, Azerbaijan}

\author{Himanshu Chaudhary}
\email{himanshu.chaudhary@ubbcluj.ro}
\affiliation{Department of Physics, Babeș-Bolyai University, Kogălniceanu Street, Cluj-Napoca, 400084, Romania}

\author{M. Yousaf}
\email{myousaf.math@gmail.com}
\affiliation{Department of Mathematics, Virtual University of Pakistan, 54-Lawrence Road, Lahore 54000, Pakistan}

\author{Abdelmalek Bouzenada}
\email{abdelmalekbouzenada@gmail.com}
\affiliation{Laboratory of Theoretical and Applied Physics, Echahid Cheikh Larbi Tebessi University 12001, Algeria}

\author{Allah Ditta}
\email{mradshahid01@gmail.com}
\affiliation{Department of Mathematics, School of Science, University of Management and Technology,  Lahore, 54000, Pakistan.}

\author{Farruh Atamurotov}
\email{atamurotov@yahoo.com}
\affiliation{Urgench State University, Urgench 220100, Uzbekistan}

\date{\today}
\begin{abstract}

In this study, we uncover the accretion dynamics and oscillatory behavior around rotating black holes within the EEH nonlinear electrodynamic framework by analyzing both the motion of test particles and numerically solving the general relativistic hydrodynamic equations. Using EEH geometry, we compute the structure of circular motion, the effective potential and force, and we evaluate the orbital, radial, and vertical epicyclic frequencies together with the Lense–Thirring and periastron precession rates. Our calculations show that, compared to the Kerr model, the charge parameter $Q$ and the spin parameter $a$ significantly modify the strong gravitational field and shift the characteristic frequencies. We then model the dynamical structure formed by matter accreting toward the EEH black hole through the BHL mechanism, finding that the parameter $Q$ increases the amount of infalling matter and strengthens shock–cone instabilities near the horizon, while farther from the black hole it suppresses accretion and reduces turbulence. Time-series analysis of the accretion rate reveals robust QPOs, whose low-frequency components arise from the precession of the shock cone, while high-frequency components appear as a consequence of strong-field instabilities modified by $Q$ and $a$. A systematic parameter-space exploration identifies the regions where EEH corrections maximize QPO activity, indicating that nonlinear electrodynamics can leave observable imprints on accretion flows and may be testable with QPO and horizon-scale observations.

\end{abstract}

\maketitle

\tableofcontents

\section{Introduction}\label{S1}
Einstein’s theory of General Relativity (GR) \cite{BHQD1, BHQD2, BHQD3} predicts the existence of black holes (BHs) extraordinary regions of spacetime characterized by the presence of an event horizon, a boundary beyond which no form of matter or radiation can escape. These compact objects naturally emerge as the inevitable outcome of complete gravitational collapse and represent one of the most profound predictions of relativistic gravitation. Beyond their astrophysical importance, BHs serve as unique theoretical laboratories for exploring the unification of gravitation and quantum mechanics one of the central challenges in modern physics that remains unresolved \cite{BHQD4}. 

Recent advancements in observational astronomy opened direct pathways to probe these exotic regions of spacetime, in particular, the detection of gravitational waves and the development of very long baseline interferometry techniques enabled studies of BH environments with a level of precision unattainable under terrestrial conditions \cite{BHQD5, BHQD6}. Among the most remarkable achievements in this direction is the imaging of BH shadows through the coordinated global effort of the Event Horizon Telescope (EHT) a network of radio observatories that, by operating at a wavelength of $1.3$ millimeters, achieves an angular resolution of approximately $25~\mu$as, limited only by diffraction \cite{BHQD7}. 

In April 2019, the EHT collaboration released the first image of the shadow of the supermassive BH located in the center of the galaxy $M87^*$, an achievement documented in a series of pioneering studies \cite{BHQD7, BHQD8, BHQD9, BHQD10, BHQD11, BHQD12, BHQD13, BHQD14, BHQD15}. Subsequent investigations analyzed the rotational characteristics of this compact object \cite{BHQD16} and compared the observational findings with the theoretical predictions of the Schwarzschild solution \cite{BHQD17}. Following these results, in May 2022, the same collaboration announced the first direct observation of the shadow of $Sgr~A^*$ the supermassive BH at the center of the Milky Way with results presented comprehensively in multiple publications \cite{BHQD18, BHQD19, BHQD20, BHQD21, BHQD22,BHQD23}. Collectively, these achievements provide direct empirical validation of the predictions of relativistic gravitation and offer unprecedented insights into the nature of spacetime and gravity. Some other works related to the BH study are: Scalar Hawking radiation from regular black holes are studied in reference \cite{Sun1}; Covariant canonical quantum gravity models are discussed in reference \cite{Sun2}; the possibility of nonsingular objects, considering three phenomenological, regular tr (time-radial)-symmetric space-times (including the well-known Bardeen and Hayward ones), featuring either de Sitter or Minkowski cores are discussed in the reference \cite{Sun3}; implications of cosmologically coupled black holes for pulsar timing arrays has been studied in reference \cite{Sun4}; authors argue that the Event Horizon Telescope images of M87 and Sgr A rule out the baseline version of mimetic gravity, preventing the theory from successfully accounting for the dark sector on cosmological scales in reference \cite{Sun5}; superradiant evolution of the shadow and photon ring of Sgr $A^{*}$ \cite{Sun6}; black holes with scalar hair in light of the Event Horizon Telescope have been discussed in \cite{Sun7}. 

The framework of nonlinear electrodynamics (NLED) has emerged as a fundamental tool in the study of BH physics, particularly in relation to BH shadows, which currently stand at the forefront of observational astrophysics \cite{HHTR1}. Within NLED, the propagation of electromagnetic radiation is effectively modified, so that photons no longer follow the standard null geodesics of Minkowski spacetime, but instead travel along null curves of an effective geometry, rendering the medium classically dispersive \cite{HHTR2, HHTR3}. A major milestone in this direction was achieved by Novello et al., who investigated photon propagation within the Euler–Heisenberg (EH) model under regular BH backgrounds, thus unveiling the concept of an effective geometry \cite{HHTR3}. 

With the advent of nonlinear quantum electrodynamics, this area gained renewed attention, inspiring numerous studies on light propagation, effective metrics, and photon surfaces across various nonlinear frameworks \cite{HHTR4, HHTR5, HHTR6}, as well as investigations of generalized Born-Infeld electrodynamics \cite{HHTR7}. Among these developments, the Einstein-Euler-Heisenberg (EEH) theory which couples EH type NLED to GR has proven particularly significant, yielding exact BH solutions explored from multiple perspectives, including their thermodynamic behavior \cite{HHTR8}, the influence of thermal fluctuations \cite{HHTR9}, gravitational lensing properties \cite{HHTR10}, energy extraction mechanisms \cite{PRD2025Mustafa}, accretion disk dynamics \cite{HHTR11}, and shadow and quasinormal mode spectra \cite{HHTR12}. 

For example, the analysis in \cite{HHTR13} examined the shadows of charged EEH BHs by tracing null geodesics in the background geometry and comparing the predicted shadow radii with the observations of the Event Horizon Telescope (EHT) of $Sgr~A^\ast$, finding consistency within a narrow range of electric charge values. A complementary approach presented in \cite{HHTR14} revisited the EEH BH shadow using null geodesics of the effective metric and compared theoretical predictions with the EHT images of both $Sgr~A^\ast$ and $M87^\ast$. Furthermore, it demonstrated that thermal and quantum fluctuations substantially influence the thermodynamics of EEH BHs, with vacuum polarization effects becoming increasingly dominant at high charge values and for large EH parameters \cite{HHTR9}. Additionally, analysis of the phase structure of EEH AdS BHs in canonical and grand canonical ensembles revealed intricate phase transitions governed by nonlinear interactions \cite{HHTR8}. Consequently, these findings underscore that NLED profoundly reshapes the effective geometry governing photon trajectories and significantly alters the thermodynamic and observational characteristics of BHs, thereby bridging deep theoretical insights with astrophysical observations.

The quasi-periodic oscillations (QPOs) observed in X-ray binaries provide one of the most compelling observational probes of strong field gravity and accretion dynamics around compact objects; in particular, the detection of twin high frequency QPOs in microquasars consistently appearing in the characteristic $3{:}2$ ratio suggests that these signals originate from nonlinear resonances between the fundamental oscillation modes of nearly Keplerian accretion disks in curved spacetime \cite{QPOs1, QPOs2, QPOs3}. The empirical scaling relation $\nu \sim 1/M$, linking the observed frequencies to the BH mass, strongly supports their relativistic nature \cite{QPOs4, QPOs5}. 

Several theoretical models have been proposed to explain these oscillations, including trapped $g$ modes \cite{QPOs6, QPOs7, QPOs8}, corrugation $c$ modes associated with Lense Thirring precession \cite{QPOs9, QPOs10}, and various orbital resonance mechanisms \cite{QPOs11}. Among these, the parametric resonance $3{:}2$ between vertical and radial epicyclic oscillations has emerged as the most consistent interpretation, supported by both analytical treatments and high resolution numerical simulations \cite{QPOs12, QPOs13, QPOs14}. 

By fitting this resonance model to observed QPOs frequencies, one can place constraints on the spin parameters of stellar mass BHs, notably, sources such as $GRO~1655{-}40$ and $XTE~1550{-}564$ exhibit high rotation rates inferred from such analyzes \cite{QPOs15, QPOs16}. Beyond galactic binaries, the same resonance framework has been successfully employed to estimate BH masses in ultraluminous X-ray sources and active galactic nuclei, reaffirming the robustness of the $1/M$ scaling relation \cite{QPOs17, QPOs18}.

In this work, we explore the dynamics of accretion processes around rotating BHs within the framework of the EEH theory. Using high resolution numerical simulations, we examine how the interplay between the charge parameter ($Q$) and the spin parameter ($a$) influences the global structure of the accretion flow, with particular emphasis on the emergence of plasma configurations and the development of shock cones. The temporal evolution of these shock structures is closely monitored to evaluate their stability and to uncover distinctive nonlinear behaviors induced by strong electromagnetic fields. 

Furthermore, oscillatory features extracted from the numerical data reveal QPOs, which serve as sensitive diagnostics of matter field interactions in extreme gravitational environments. Through a systematic exploration of the parameter space, we identify critical regimes where QPOs activity is most prominent, offering new insights into the mechanisms governing oscillatory phenomena in EEH modified spacetimes. The astrophysical relevance of these findings is highlighted by their observational implications, suggesting that NLED effects could imprint detectable signatures on the variability patterns of realistic accreting systems. Consequently, our findings demonstrate that the study of QPOs provides a reliable avenue for probing strong field gravity, testing departures from classical electrodynamics, and deepening our understanding of fundamental interactions near compact astrophysical objects.

The structure of this paper is organized as follows. In Sec.~\ref{S1}, we provide a general introduction and motivation for the study. Section~\ref{S2} outlines the theoretical foundations of EEH theory, which form the basis of our analysis. In Sec.~\ref{S3}, we present and discuss the simulation results in detail, beginning with a numerical investigation of the combined influence of the charge parameter $Q$ and the spin parameter $a$ on accretion dynamics (Sec.~\ref{S3-1}), followed by an examination of plasma formation and the emergence of the shock cone (Sec.~\ref{S3-2}), and concluding with the analysis of the shock cone evolution (Sec.~\ref{S3-3}). Section~\ref{S4} is devoted to the extraction and interpretation of QPOs from the numerical data. In Sec.~\ref{S5}, we carry out a systematic parameter space exploration to identify the critical regimes of the system. Section~\ref{S6} discusses the astrophysical relevance of our findings and their potential observational signatures. Finally, the main results and concluding remarks are summarized in Sec.~\ref{S7}.

\section{THE EINSTEIN-EULER-HEISENBERG THEORY}\label{S2}

We revisit the fundamental aspects of Einstein's gravity in the presence of Euler-Heisenberg nonlinear electrodynamics (EH-NLED)~\cite{1936ZPhy...98..714H}, formulated within the theoretical framework introduced by Plebański~\cite{Plebański}. In this context, the action describing Einstein’s gravity minimally coupled to the linear Maxwell theory or its nonlinear extensions can be expressed as~\cite{1936ZPhy...98..714H, 2001CQGra..18.1677G}:
\begin{equation}
    W=\frac{1}{16\pi G}\int_{M_4}d^4x\sqrt{-g}R+W_M(X,Y),
\end{equation}
where $R$ denotes the Ricci scalar curvature, $g$ is the determinant of the spacetime metric $g_{\mu\nu}$, and $G$ represents Newton's gravitational constant, which is set to unity for simplicity. The quantities $X$ and $Y$ denote the two independent electromagnetic invariants constructed from the Maxwell field tensor in four dimensions, namely,
\begin{equation}
    X=\frac{1}{4}F_{\mu\nu}F^{\mu\nu},
    \qquad
    Y=\frac{1}{4}F_{\mu\nu}{^*}F^{\mu\nu}.\label{Maxwell inv},
\end{equation}
here, the Faraday tensor is defined as $F_{\mu\nu}=\partial_\nu A_\mu-\partial_\mu A_\nu$, where $A_\mu$ denotes the electromagnetic four-potential. Its dual is given by ${^*}F^{\mu\nu}=\tfrac{1}{2\sqrt{-g}}\epsilon^{\mu\nu\sigma\rho}F_{\sigma\rho}$, with $\epsilon_{\mu\nu\sigma\rho}$ being the completely antisymmetric Levi-Civita tensor satisfying $\epsilon_{\mu\nu\sigma\rho}\epsilon^{\mu\nu\sigma\rho}=-4!$. The components of $F_{\mu\nu}$ correspond to the electric field $\boldsymbol{E}$ and the magnetic field $\boldsymbol{B}$, from which the electromagnetic invariants are expressed as
\[
X=\tfrac{1}{2}\left(\boldsymbol{E}^2-\boldsymbol{B}^2\right), 
\qquad 
Y=-\boldsymbol{E}\cdot\boldsymbol{B}.
\]
For the EH-NLED, the action takes the form:
\begin{eqnarray}
W_M=\frac{1}{4\pi}\int_{M_4}d^4x\sqrt{-g}\left(-X+\frac{2\alpha^2}{45m^4_e}\big\{4X^2+7Y^2\big\}\right),
\end{eqnarray}
where $m_e$ is the electron mass and $\alpha$ denotes the fine-structure constant. The parameter $\alpha$ characterizes the strength of quantum electrodynamical (QED) corrections in the presence of external electromagnetic fields~\cite{1916AnP...356....1S}. Its physical significance can be summarized as follows:
\begin{itemize}
    \item In the EH theory, $\alpha=e^2/(4\pi\epsilon_0\hbar c)\approx 1/137$, quantifying the strength of the electromagnetic interaction~\cite{1951PhRv...82..664S}.
    \item In the extended EEH framework, $\alpha$ governs the nonlinear corrections to Maxwell’s electrodynamics that stem from the production of virtual electron positron pairs. These effects become significant with field strengths approaching or exceeding the Schwinger critical value $E_c\approx 1.32\times10^{18}~\mathrm{V/m}$.
    \item More generally, $\alpha$ appears in the perturbative expansion of the QED effective Lagrangian, determining the contribution of higher-order nonlinear terms.
\end{itemize}
From the QED perspective, the EH-NLED encapsulates vacuum polarization effects, wherein virtual charged particle pairs act to screen the physical electric charge and, consequently, the associated rotation induced magnetic moment. This screening modifies the spacetime geometry only through the effective values, or ``screened,'' of the charges~\cite{2013PhRvD..88h5004R}. As discussed in~\cite{2022PhRvD.105j4046B}, the vacuum polarization contributions remain nearly constant, affecting primarily the effective electric charge in a manner analogous to the flat spacetime case~\cite{2010PhR...487....1R}. 
The corresponding solution to the Einstein field equations within this framework describes a rotating EEH BH), whose line element can be written as:

\begin{eqnarray}
    ds^2&=&-\left(1-\frac{2Mr-\Tilde{Q}^2}{\Sigma}\right)dt^2+\frac{\Sigma}{\Delta}dr^2 \nonumber \\
    &&-\frac{(2Mr-\Tilde{Q}^2)2a\sin^2{\theta}}{\Sigma}\,dt\,d\phi+\Sigma\,d\theta^2 \nonumber\\
    &&+\left(r^2+a^2+\frac{(2Mr-\Tilde{Q}^2)a^2\sin^2{\theta}}{\Sigma}\right)\sin^2{\theta}\,d\phi^2, \nonumber \\
    \Sigma&=&r^2+a^2\cos^2{\theta}, \nonumber\\
    \Delta&=&r^2+a^2-2Mr+\Tilde{Q}^2, \label{main metric}
\end{eqnarray}
which represents a Kerr–Newman-like BH with screened charge $\Tilde{Q}$. The effective charge is defined as:
\begin{eqnarray}
    \Tilde{Q}^2&=&Q^2\Bigg\{1-\beta\frac{Q^2 M^2}{\Sigma^2}\Bigg[1-\frac{4a^2\cos^2{\theta}}{\Sigma}\left(1-\frac{a^2\cos^2{\theta}}{\Sigma}\right)\nonumber\\
    &&\times\left(7-12\frac{a^2\cos^2{\theta}}{\Sigma}+12\frac{a^4\cos^4{\theta}}{\Sigma^2} \right)\Bigg]\Bigg\}, \label{hidden charge}
\end{eqnarray}
where $\beta$ is a dimensionless parameter depending only on the BH mass, given by:
\begin{equation}
    \beta=\frac{\alpha}{45\pi E_c^2M^2}\approx 1.85\times10^{8}\left(\frac{M_{\odot}}{M}\right)^2 ,\label{beta}
\end{equation}
with $E_c=m^2_ec^3/(e\hbar)$ denoting the critical electric field. Here, $Q^2$ refers to the scale $Q^2G/(4\pi\varepsilon_0 c^4)$ associated with the electric charge $Q$, and $2M$ represents the Schwarzschild radius.

\begin{figure*}[!htp]
  \vspace{1cm}
  \center
   \includegraphics[scale=0.42]{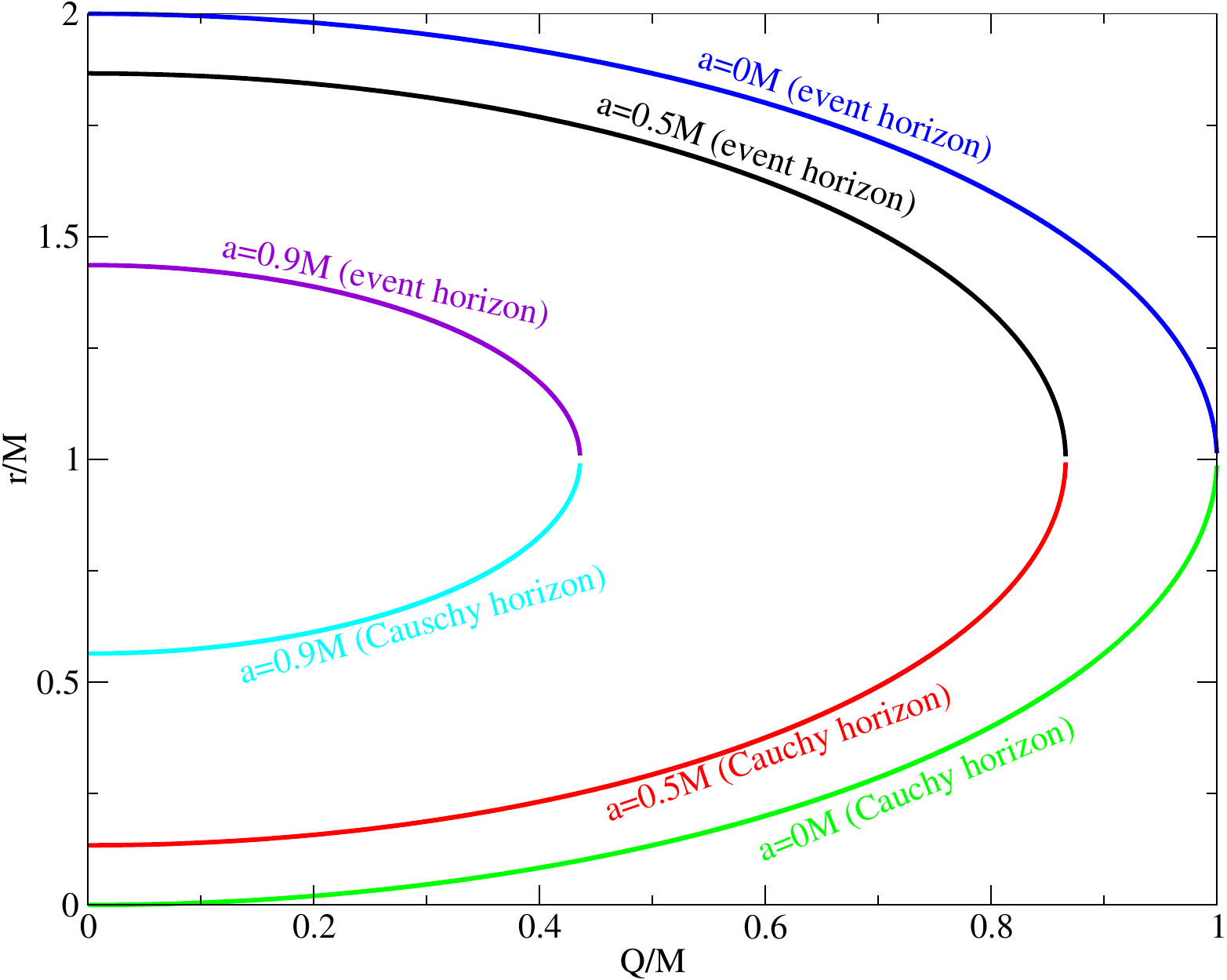}
     \caption{Variation of the horizon radii ($r_{\pm}/M$) in the EEH metric as a function of the charge-to-mass ratio ($Q$) for different values of the BH spin parameter ($a$). As $a$ increases, the admissible range of $Q$ that yields a regular BH solution becomes narrower, whereas values of $Q$ outside this range correspond to the formation of a naked singularity \label{fig1}.}
\vspace{1cm}
\label{QCBH_horizon}
\end{figure*}

For the EEH geometry, the location of the event horizon is determined by solving $\Delta(r,\Tilde{Q})=r^2+a^2-2Mr+\Tilde{Q}^2=0$. In case $M=1$, substituting $\Tilde{Q}$ from Eq.~\ref{hidden charge} and solving $\Delta(r,\Tilde{Q})=0$ gives the inner and outer horizons shown in Fig.~\ref{QCBH_horizon} for different values of the spin parameter $a$. As illustrated in Fig.~\ref{QCBH_horizon}, the existence of a classical BH depends sensitively on the value of the charge parameter $Q$. Specifically, the maximum allowed charge decreases with increasing rotation: for $a=0M$, one finds $Q_{\mathrm{max}}=1M$; for $a=0.5M$, the limit becomes $Q_{\mathrm{max}}=0.866M$, and for $a=0.9M$, it reduces further to $Q_{\mathrm{max}}=0.435M$. For each $a$, the values of $Q$ that exceed these thresholds correspond to naked singularities rather than BH.

\subsection{Circular orbits around rotating EEH BH} 

\begin{figure*}
\centering 
\includegraphics[width=\hsize]{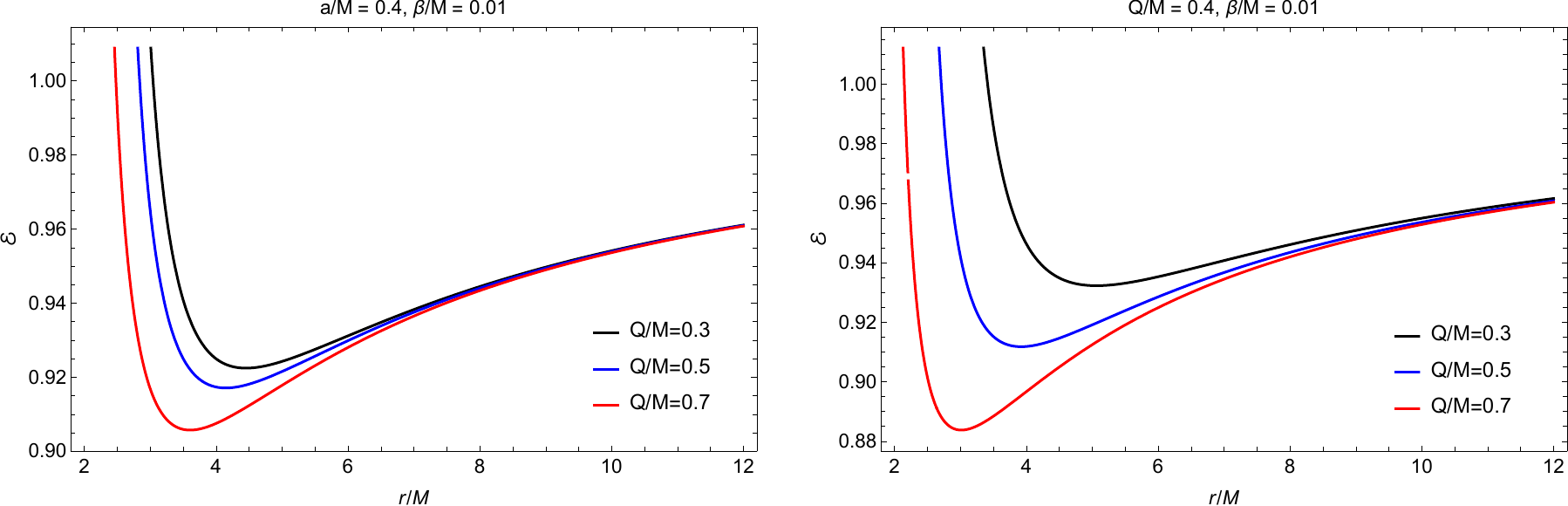}
\caption{Pictorial representation of energy of particles around rotating EEH BH\label{fig2}.}
\label{figENG}
\end{figure*}

\begin{figure*}
\centering 
\includegraphics[width=\hsize]{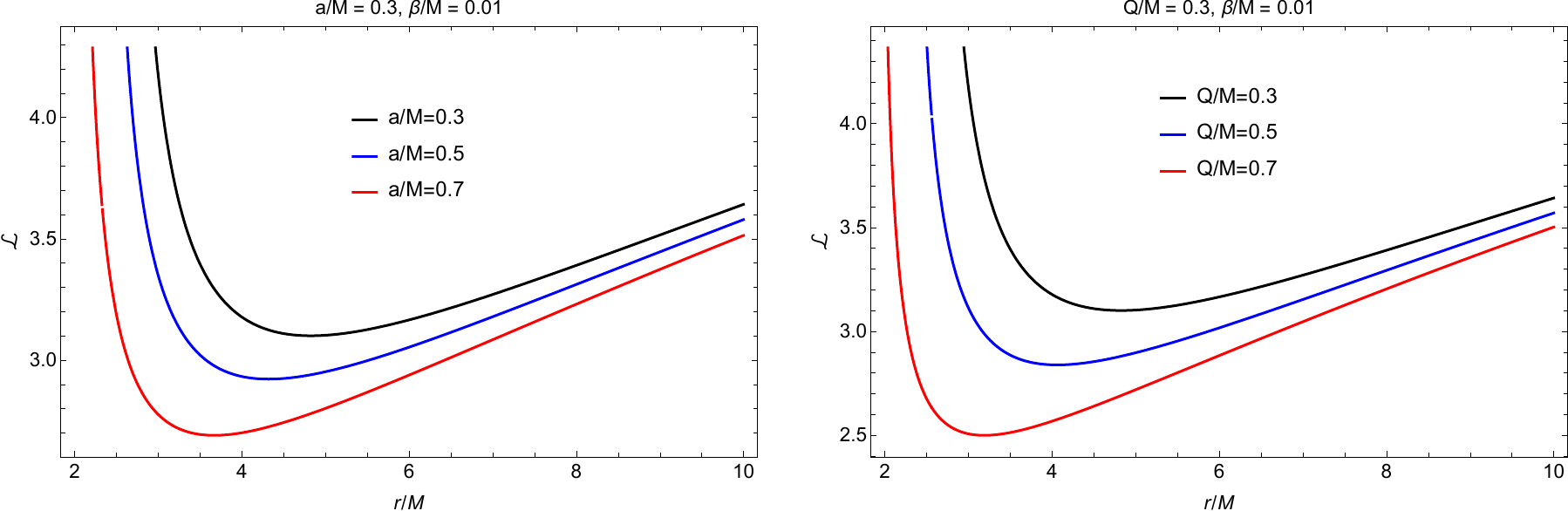}
\caption{Plots of angular momentum of particles around rotating EEH BH\label{fig3}.}
\label{fig_ANG}
\end{figure*}

The motion of a neutral particle can be discussed by the Hamiltonian, which is given as:
\beq\label{Ham}
H=\frac{1}{2}g^{\alpha \beta} p_{\alpha} p_{\beta} + \frac{1}{2}m^2,
\eeq
where $m$ implies the mass of the particle, $p^{\gamma}=m u^{\gamma}$ implies the four-momentum, $u^{\gamma}= d x^{\gamma}/d\tau$ denotes the four-velocity and $\tau$ is the appropriate time of the test particle. The equations governing the Hamiltonian dynamics can be formulated as:
\beq
\frac{dx^{\gamma}}{d\zeta}\equiv m u^{\gamma}=\frac{\partial H}{\partial p_{\gamma}}, \quad
\frac{d p_{\gamma}}{d\zeta} = -\frac{\partial H}{\partial x^{\gamma}},
\eeq
where $\zeta=\tau/m$ denotes the affine parameter. The inherent symmetries of the BH spacetime give rise to two conserved quantities, namely the specific energy $E$ and the specific angular momentum $L$, which are defined as:
\bea\label{EE}
\frac{p_{t}}{m}&=&g_{tt}u^{t}+g_{t\phi}u^{\phi}=-\mathcal{E}, \\\label{LL}
\frac{p_{\phi}}{m}&=&g_{\phi\phi}u^{\phi}+g_{t\phi}u^{t}=\mathcal{L},
\eea
where $\mathcal{E} = E/m$ and $\mathcal{L} = L/m$ represent the specific energy and specific angular momentum, respectively. The Hamiltonian (\ref{Ham}) describing the equatorial motion in the spacetime of a rotating EEH BH can thus be expressed as:
\begin{eqnarray}
H&=&\frac{1}{2 r^6 \left(a^2 r^4-\beta  Q^6+Q^2 r^4+(r-2) r^5\right)}\Big[a^4 p^{2}_{r} r^8\nonumber\\&+&a^2 r^4 \Big[-\mathcal{E}^2 \left(\beta  Q^6-Q^2 r^4+r^5 (r+2)\right)+p^{2}_{\theta } r^4\nonumber\\&-&2 \beta  p^{2}_{r} Q^6+2 p^{2}_{r} Q^2 r^4+2 p^{2}_{r} r^6-4 p^{2}_{r} r^5+r^6\Big]\nonumber\\&+&2 a \mathcal{E} \mathcal{L} r^4 \left(\beta  Q^6-Q^2 r^4+2 r^5\right)-\mathcal{E}^2 r^{12}\nonumber\\&+&\mathcal{L}^2 \left(-\beta  Q^6 r^4+Q^2 r^8+(r-2) r^9\right)-\beta  p^{2}_{\theta } Q^6 r^4\nonumber\\&+&p^{2}_{\theta } Q^2 r^8+p^{2}_{\theta } r^{10}-2 p^{2}_{\theta } r^9+\beta ^2 p^{2}_{r} Q^{12}\nonumber\\&-&2 \beta  p^{2}_{r} Q^8 r^4-2 \beta  p^{2}_{r} Q^6 r^6+4 \beta  p^{2}_{r} Q^6 r^5+p^{2}_{r} Q^4 r^8\nonumber\\&+&2 p^{2}_{r} Q^2 r^{10}-4 p^{2}_{r} Q^2 r^9+p^{2}_{r} r^{12}-4 p^{2}_{r} r^{11}\nonumber\\&+&4 p^{2}_{r} r^{10}-\beta  Q^6 r^6+Q^2 r^{10}+r^{12}-2 r^{11}\Big].
\end{eqnarray}
Moreover, the Hamiltonian formalism leads to the following equations of motion:

\bea
\frac{\mathrm{d} r}{\mathrm{d} \tau} &=& \frac{p_{r} \left(a^2-\frac{\beta  Q^6}{r^4}+Q^2+(r-2) r\right)}{r^2},\\\
\frac{\mathrm{d} \theta}{\mathrm{d} \tau} &=& \frac{p_\theta}{r^2},\\\
\frac{\mathrm{d} p_\theta}{\mathrm{d} \tau} &=& 0,\\\
\frac{\mathrm{d} \phi}{\mathrm{d} \tau} &=&\frac{1}{a^2 r^6-\beta  Q^6 r^2+Q^2 r^6+(r-2) r^7}\nonumber\\&\times&\Big[a \mathcal{E} \Big[\beta  Q^6-Q^2 r^4+2 r^5\Big]+\mathcal{L} \Big[-\beta  Q^6\nonumber\\&+&Q^2 r^4+(r-2) r^5\Big]\Big],\\\
        \frac{\mathrm{d} p_r}{\mathrm{d} \tau} &=&\frac{\hat{p}_{1}(r)}{r^7 \left(-\beta  Q^6+r^4 Q^2+(r-2) r^5+a^2 r^4\right)^2} ,
\eea

where $\hat{p}_{1}(r)$ is given in Appendix.

The specific energy and specific angular momentum of a test particle orbiting the rotating EEH BH are expressed as:

\begin{eqnarray}
\label{r19}
\mathcal{E}&&=\frac{1}{r^3 \left(r^8-a^2 \left(3 \beta  Q^6-Q^2 r^4+r^5\right)\right) \sqrt{\hat{p}_{3}(r)}}\nonumber\\&\times&\Big[-a^2 \left(3 \beta  Q^6 r^2-Q^2 r^6+r^7\right)+a \Big[\beta  Q^6-Q^2 r^4\nonumber\\&+&2 r^5\Big] \sqrt{3 \beta  Q^6-Q^2 r^4+r^5}-\beta  Q^6 r^4+Q^2 r^8\nonumber\\&+&(r-2) r^9\Big],\\\label{r20}
\mathcal{L}&&=\frac{1}{r^3 \left(r^8-a^2 \left(3 \beta  Q^6-Q^2 r^4+r^5\right)\right) \sqrt{\hat{p}_{3}(r)}}\nonumber\\&\times&\Big[-a^3 \left(3 \beta  Q^6 r^2-Q^2 r^6+r^7\right)+a^2 \Big[\beta  Q^6-Q^2 r^4\nonumber\\&+&r^5 (r+2)\Big] \sqrt{3 \beta  Q^6-Q^2 r^4+r^5}+a \Big[-4 \beta  Q^6 r^4\nonumber\\&+&2 Q^2 r^8-3 r^9\Big]+r^8 \sqrt{3 \beta  Q^6-Q^2 r^4+r^5}\Big],
\end{eqnarray}
where $\hat{p}_{3}(r)$ is given in Appendix.

Figure \ref{figENG} illustrates the variation of the specific energy of circular equatorial orbits around a rotating EEH BH. In the first column, the influence of the magnetic parameter $B$ is displayed, while the second column shows the effect of the BH spin parameter $a$. The results indicate that the energy of circular orbits increases with a growing magnetic parameter $B$, whereas the opposite trend is observed for increasing $a$. Specifically, circular orbits possess lower energy when the spin parameter $a$ is small. Furthermore, a comparison between the Kerr and rotating EEH BHs reveals that particles orbiting a Kerr BH exhibit smaller energy than those around a rotating EEH BH.

Figure \ref{fig_ANG} presents the behavior of the angular momentum of particles in circular orbits around a rotating EEH BH. The first column demonstrates the dependence on the magnetic parameter $B$, while the second column shows the influence of the spin parameter $a$. The results indicate that the angular momentum of circular orbits increases with increasing $B$, while an opposite trend is observed for increasing $a$. In particular, circular orbits exhibit smaller angular momentum when the spin parameter $a$ is large. Additionally, the angular momentum grows with the radial distance $r$. A comparative analysis further shows that particles orbiting a rotating Kerr BH possess lower angular momentum than those in the rotating EEH BH.

\begin{figure*}
\centering 
\includegraphics[width=\hsize]{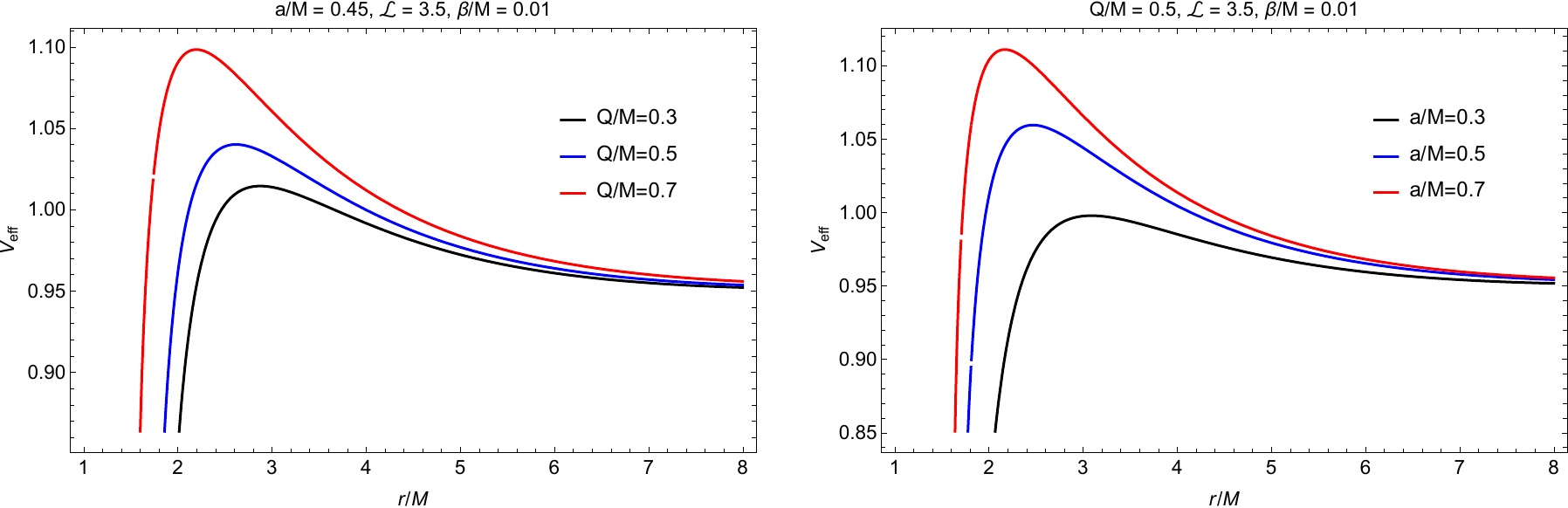}
\caption{Effective potential for particles around rotating EEH BH\label{fig4}.
}\label{fig_eff}
\end{figure*}

\subsection{Effective potential}

Using the normalization condition $g_{\nu \sigma} u^\nu u^\sigma = -1$, one can write
\beq
V_{eff}(r, \theta) = g_{rr}\, \Dot{r}^{2} + g_{\theta \theta}\, \Dot{\theta}^{2},
\eeq
where $\Dot{r} = dr/d\tau$, $\Dot{\theta} = d\theta/d\tau$, and $V_{eff}$ denotes the effective potential defined by the following relation 
\beq
V_{eff}(r, \theta) = \frac{\EE^2 g_{\phi \phi} + 2 \EE \LL g_{t \phi} + \LL^2 g_{tt}}{g_{t\phi}^{2} - g_{tt} g_{\phi\phi}} - 1.
\eeq
For the current study, $V_{eff}(r, \theta) $ is calculated as
\bea\non
V_{eff}(r) &=&\frac{1}{a^2 \left(\beta  Q^6-Q^2 r^4+r^5 (r+2)\right)+r^8}\Big[a^2 r^6 \hat{p}_{2}(r)\nonumber\\&+&r^2 \left(-\beta  Q^6+Q^2 r^4+(r-2) r^5\right) \hat{p}_{2}(r)\nonumber\\&+&a \mathcal{L} \left(\beta  Q^6-Q^2 r^4+2 r^5\right)\Big],
\eea
where $\hat{p}_{2}(r)$ is given in Appendix.

To analyze the motion of test particles, the effective potential $V_{\text{eff}}(r, \theta)$ serves as an essential tool. It characterizes the dynamics of particle trajectories without the explicit use of the equations of motion. The conditions for circular orbits confined to the equatorial plane $(\theta = \pi/2)$ can be obtained as follows:
\beq
V_{\rm eff}(r) = 0, \quad \frac{\d V_{\rm eff} (r)}{\d r} = 0.\label{Veff-1}
\eeq
The extrema of the effective potential correspond to the locations of circular orbits, where the minima represent stable configurations, and the maxima indicate unstable ones. In Newtonian mechanics, for a given value of angular momentum, the effective potential possesses a single minimum, corresponding to a stable ISCO. However, in more intricate scenarios, where additional factors such as the particle's spin, rotational momentum, or other physical parameters affect the potential, the position of these orbits may shift accordingly. Within the framework of GR, for a fixed angular momentum, the effective potential near a Schwarzschild BH admits two extremal points that define the possible circular trajectories. In the present analysis, the evaluated angular momentum plays a key role in determining the position of the ISCO. Figure~\ref{fig_eff} illustrates the variation of the effective potential $V_{\text{eff}}$ as a function of the radial coordinate $r$ for different values of the BH parameters. The first column displays the effect of the magnetic parameter $B$, while the second column shows the influence of the rotation parameter $a$. Interestingly, these parameters exhibit opposite trends: an increase in the magnetic parameter $B$ lowers the minima of $V_{\text{eff}}$, whereas an increase in the rotation parameter $a$ raises them. Furthermore, the minimum of $V_{\text{eff}}$ for a rotating Kerr BH lies above that of a rotating EEH BH, highlighting the distinct influence of non-linear electrodynamic effects.

\subsection{Effective force} 

\begin{figure*}
\centering 
\includegraphics[width=\hsize]{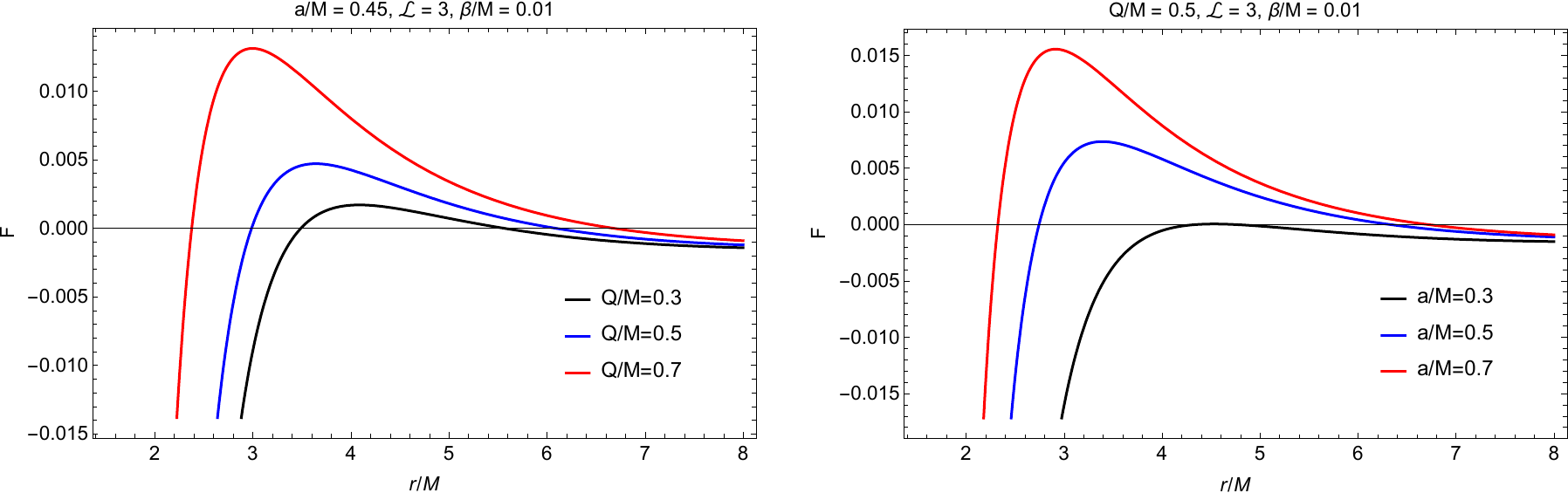}
\caption{Variation of the effective force acting on test particles around a rotating EEH BH for different values of the magnetic parameter $B$ and spin parameter $a$.}\label{figForce}
\end{figure*}

The effective force acting on a test particle plays a crucial role in determining its motion relative to the BH, indicating whether the particle experiences attraction toward or repulsion away from it. In this work, we analyze the particle dynamics in the spacetime of a rotating EEH BH, where both attractive and repulsive gravitational effects may arise. The expression for the effective force is obtained from Eq.~(\ref{Veff-1}) as   
\beq
F = -\frac{1}{2}\frac{dV_{\text{eff}}}{dr}.
\eeq  
Figure~\ref{figForce} shows the variation of the effective force with respect to the radial coordinate $r$ for different values of the BH parameters. The first column illustrates the influence of the magnetic parameter $B$, while the second column illustrates the effect of the spin parameter $a$. For lower values of $B$, the effective force remains predominantly attractive, while increasing $B$ improves the overall strength of the force. In contrast, the spin parameter $a$ exhibits the opposite behavior. Moreover, the effective force acting on particles in the spacetime of a rotating Kerr BH is weaker than that in the rotating EEH case, emphasizing the significance of NLED corrections.

\begin{figure*}
\centering 
\includegraphics[width=\hsize]{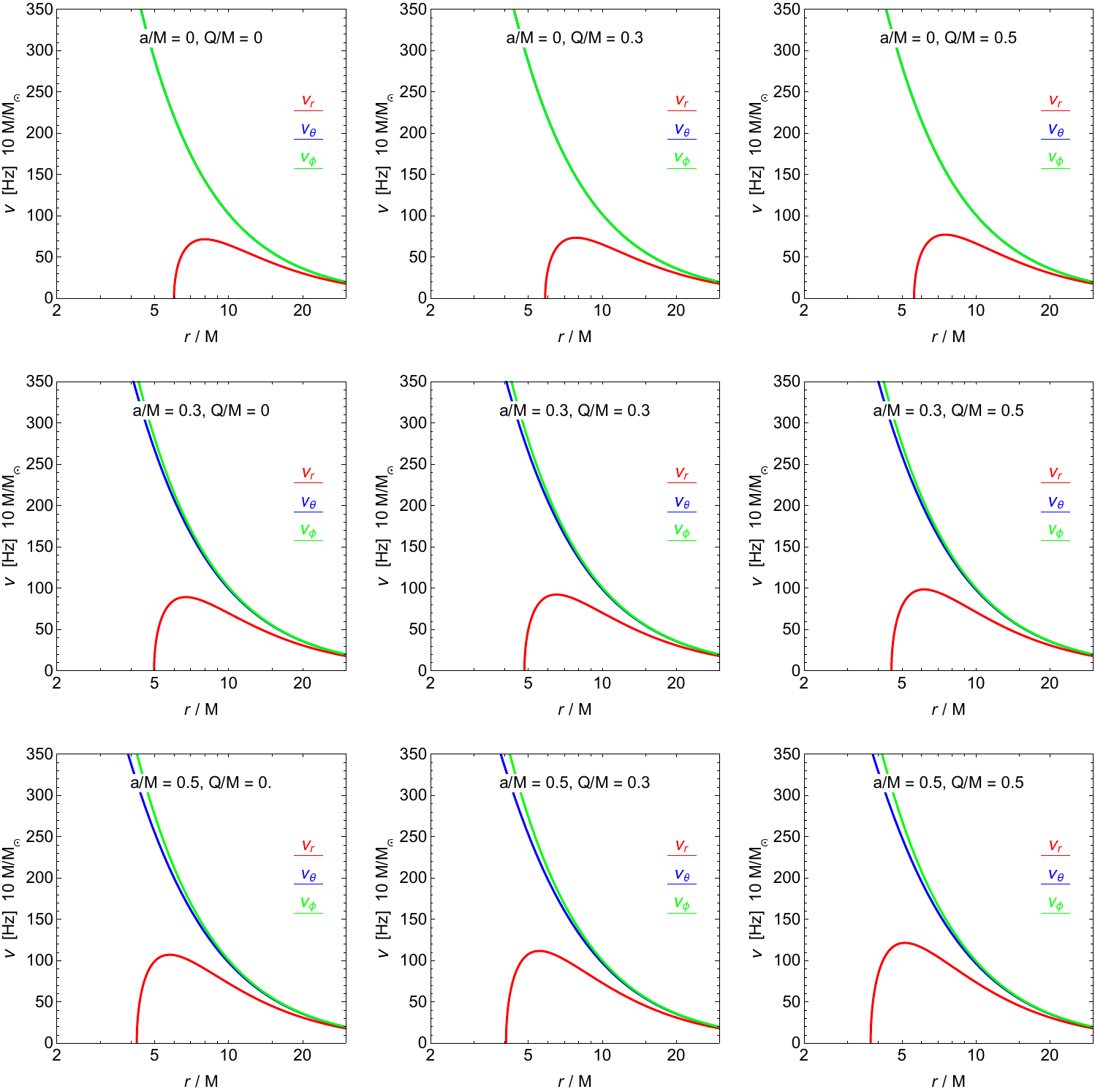}
\caption{Plot for fundamental frequencies of particles moving around a rotating EEH BH\label{fig6}.
}\label{figFRQ}
\end{figure*}

\begin{figure*}
\centering 
\includegraphics[width=\hsize]{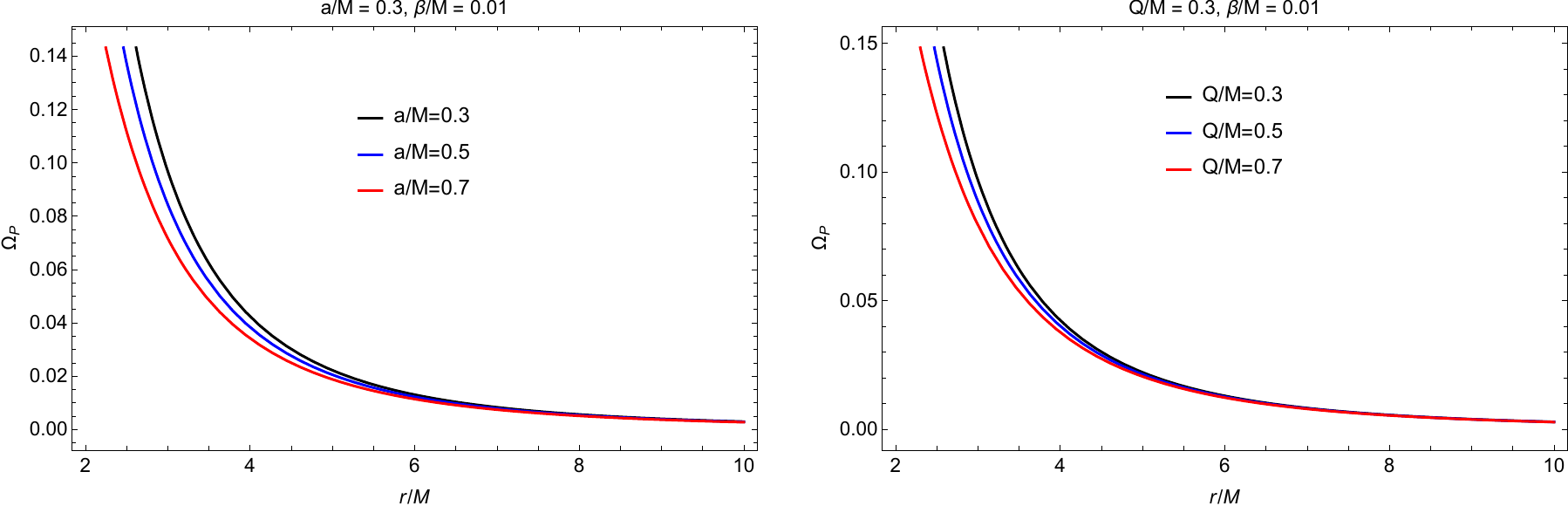}
\caption{Plots for periastron frequency of particles around rotating EEH BH\label{fig7}.}\label{fig_periastron}
\end{figure*}

\begin{figure*}
\centering 
\includegraphics[width=\hsize]{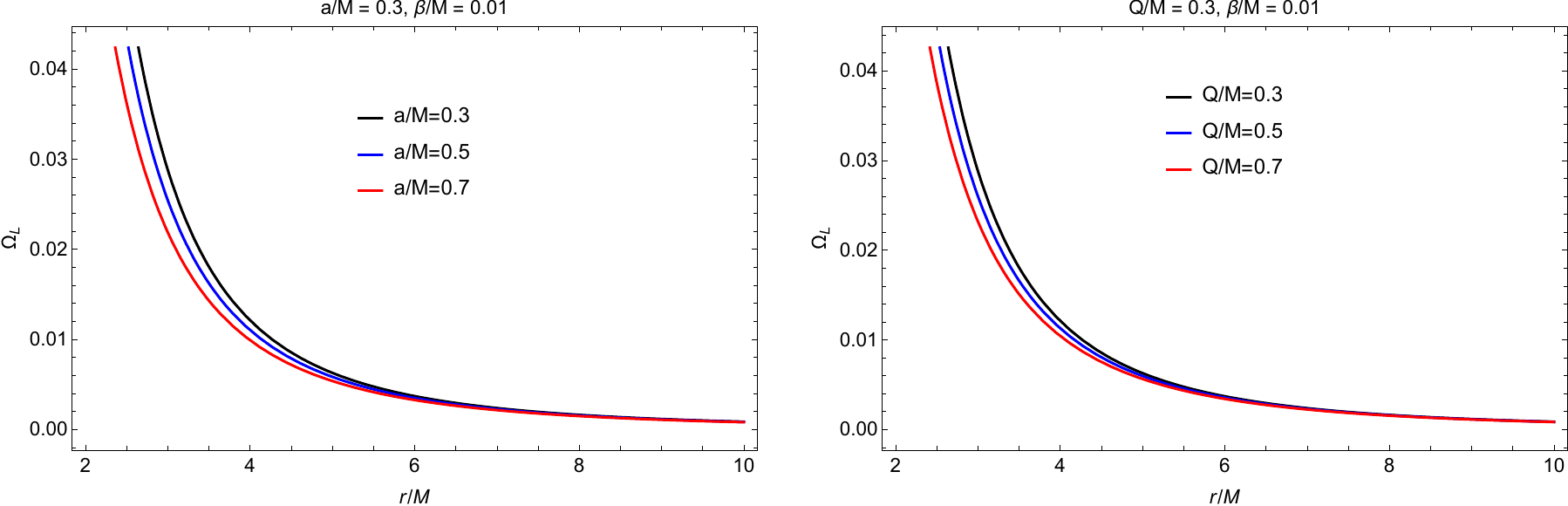}
\caption{Plots for Lense-Thirring precession frequency around rotating EEH BH\label{fig8}.}\label{fig_LT}
\end{figure*}

\section{Harmonic oscillations as perturbation of circular orbits} \label{oscillations}

To investigate the oscillatory motion of neutral test particles, we perturb the equations of motion in the vicinity of stable circular orbits. When a particle is slightly displaced from its equilibrium position along a circular trajectory in the equatorial plane, it undergoes an epicyclic motion characterized by small harmonic oscillations. The corresponding oscillation frequencies, as measured by a local observer, are given by:

\bea\label{Freq-2}
\omega_{r}^{2} &=&  \frac{-1}{2\, g_{rr}} \frac{\partial^{2} V_{\rm eff} (r, \theta)}{\partial r^{2}},\\\label{Freq-3}
\omega_{\theta}^{2} &=& \frac{-1}{2\, g_{\theta \theta}} \frac{\partial^{2} V_{\rm eff}(r, \theta)}{\partial \theta^{2}},\\\label{Freq-4}
\omega_\phi &=& \frac{\d \phi}{\d \tau}.
\eea

Analyzing the behavior of the fundamental frequencies \(\omega_r\), \(\omega_\theta\), and \(\omega_\phi\), as well as their ratios, provides deep insight into the geometry of the epicyclic motion of particles around stable circular orbits. In Newtonian gravity, all three frequencies coincide, leading to perfectly elliptical trajectories for particles orbiting spherically symmetric masses. However, for a Schwarzschild BH, the relationship \(\omega_r < \omega_\theta = \omega_\phi\) is valid. This disparity leads to a periapsis shift and induces relativistic precession as the orbital radius decreases, bringing the particle into regions of stronger gravitational influence.

\subsection{Frequencies measured by distant observer} 

The locally measured angular frequencies $\omega_\alpha$ are presented in Eqs.~(\ref{Freq-2})--(\ref{Freq-4}), while the corresponding angular frequencies as measured by a distant static observer, denoted by $\Omega$, can be expressed as:
\beq\label{frequencies}
\Omega_{\alpha} = \omega_{\alpha} \frac{\d \tau}{\d t},
\eeq
where $\d \tau/\d t$ is the redshift coefficient. Using Eqs. (\ref{EE}) and (\ref{LL}), we found
\begin{equation}
    \frac{\d t}{\d \tau} = - \frac{E g_{\phi \phi} + L g_{t \phi}}{g_{tt} g_{\phi \phi} - g_{t \phi}^2}.
\end{equation}
When the frequencies of small harmonic oscillations are expressed in physical units as measured by a distant observer, their corresponding dimensionless forms are obtained by scaling with the factor $c^{3}/(G M)$, where $G$ denotes the gravitational constant, $c$ is the speed of light, and $M$ represents the mass of the BH. Accordingly, the oscillation frequencies of neutral particles, as perceived by distant observers, can be written as:
\beq\label{nu_rel}
\nu_{j}=\frac{1}{2\pi}\frac{c^{3}}{GM} \, \Omega_{j}[{\rm Hz}].
\eeq
Here, \(j \in \{r, \theta, \phi\}\), \(\Omega_r\), \(\Omega_\theta\), and \(\Omega_\phi\) represent the dimensionless angular frequencies, as measured by a distant observer, for the radial, latitudinal, and axial components, respectively. For rotating EEH BH, the expressions for $\Omega_{\alpha}$ take the form
\bea\label{nu_r}
\Omega_{r}^{2} &=&\frac{1}{\hat{p}_{4}(r)}\Big[a^6 \mathcal{E}^2 r^8 \left(-21 \beta  Q^6+3 Q^2 r^4-2 r^5\right)\nonumber\\&+&2 a^5 \mathcal{E} \mathcal{L} r^8 \left(21 \beta  Q^6-3 Q^2 r^4+2 r^5\right)\nonumber\\&+&a^4 r^4 \hat{p}_{5}(r)-2 a^3 \mathcal{E} \mathcal{L} r^4 \Big[8 \beta ^2 Q^{12}-30 \beta  Q^8 r^4\nonumber\\&+&\beta  Q^6 r^5 (72-53 r)+6 Q^4 r^8+Q^2 r^9 (9 r-20)\nonumber\\&-&6 (r-2) r^{10}\Big]+a^2 \hat{p}_{6}(r)+2 a \mathcal{E} \mathcal{L} \hat{p}_{7}(r)-\mathcal{E}^2 r^8 \nonumber\\&\times& \hat{p}_{8}(r)-3 \mathcal{L}^2 \left(\beta  Q^6-Q^2 r^4-(r-2) r^5\right)^3\Big]
,\\\label{nu_theta}
\Omega_{\theta}^{2} &=&\frac{1}{\hat{p}_{9}(r)}\Big[a^6 \mathcal{E}^2 r^4 \left(31 \beta  Q^6-Q^2 r^4+2 r^5\right)\nonumber\\&+&2 a^5 \mathcal{E} \mathcal{L} r^4 \left(-31 \beta  Q^6+Q^2 r^4-2 r^5\right)+a^4 \hat{p}_{10}(r)\nonumber\\&+&2 a^3 \mathcal{E} \mathcal{L} \hat{p}_{11}(r)-a^2 \hat{p}_{12}(r)+\mathcal{L}^2 r^2 \Big[-\beta  Q^6\nonumber\\&+&Q^2 r^4+(r-2) r^5\Big]^2\Big]
,\\\label{nu_phi}
\Omega_\phi &=& \Big[\frac{1}{r^8-a^2 \left(3 \beta  Q^6-Q^2 r^4+r^5\right)}\nonumber\\&\times&\Big[r^4 \sqrt{3 \beta  Q^6-Q^2 r^4+r^5}-a \Big[3 \beta  Q^6\nonumber\\&-&Q^2 r^4+r^5\Big]\Big]\Big]^2,
\eea

where $\hat{p}_{i}(r),\;i=4 \dots 12$ are given in Appendix.

Figure~\ref{figFRQ} illustrates the radial variation of the oscillation frequencies $\nu_j$ corresponding to small harmonic motions of neutral particles around a rotating EEH BH for different values of the spin parameter $a$ and the magnetic parameter $B$, measured by a distant observer. The frequency profiles shift closer to the event horizon as the magnetic parameter $B$ increases. In contrast, the BH’s rotation parameter $a$ exhibits the opposite effect: as the rotation rate increases, the frequency profiles move outward, away from the event horizon.

\subsection{Periastron and Lense-Thirring precession} 

In this subsection, we discuss the Lense-Thirring precession frequency and the periapsis precession of a neutral test particle that is slightly perturbed from the equatorial plane $(\theta = \pi/2)$ while orbiting a rotating EEH BH. When the particle experiences a small deviation from its stable circular orbit, it undergoes oscillations about the equilibrium position with a characteristic radial frequency $\Omega_r$, which enables the determination of the periapsis precession. The Lense-Thirring precession frequency $\Omega_{LT}$ is defined as the difference between the orbital frequency $\Omega_\phi$ and the vertical (latitudinal) frequency $\Omega_\theta$, whereas the periapsis precession frequency $\Omega_P$ is given by the difference between the orbital frequency $\Omega_\phi$ and the radial frequency $\Omega_r$, expressed as:
\beq
\Omega_{P} = \Omega_{\phi} - \Omega_{r},
\eeq
\beq
\Omega_{L} = \Omega_{\phi} - \Omega_{\theta}.
\eeq
Figure~\ref{fig_periastron} presents the radial profiles of the periapsis precession frequency for a rotating EEH BH as a function of the radial coordinate $r$, considering different values of the BH parameters. The first row illustrates the variation of the periapsis precession frequency for different values of the magnetic parameter $B$, while the second column corresponds to different values of the rotation parameter $a$. As the magnetic parameter $B$ increases, the frequency of the periapsis precession decreases. In contrast, an increase in the rotation parameter $a$ leads to a higher frequency of periapsis precession. It is also observed that the frequency of periapsis around a rotating EEH BH is greater than that corresponding to a Kerr BH.  

Figure~\ref{fig_LT} depicts the behavior of the lense-three precession frequency around a rotating EEH BH. The first column illustrates its dependence on the BH's rotation parameter $a$, whereas the second column shows the variation with respect to the magnetic parameter $B$. The Lense-Thirring precession frequency decreases with both increasing rotation parameter $a$ and magnetic parameter $B$. Moreover, the precession frequency in the rotating EEH BH spacetime is found to be smaller than that in the Kerr geometry.


\section{Simulation Results}\label{S3}

The investigation of physical phenomena that may occur around an EEH BH, the analytical characterization of the parameters influencing these processes, and the theoretical calculation of the epicyclic frequencies are of great significance in astrophysics. In particular, revealing the impact of the BH spin parameter and charge parameter on these frequencies and on the structure of stable-unstable orbits is a critical study, both for testing the predictions of GR and for comparing theoretical results with observational data.

In addition to these theoretical analyzes, in this work we numerically examine the plasma structures formed by Bondi-Hoyle-Lyttleton (BHL) accretion around a rotating EEH BH, as well as the dynamical features of the resulting shock cone. By investigating the influence of the spin parameter and the charge parameter on the morphology of the system within the shock cone region, we compute the cavities formed inside the cone and the fundamental oscillation modes trapped within these cavities. Furthermore, the nonlinear coupling mechanisms among these modes are analyzed numerically, thereby providing theoretical predictions that can be directly compared with oscillatory behaviors observed in real astrophysical systems. This approach makes a direct contribution to testing the EEH gravity theory in an observational astrophysics context. 

To perform the numerical simulations, the General Relativistic Hydrodynamics (GRH) equations are solved using the fixed spacetime metric of the EEH BH. In this way, the systematic effects of different values of $a$ and $Q$ on the dynamics of the accreting fluid and on the evolution of the shock cone are revealed. For the complete numerical solution of the GRH equations, we employ high-resolution shock capturing (HRSC) schemes (see for details \cite{Font2000LRR,Orh1,Donmez2}). These methods enable the accurate modeling of sharp gradients that occur near shock fronts, ensuring a reliable analysis of plasma dynamics. Moreover, to carry out the solution, the fixed spacetime metric of EEH gravity given in Eq.\ref{main metric} is used, and the GRH equations are solved numerically by applying the appropriate initial and boundary conditions provided in \cite{Donmez2024MPLA}. 

Hence, the numerically obtained results below can then be compared with observational data, and the theoretical calculations presented here thereby contribute to the astrophysical testing of EEH gravity.

\subsection{Numerical Investigation of $Q$-$a$ Effects on Accretion Dynamics}\label{S3-1}

Here, we reveal how the EEH charge parameter $Q$ and the BH spin parameter $a$ influence the BHL accretion mechanism. Since the accretion process not only governs the formation of a disk around the BH but also determines the physical properties of possible phenomena occurring in its vicinity, understanding this interplay is crucial. By numerically modeling BHL accretion and the behavior of matter around the BH, we demonstrate how the parameters of the EEH BH affect the morphology of the resulting shock cone, its dynamical structure, and the instabilities that arise within it.

\begin{figure*}[!htp]
  \vspace{1cm}
  \center
     \includegraphics[width=10.0cm,height=5.8cm]{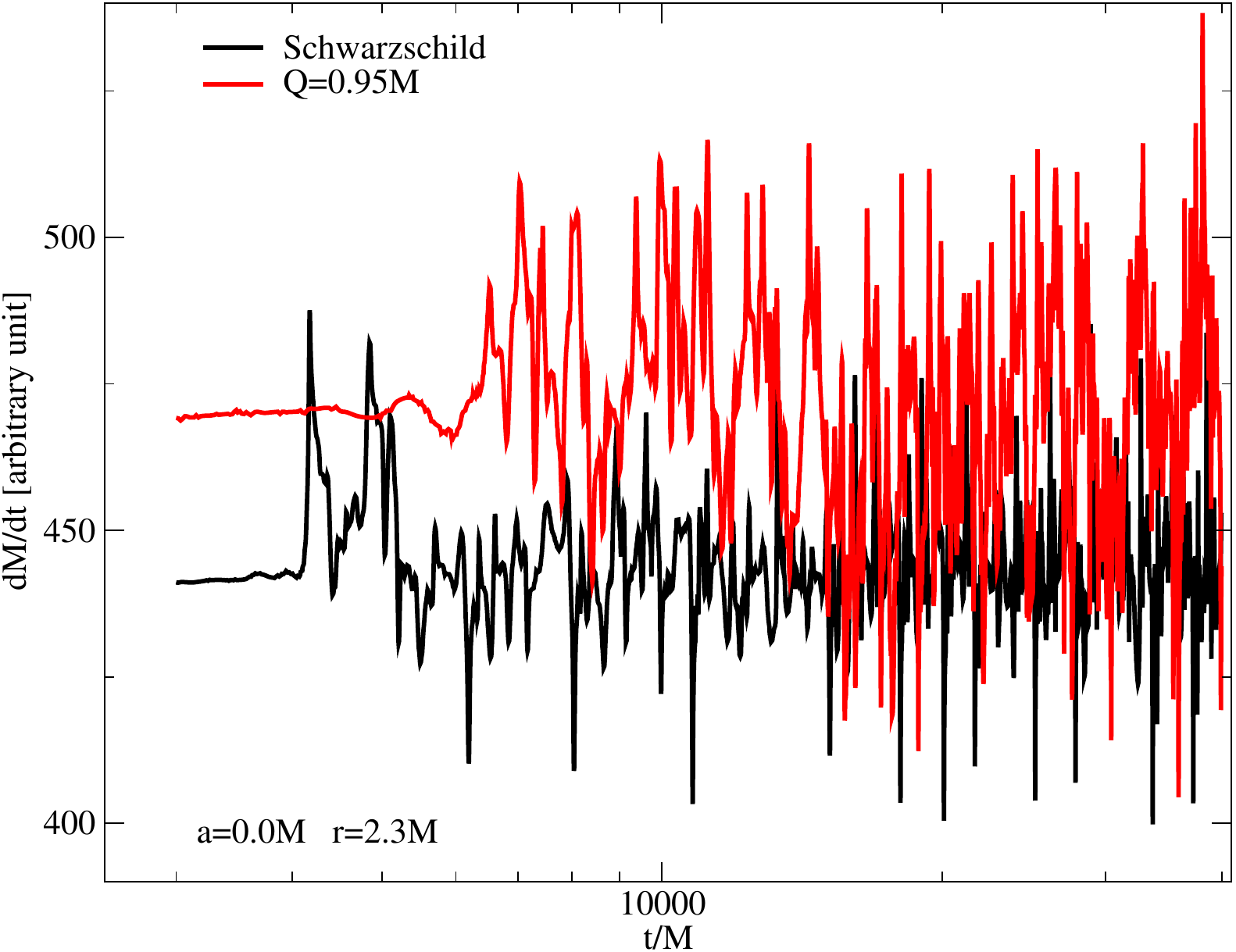} \\
  \vspace{0.3cm}
     \includegraphics[width=10.0cm,height=5.8cm]{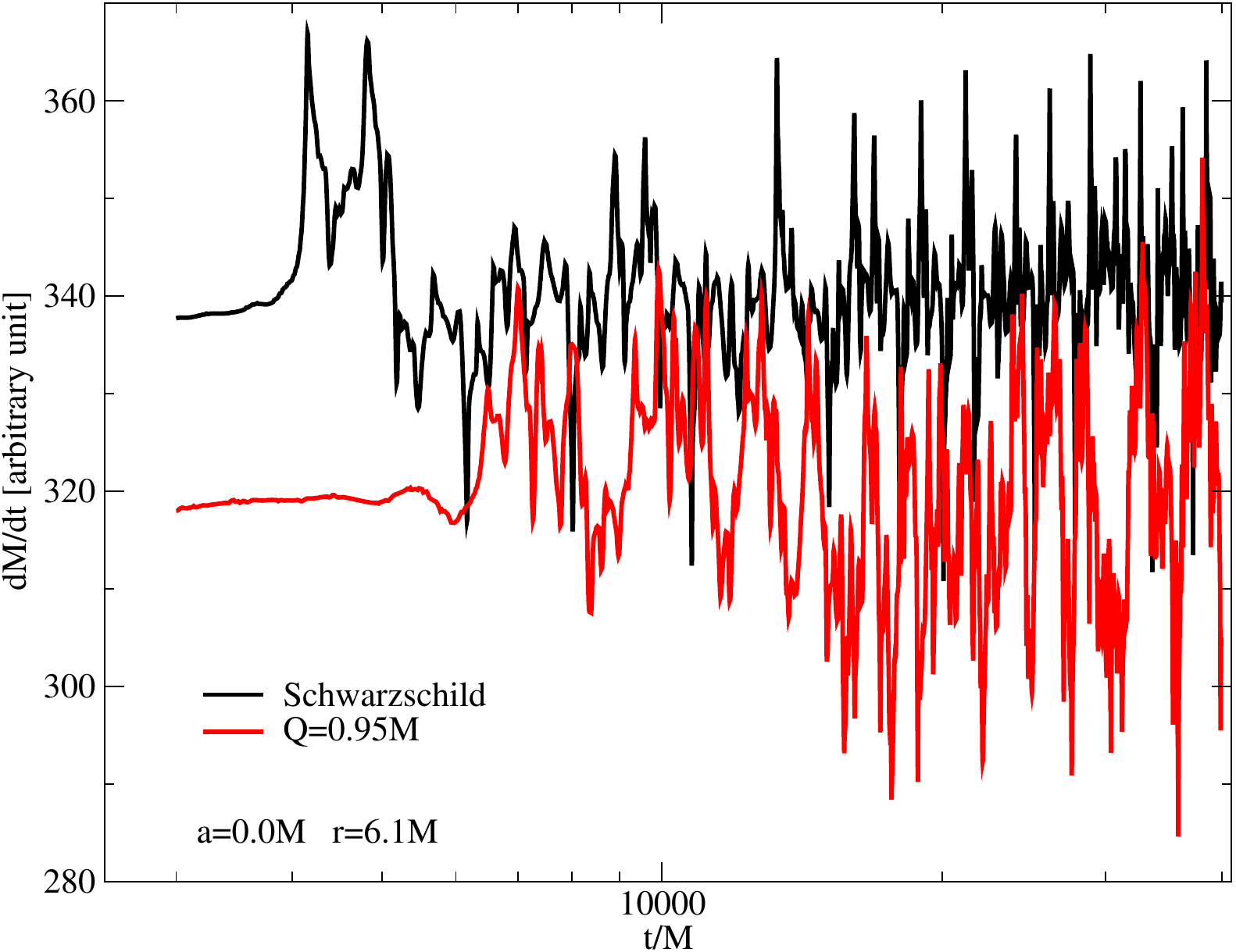}\\
  \vspace{0.3cm}  
     \includegraphics[width=10.0cm,height=5.8cm]{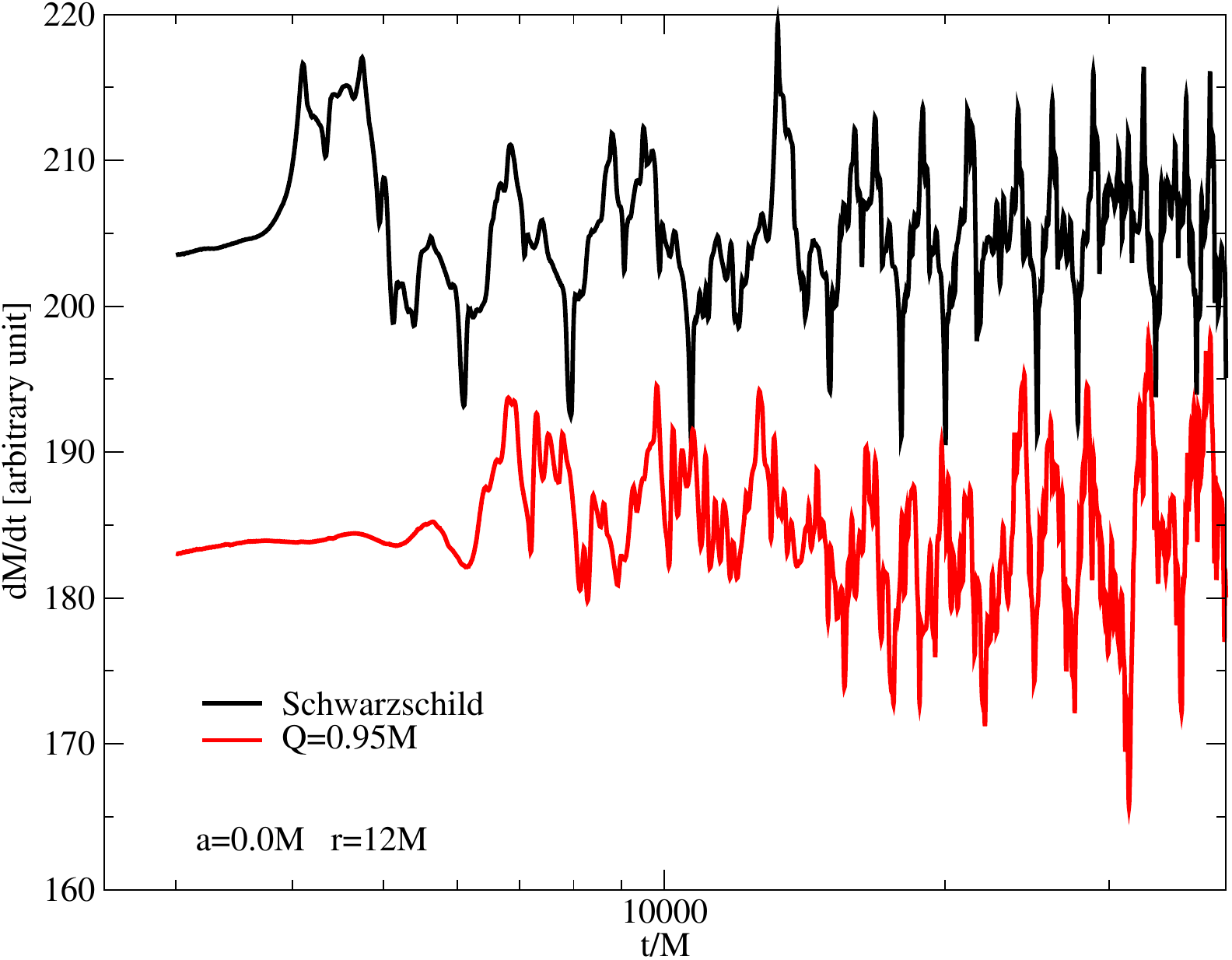}
     \caption{Time evolution of the mass accretion rate at three radial positions near the EEH BH for the non-rotating case, shown together with the Schwarzschild reference. At all radii, the plasma morphology and the associated shock cone exhibit strong instabilities. The EEH charge parameter $Q$ enhances the accretion rate in the strong-field region ($r = 2.3M$) compared to Schwarzschild, while at larger radii ($r = 6.1M$ and $r = 12M$) it leads to a suppression of accretion efficiency\label{fig9}.
    }
\vspace{1cm}
\label{Mass_acc_a00}
\end{figure*}

In the non-rotating case, the effect of the EEH charge parameter $Q$ on the accretion mechanism, and its comparison with the Schwarzschild BH, is shown in Fig.\ref{Mass_acc_a00}. In the top panel of Fig.\ref{Mass_acc_a00}, the mass accretion rate of the matter accreted closest to the horizon is calculated, i.e., at the inner boundary of the computational domain at $r = 2.3M$. As seen here, in the strong gravitational field, the accretion rate of matter falling toward the BH for $Q = 0.95M$ is higher compared to the Schwarzschild case. This amplifies the density and velocity gradients within the shock cone, intensifying instabilities. The growth of such instabilities significantly increases the observability of fundamental QPOs modes trapped within the cavity of the shock cone. However, as shown in the middle and bottom panels of Fig.\ref{Mass_acc_a00}, at $r = 6.1M$ and $r = 12M$ regions where the gravitational field is weaker, the accretion rate is suppressed compared to the Schwarzschild case. This suppression leads to a weakening of the turbulence inside the shock cone. Consequently, the effectiveness of the resulting QPOs, that is, their amplitudes, decreases, reducing their detectability. These results indicate that even in the absence of BH spin, the observability of QPOs generated in the strong-field region is enhanced by the EEH charge parameter $Q$, while at larger radii the behavior becomes more similar to the Schwarzschild case.

\begin{figure*}[!htp]
  \vspace{1cm}
  \center
   \includegraphics[width=10.0cm,height=5.8cm]{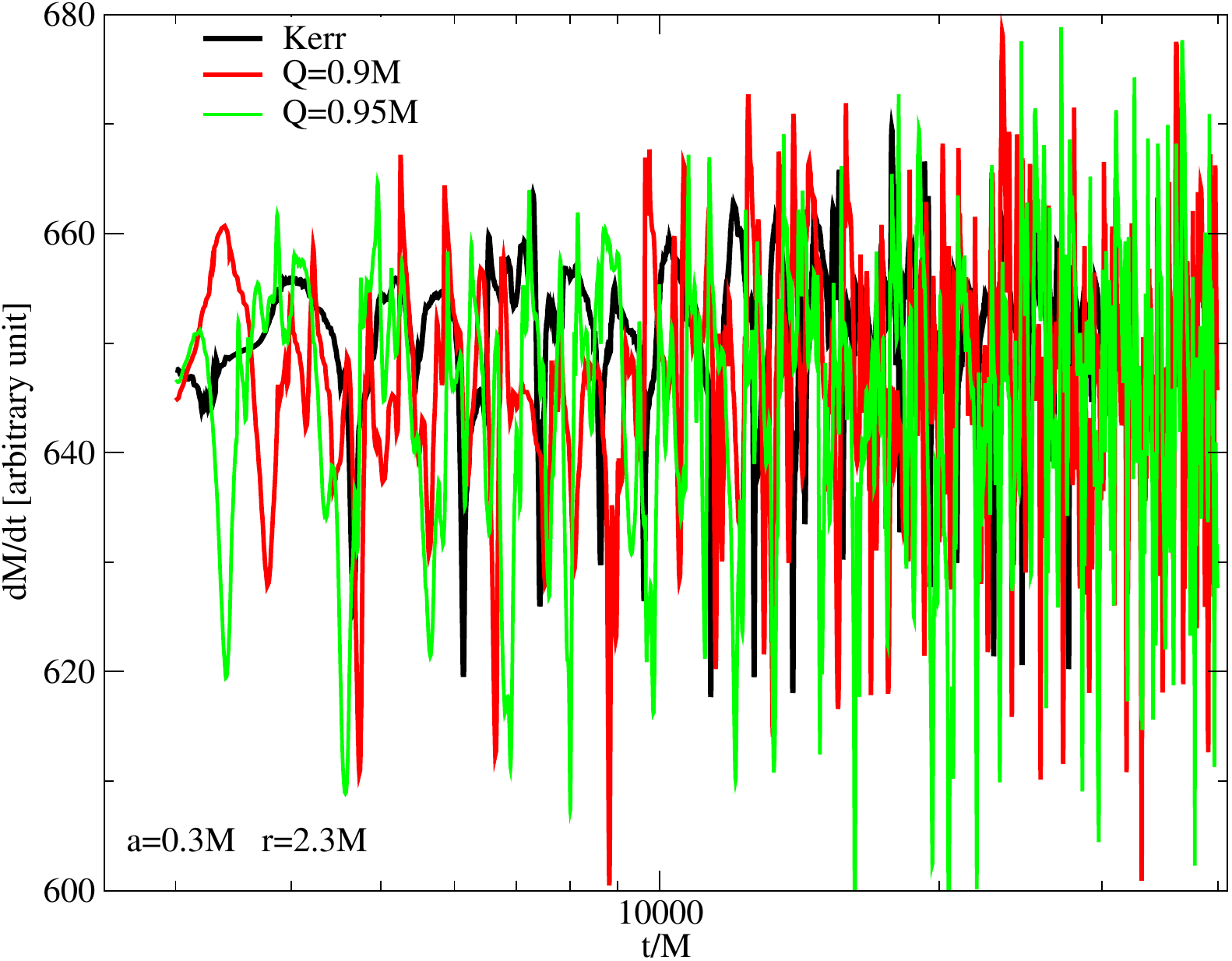} \\
  \vspace{0.3cm}  
  \includegraphics[width=10.0cm,height=5.8cm]{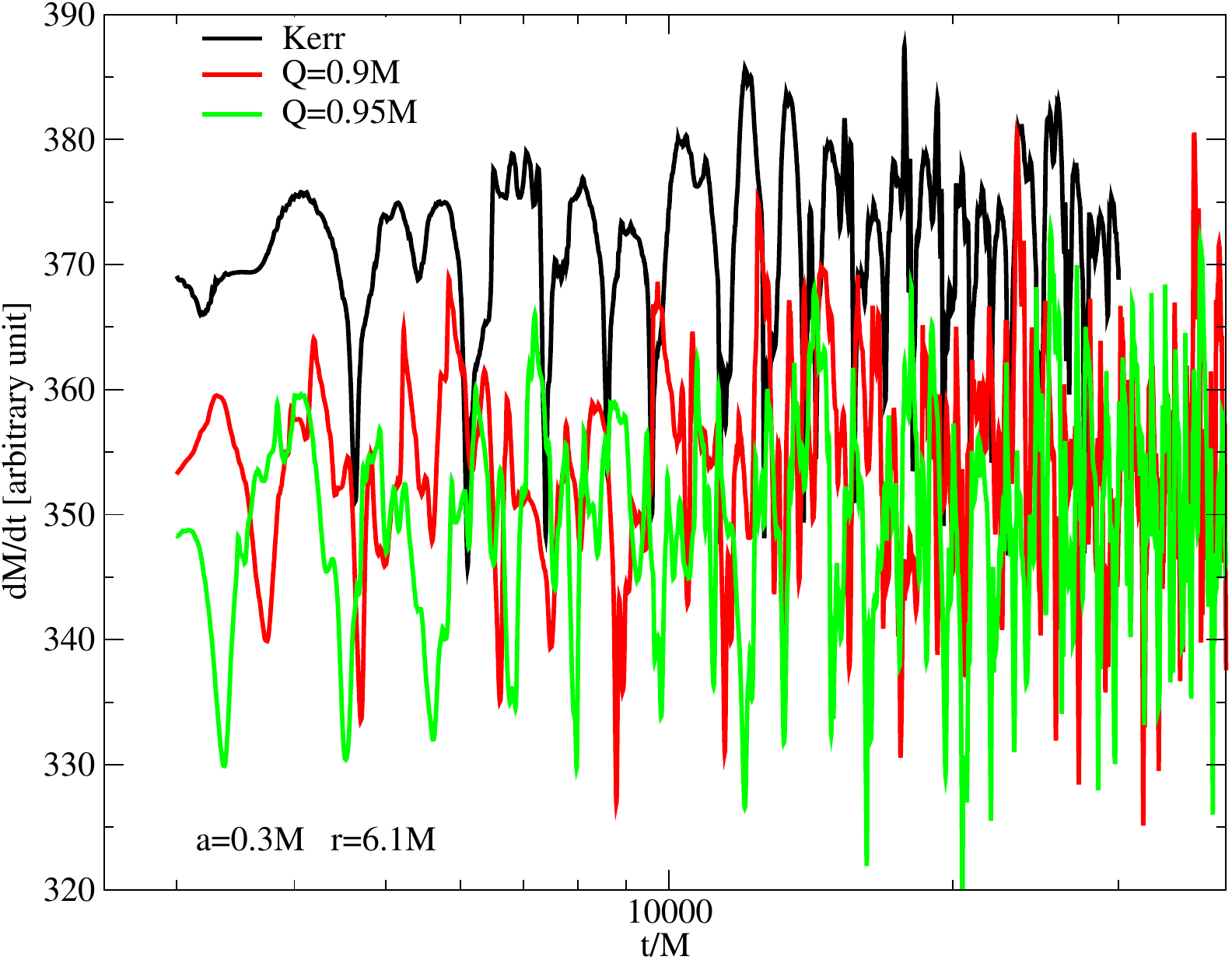} \\
  \vspace{0.3cm}
  \includegraphics[width=10.0cm,height=5.8cm]{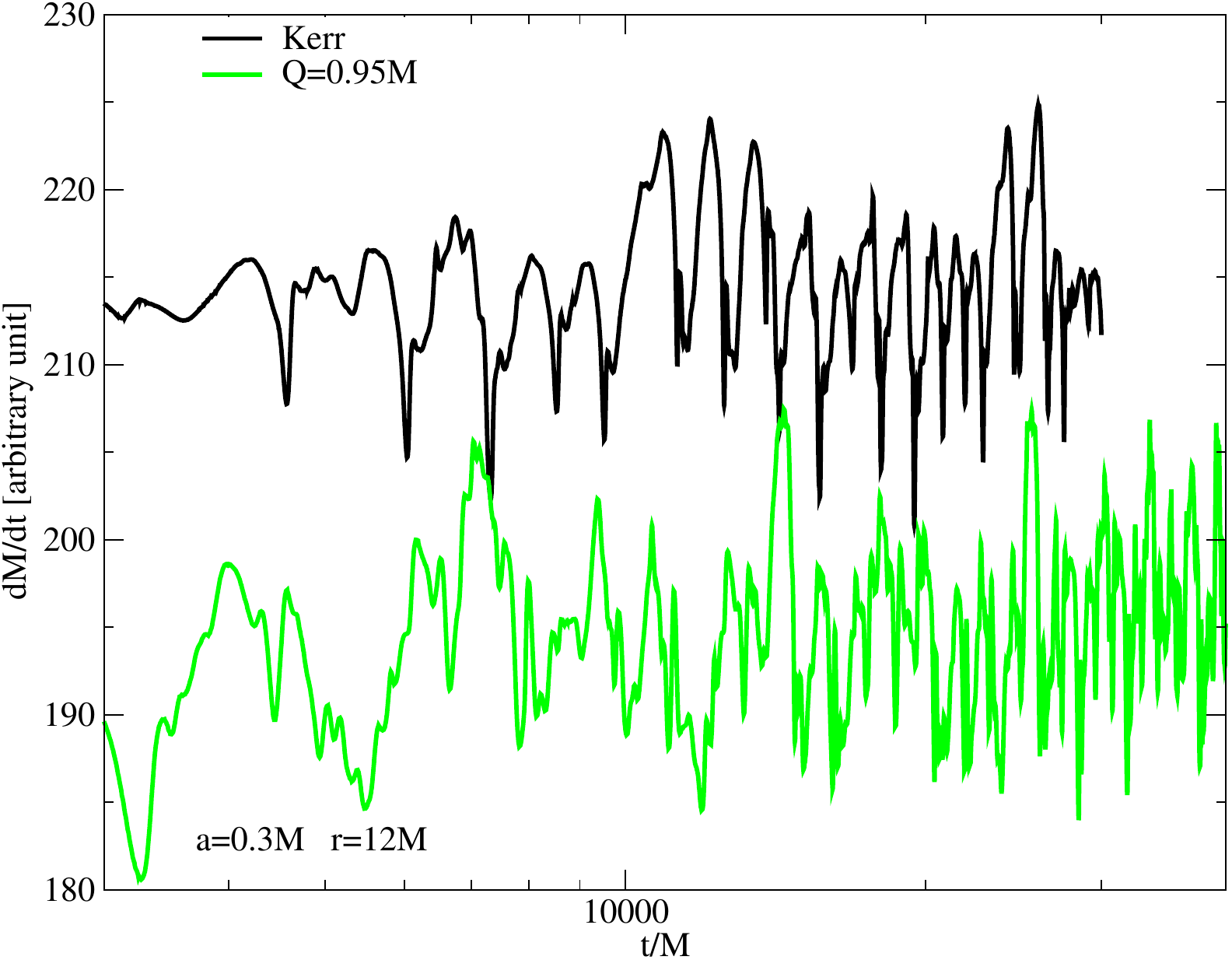}   
     \caption{Same analysis as in Fig. \ref{Mass_acc_a00}, now comparing EEH and Kerr BHs for $a = 0.3M$. The results show that, consistent with the non-rotating case, the accretion rate is modestly enhanced in the strong-field regime and progressively suppressed with increasing radial distance from the horizon\label{fig10}.
    }
\vspace{1cm}
\label{Mass_acc_a03}
\end{figure*}

In Fig.\ref{Mass_acc_a03}, we present the case where the BH is slowly rotating, so that both the spin parameter and the charge parameter contribute simultaneously to the accretion dynamics. In the top panel of Fig.\ref{Mass_acc_a03}, for $a = 0.3M$ in the strong gravitational field at $r = 2.3M$, the combined effect of $Q$ and the frame-dragging produced by the slowly rotating BH enhances the deformation of the shock cone. This deformation creates the necessary environment for the formation of low-frequency QPOs (LFQPOs). On the other hand, as seen in the middle and bottom panels of Fig.\ref{Mass_acc_a03}, moving away from the strong-field region, the accretion rate decreases as in Fig.\ref{Mass_acc_a00}. However, in this case, the reduction is much stronger than in the non-rotating configuration. The reason is that rotation injects angular momentum into the flow, preventing complete damping. This maintains mild instabilities, which act as a potential source of secondary oscillatory modes that can couple nonlinearly to the inner-region QPOs. In other words, in the low-spin regime, QPOs generation is concentrated mainly in the inner region of the disk, while in the outer region, oscillations are weaker but still present.

\begin{figure*}[!htp]
  \vspace{1cm}
  \center
    \includegraphics[width=10.0cm,height=5.8cm]{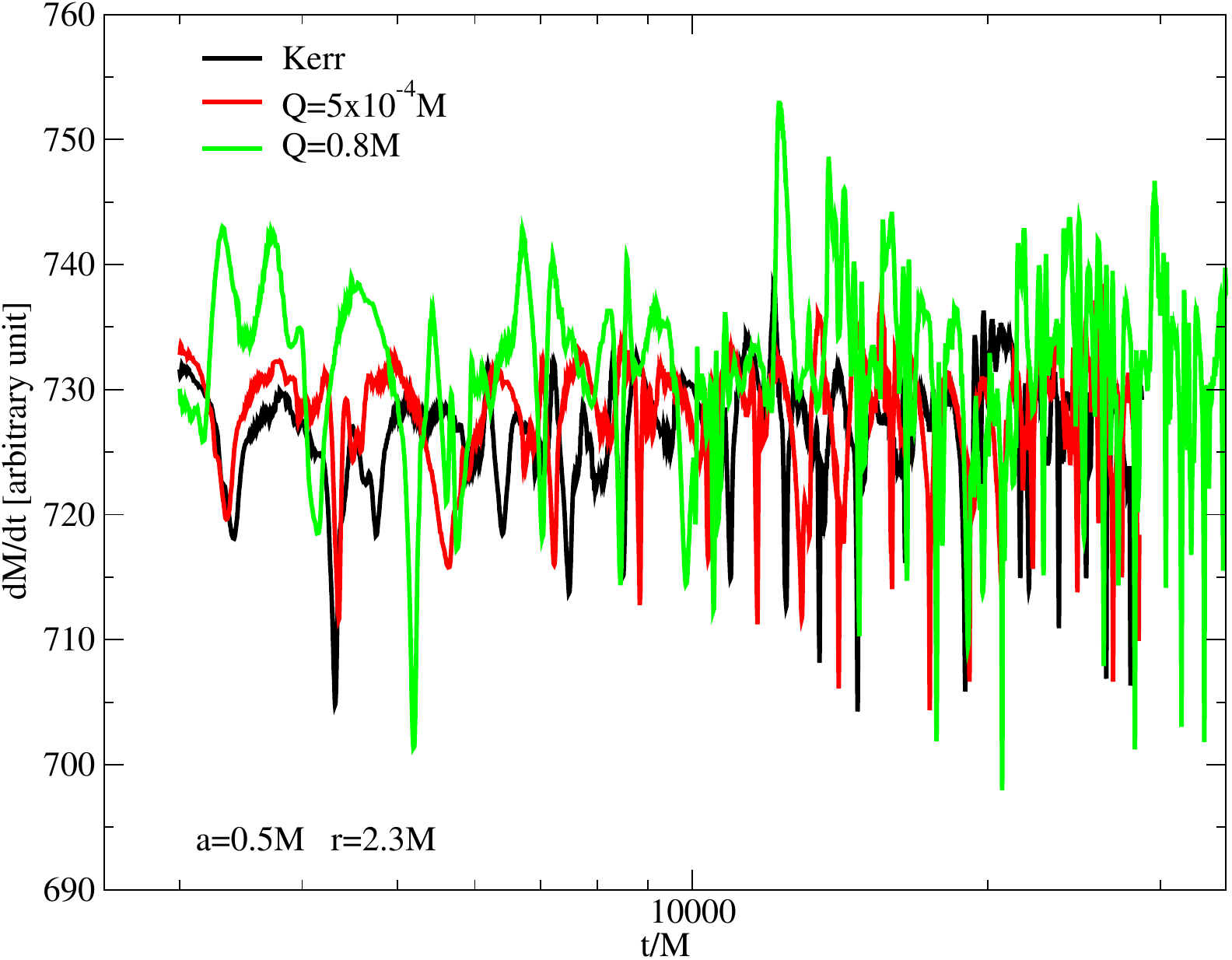} \\
    \vspace{0.3cm}
    \includegraphics[width=10.0cm,height=5.8cm]{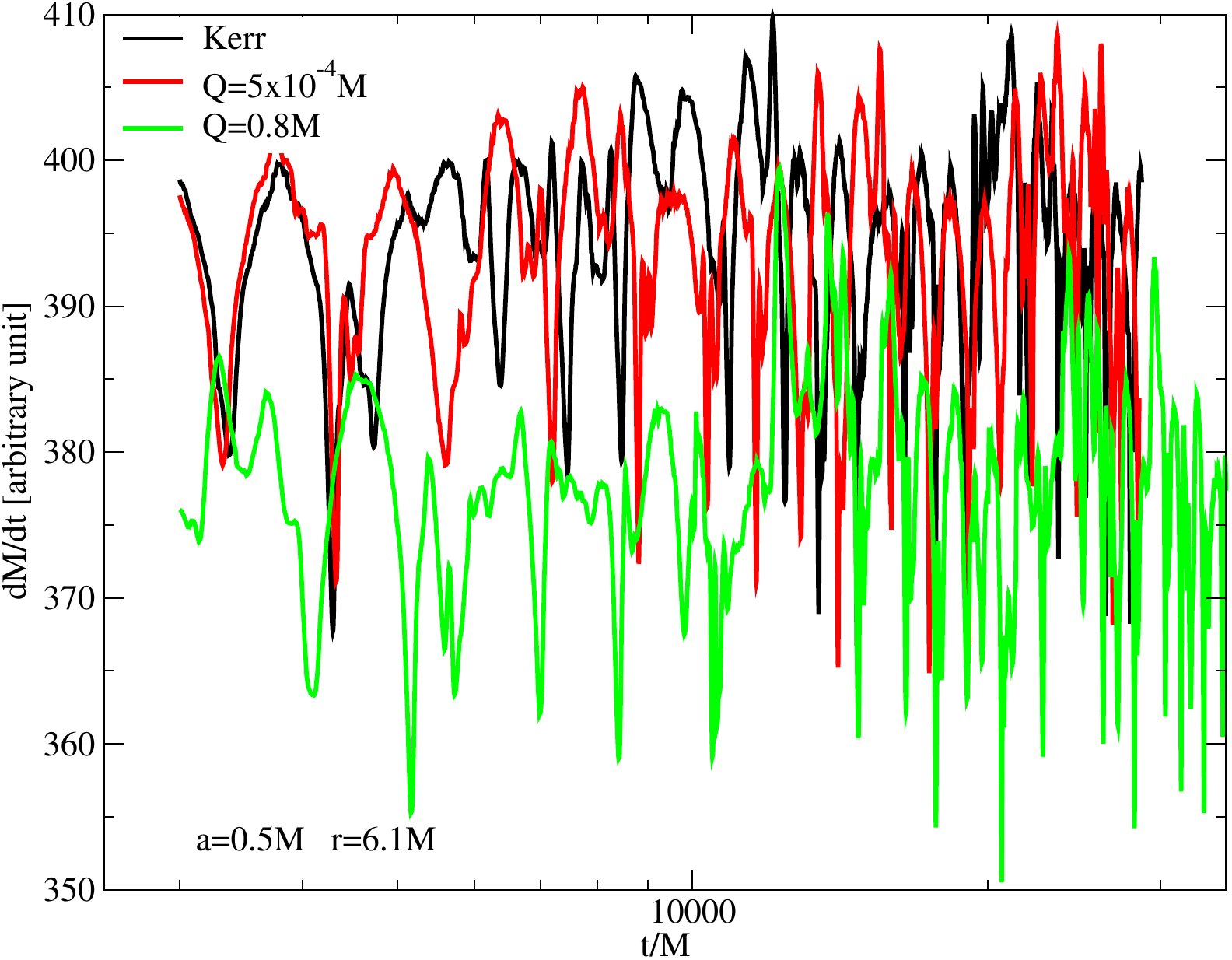} \\
     \vspace{0.3cm} 
    \includegraphics[width=10.0cm,height=5.8cm]{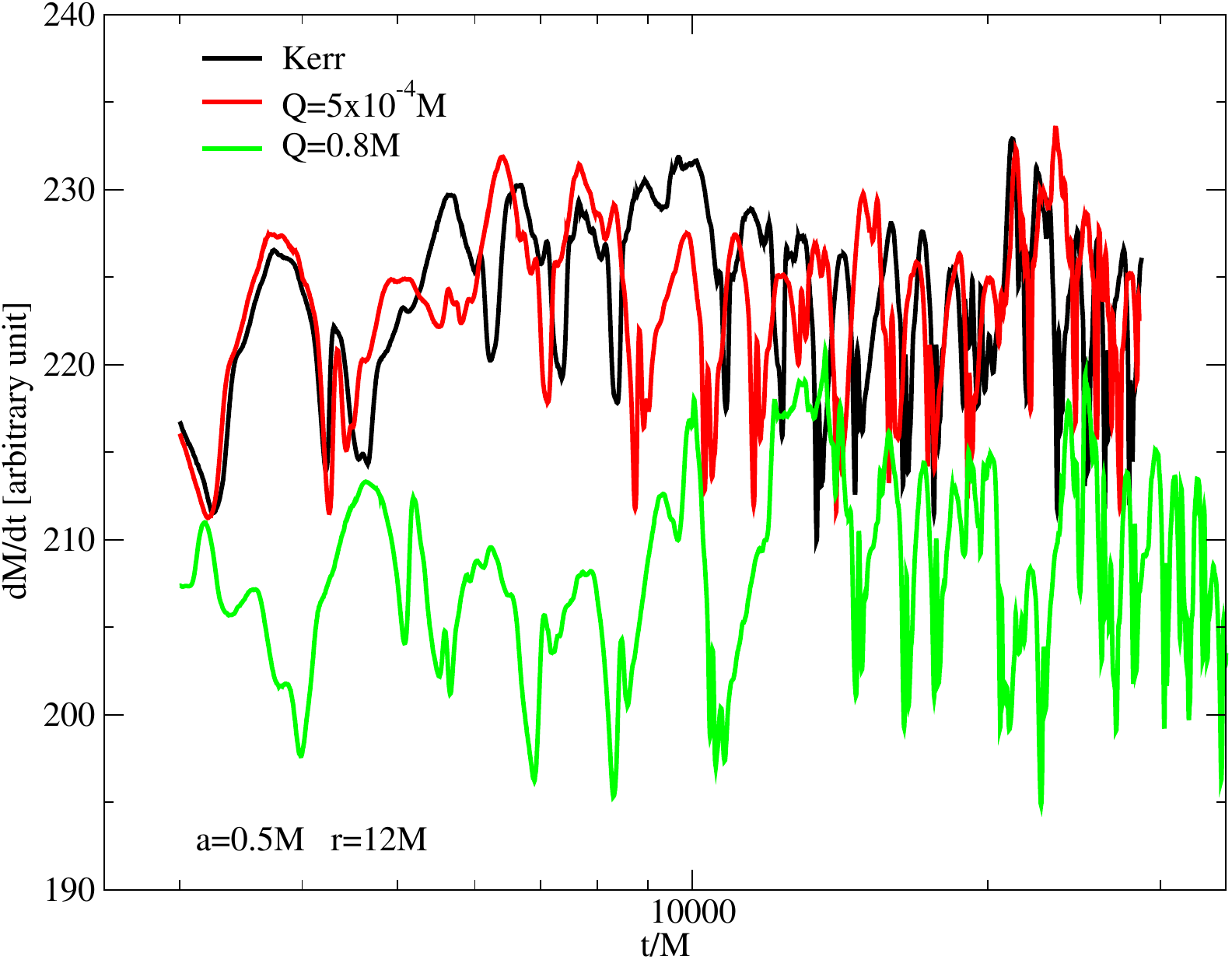} 
     \caption{Same analysis as in Figs. \ref{Mass_acc_a00} and \ref{Mass_acc_a03}, now presented for $a = 0.5M$. The results confirm that, consistent with lower-spin cases, the EEH correction enhances the mass accretion rate in the strong-field regime near the horizon, while the rate systematically decreases with increasing radial distance\label{fig11}.
    }
\vspace{1cm}
\label{Mass_acc_a05}
\end{figure*}

In the case of a moderately rotating BH (i.e., $a = 0.5M$), the effect of both $a$ and $Q$ on the shock cone formed around the BH is shown in Fig.\ref{Mass_acc_a05}. In the top panel of Fig.\ref{Mass_acc_a05}, at $r = 2.3M$ in the strong gravitational field, the mass accretion rate is presented. As a result of the combined influence of $a$ and $Q$, the shock cone is seen to be highly unstable. This instability is a key factor leading to the emergence of high-frequency QPOs (HFQPOs), since strong instabilities excite oscillation modes in the strong-field region. The LFQPOs, on other hand, arise due to the precession of the shock cone. In the weaker gravitational regions at $r = 6.1M$ and $12M$, suppression of the accretion rate still occurs, but compared to Figs.\ref{Mass_acc_a00} and \ref{Mass_acc_a03}, suppression is less pronounced. The reduced damping of the accretion flow allows oscillations generated inside the cone to propagate outward. This leads to resonant coupling between inner and outer region osc0illations along the radial direction of the cavity. Thus, through the combined effects of the moderate spin parameter and the EEH charge $Q$, both HFQPOs and LFQPOs can emerge, giving rise to a rich spectrum of nonlinear couplings.

\begin{figure*}[!htp]
  \vspace{1cm}
  \center
      \includegraphics[width=10.0cm,height=5.8cm]{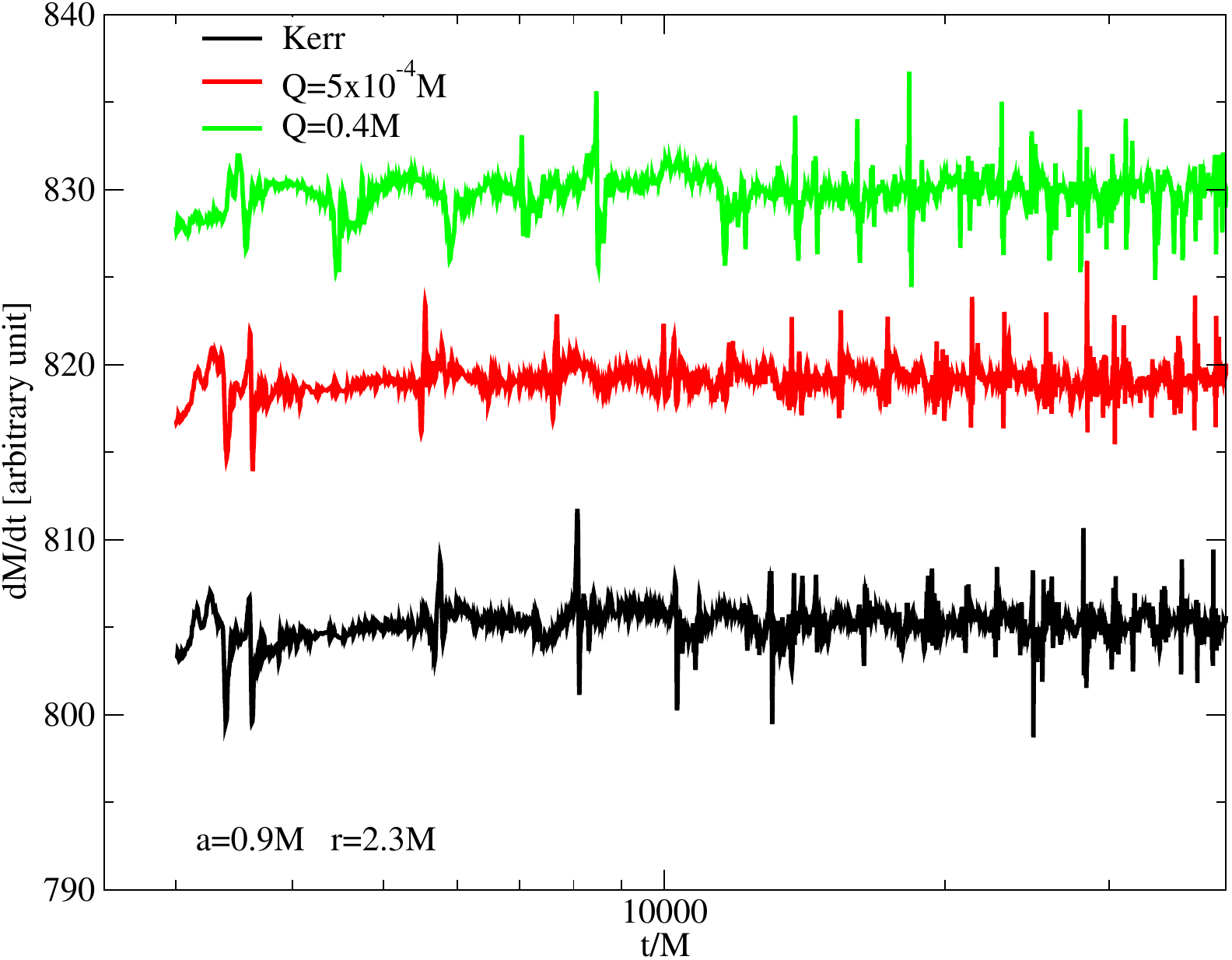} \\
    \vspace{0.3cm}
        \includegraphics[width=10.0cm,height=5.8cm]{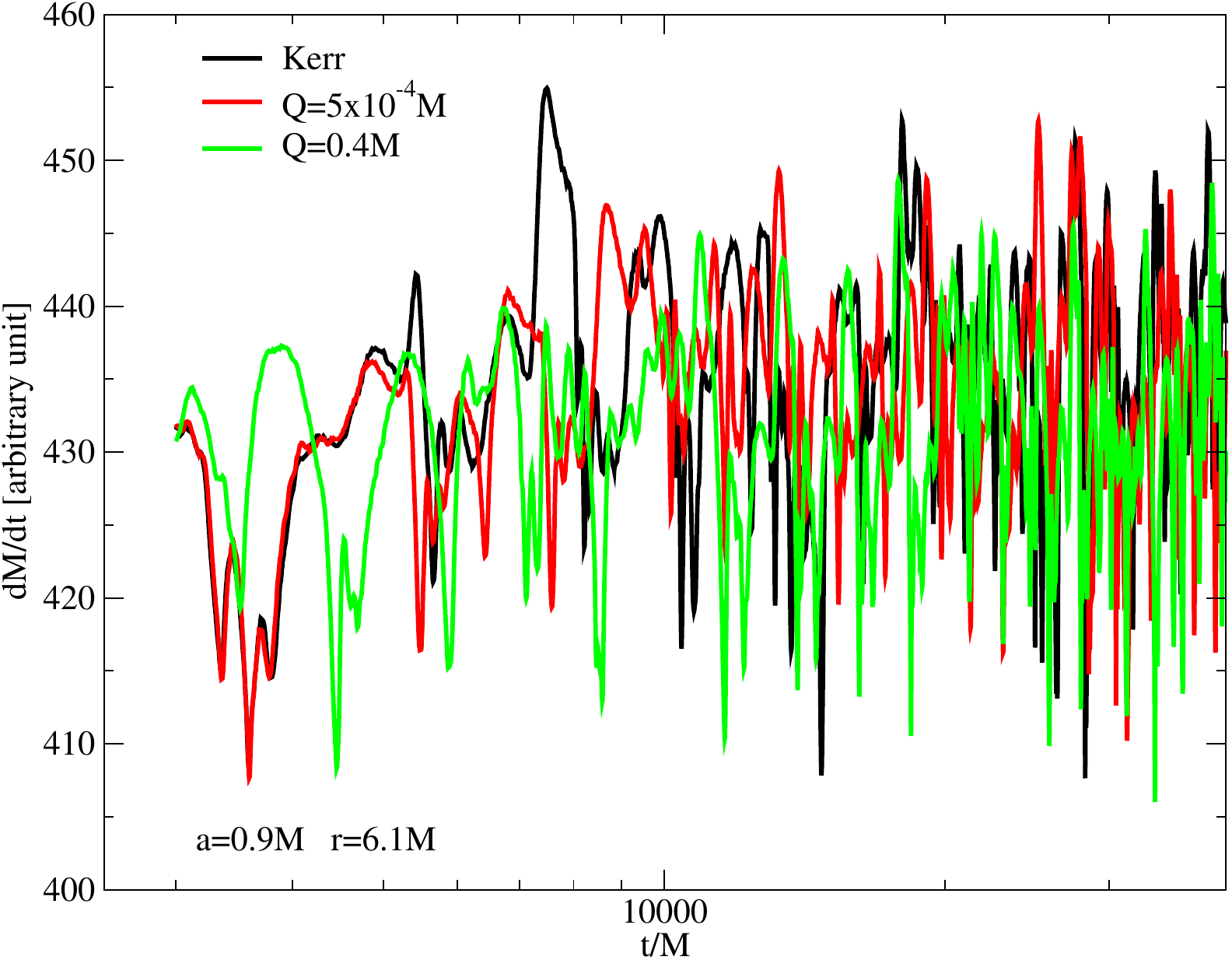} \\
     \vspace{0.3cm} 
         \includegraphics[width=10.0cm,height=5.8cm]{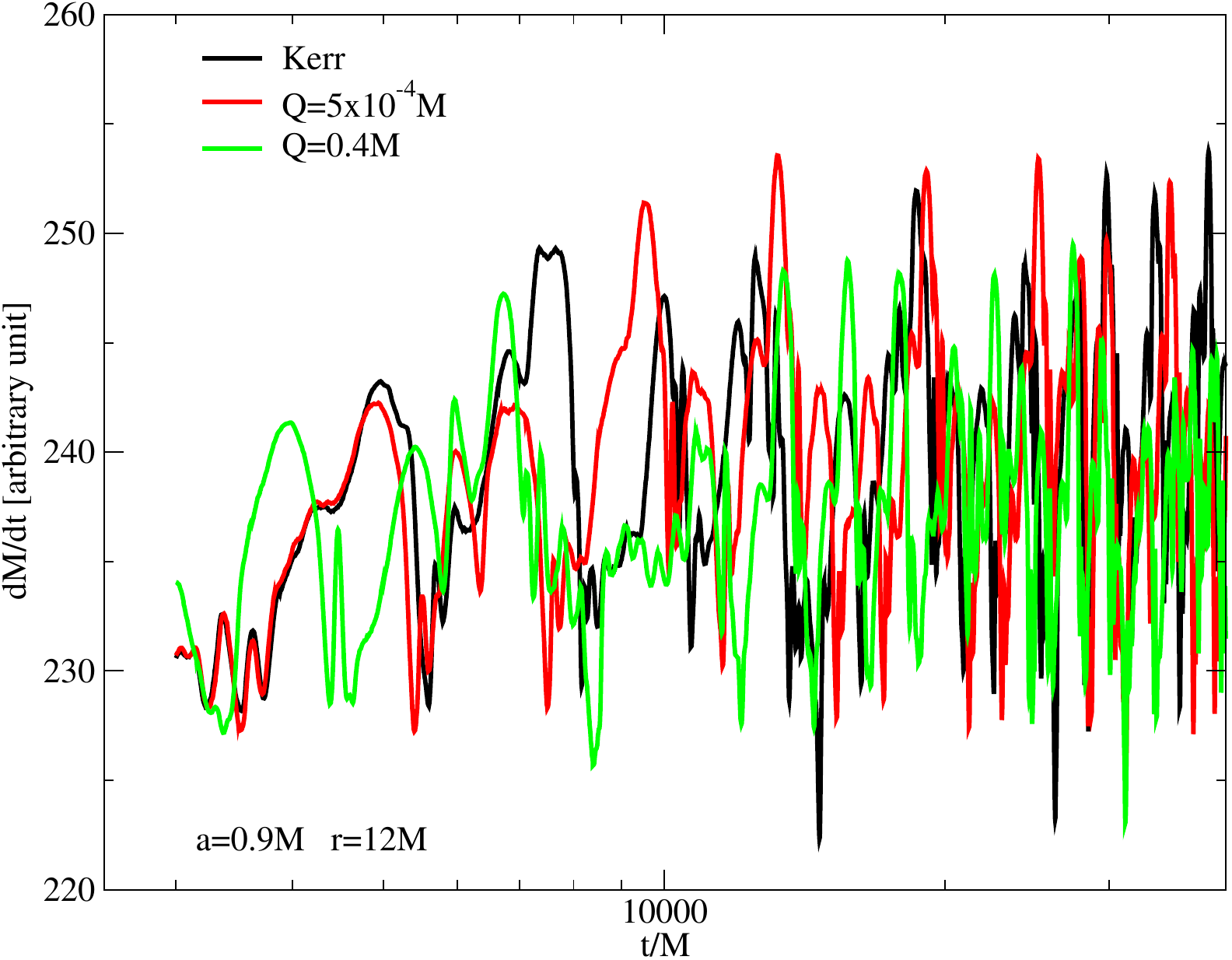} 
     \caption{Consistent with the trends shown in Figs. \ref{Mass_acc_a00}-\ref{Mass_acc_a05}, the qualitative behavior of the accretion flow remains similar across configurations. At high spin ($a = 0.9M$), however, the EEH charge parameter $Q$ significantly enhances the mass accretion rate in the strong-field region. In contrast, at larger radii the accretion rate becomes lower than in the Kerr case, though the suppression is less pronounced compared to lower-spin configurations\label{fig12}.
    }
\vspace{1cm}
\label{Mass_acc_a09}
\end{figure*}

In the extreme case where the BH rotates very rapidly with $a = 0.9M$, the changes in the accretion rate due to the combined effects of $a$ and $Q$ are shown in Fig.\ref{Mass_acc_a09}. In the region close to the horizon at $r = 2.3M$, the joint influence of $a$ and $Q$ significantly amplifies the accretion rate. As observed in previous modified gravity models \cite{2025arXiv250616405M,Donmez2025EPJC}, this change is largely dominated by the strong effect of frame dragging. This drives violent shock cone instabilities, producing strong HFQPOs and amplifying their amplitudes. In contrast, the accretion rates shown in the middle and bottom panels of Fig.\ref{Mass_acc_a09}, corresponding to regions farther from the horizon, are nearly identical to those in the Kerr case. However, the outer shock cone is less stable, meaning that oscillations generated near the horizon can survive and propagate outward, reinforcing QPOs activity across different radial zones.

Consequently, all values of $a$, the EEH charge parameter $Q$ enhances accretion in the strong gravitational field, while at larger distances from the horizon the mass accretion rate decreases compared to the Kerr case. The increase in mass inflow near the horizon indicates the formation of a highly unstable shock cone with strong oscillatory behavior, which in turn gives rise to powerful QPOs. Farther from the horizon, the suppression of the accretion rate leads to a more stable cone, resulting in weaker oscillatory activity. As discussed above, spin $a$ also plays a significant role in this process. Therefore, these results show that QPOs do not emerge as random fluctuations but rather as the outcome of the synergistic influence of $Q$ and $a$ on the excitation of modes trapped within the shock cone. While $Q$ modulates the inflow of matter toward the BH and regulates the growth of instabilities, $a$ governs angular momentum transfer, resonance conditions, nonlinear couplings, and precession effects.

\subsection{Formation of the Plasma and Shcok Cone}\label{S3-2}

\begin{figure*}[!htp]
  \vspace{1cm}
  \center
      \includegraphics[width=7.0cm,height=5.8cm]{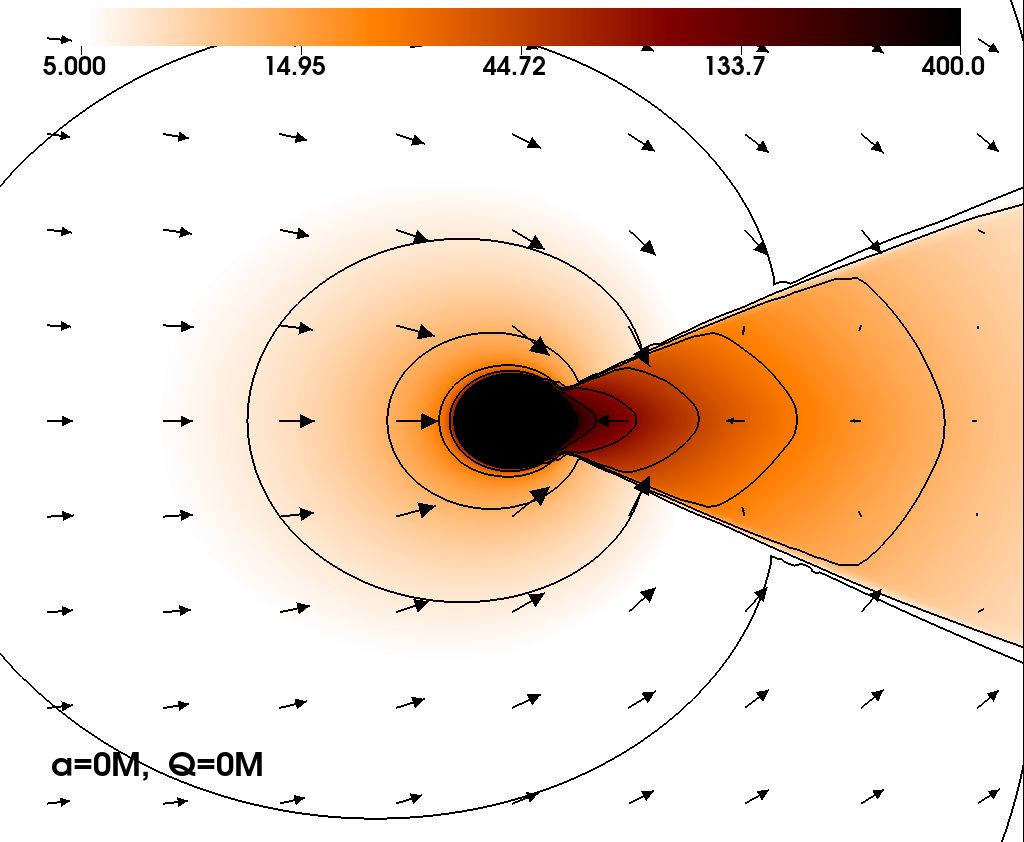} 
      \includegraphics[width=7.0cm,height=5.8cm]{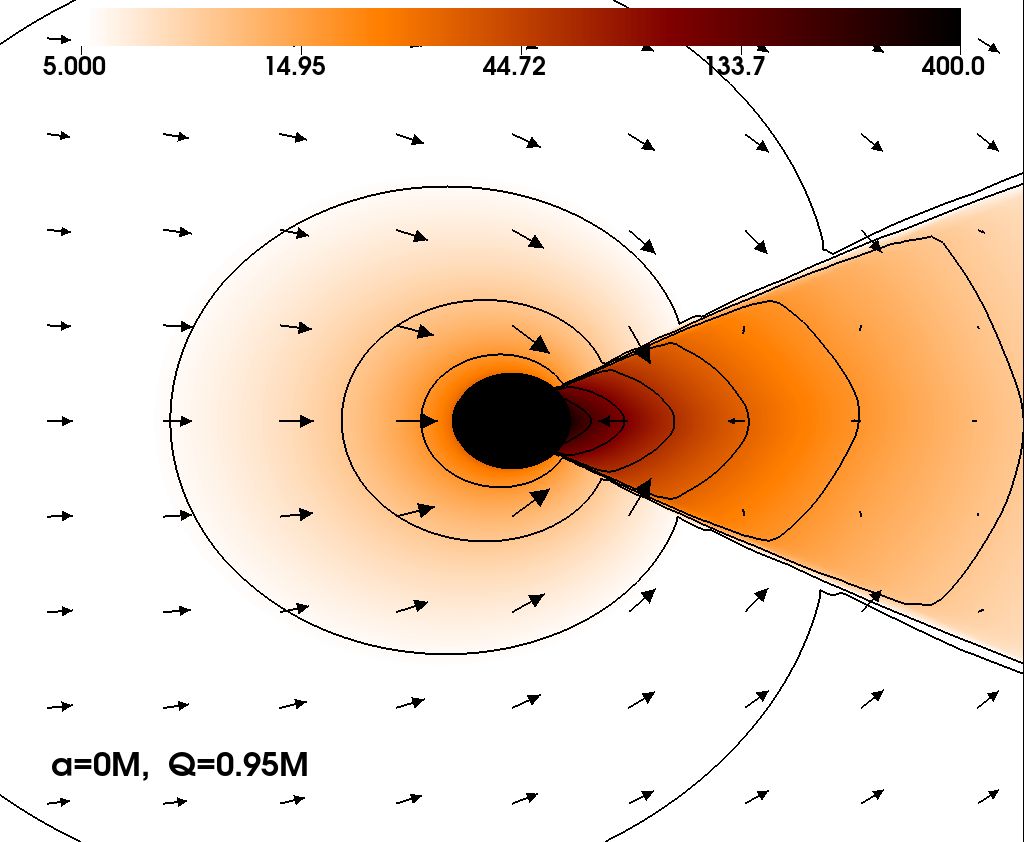}  \\
      \includegraphics[width=7.0cm,height=5.8cm]{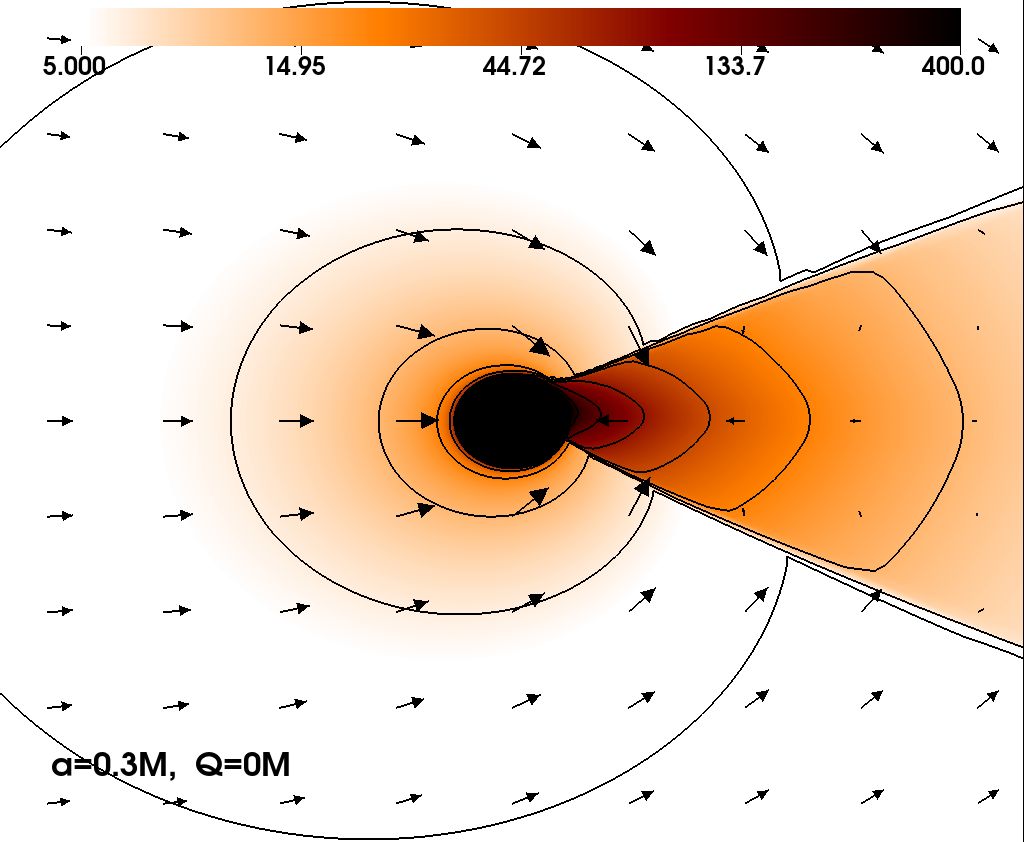} 
      \includegraphics[width=7.0cm,height=5.8cm]{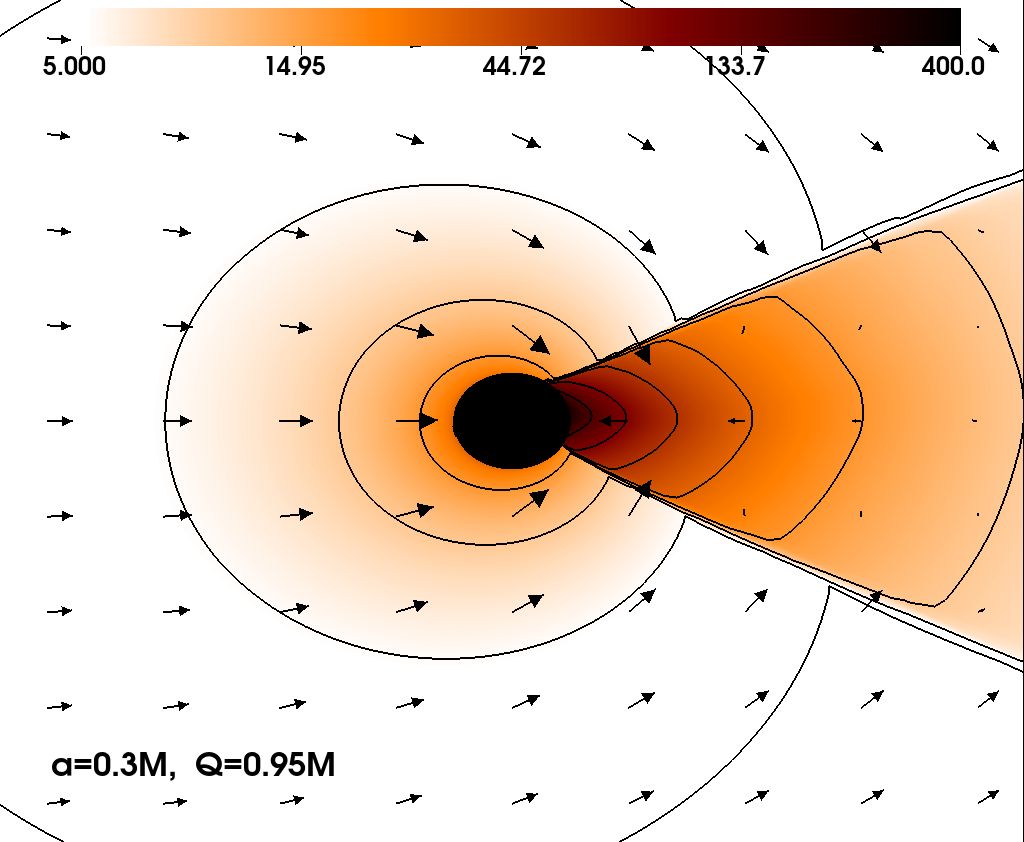}  \\
      \includegraphics[width=7.0cm,height=5.8cm]{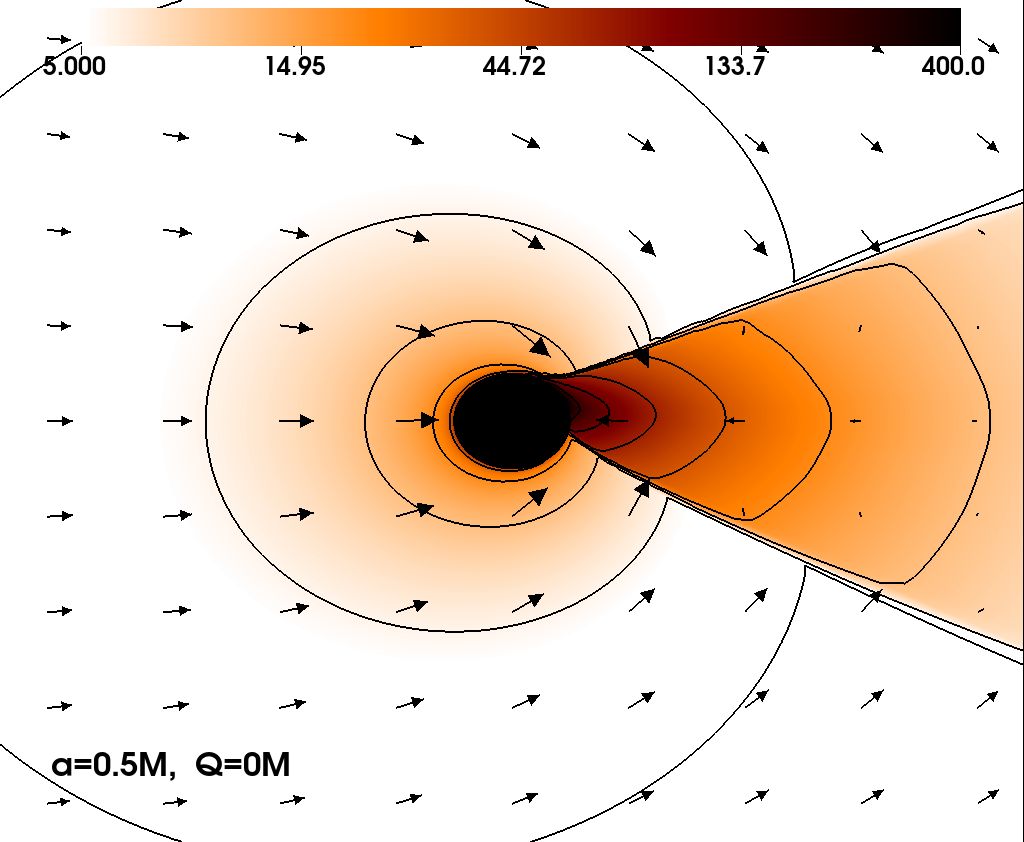} 
      \includegraphics[width=7.0cm,height=5.8cm]{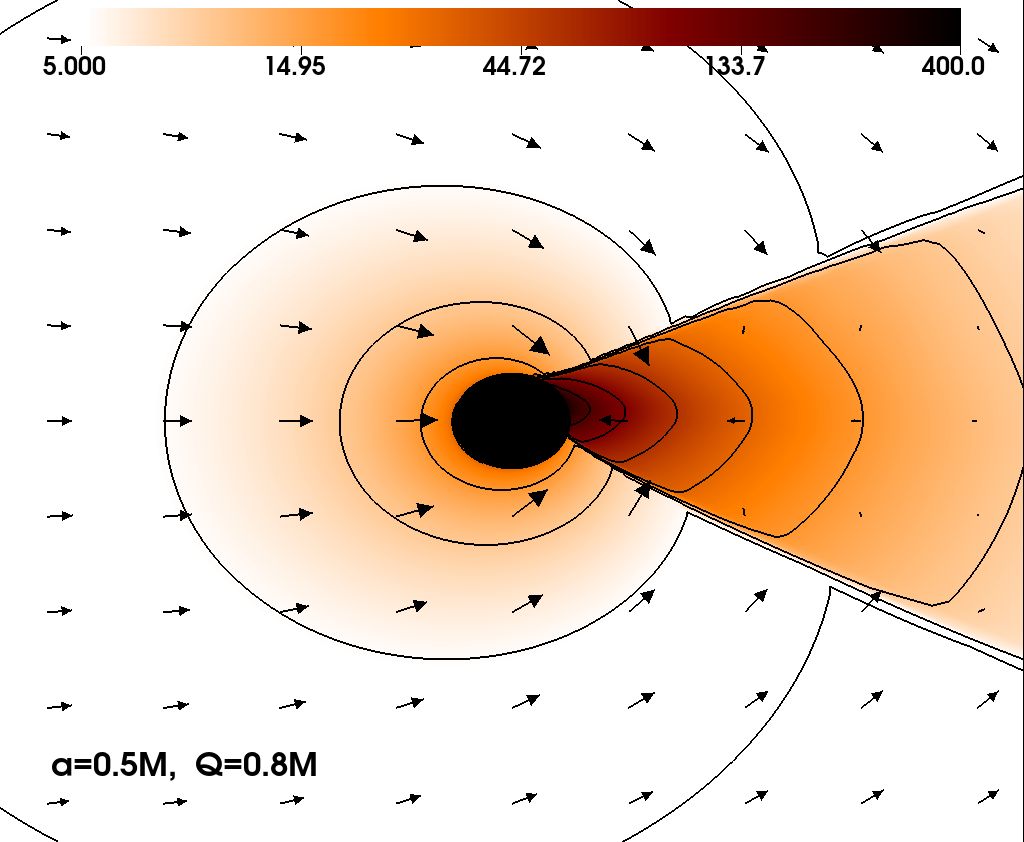}  \\
      \includegraphics[width=7.0cm,height=5.8cm]{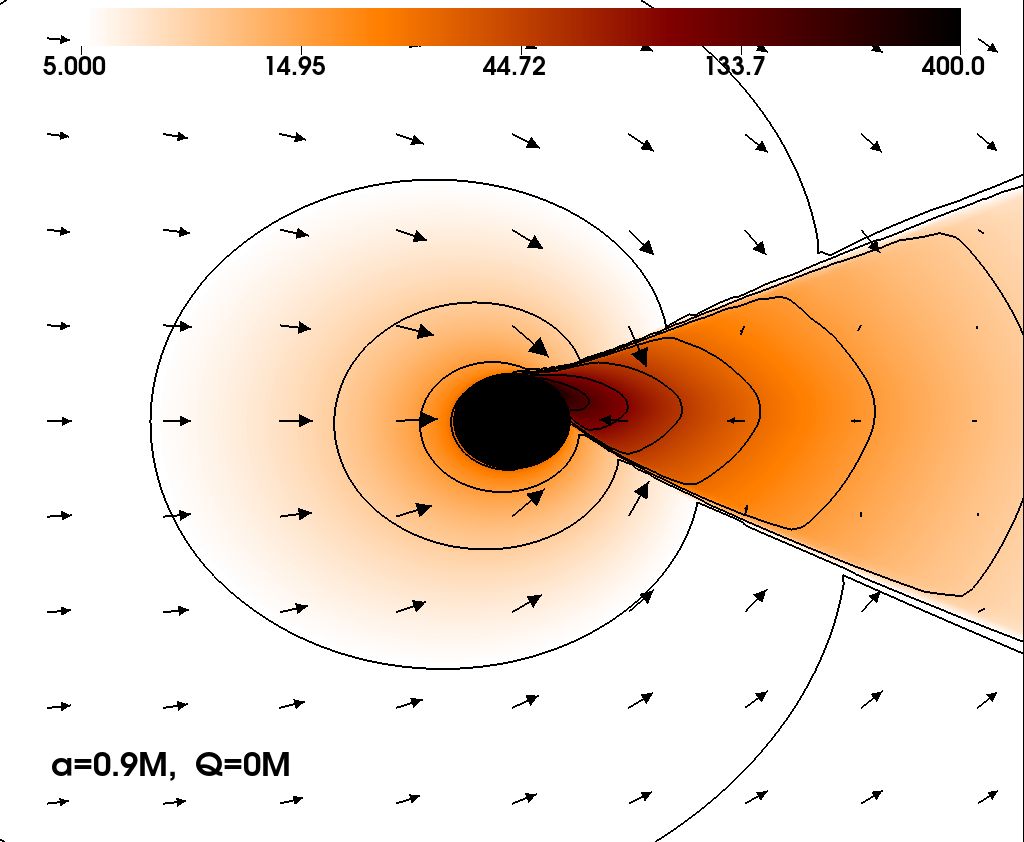} 
      \includegraphics[width=7.0cm,height=5.8cm]{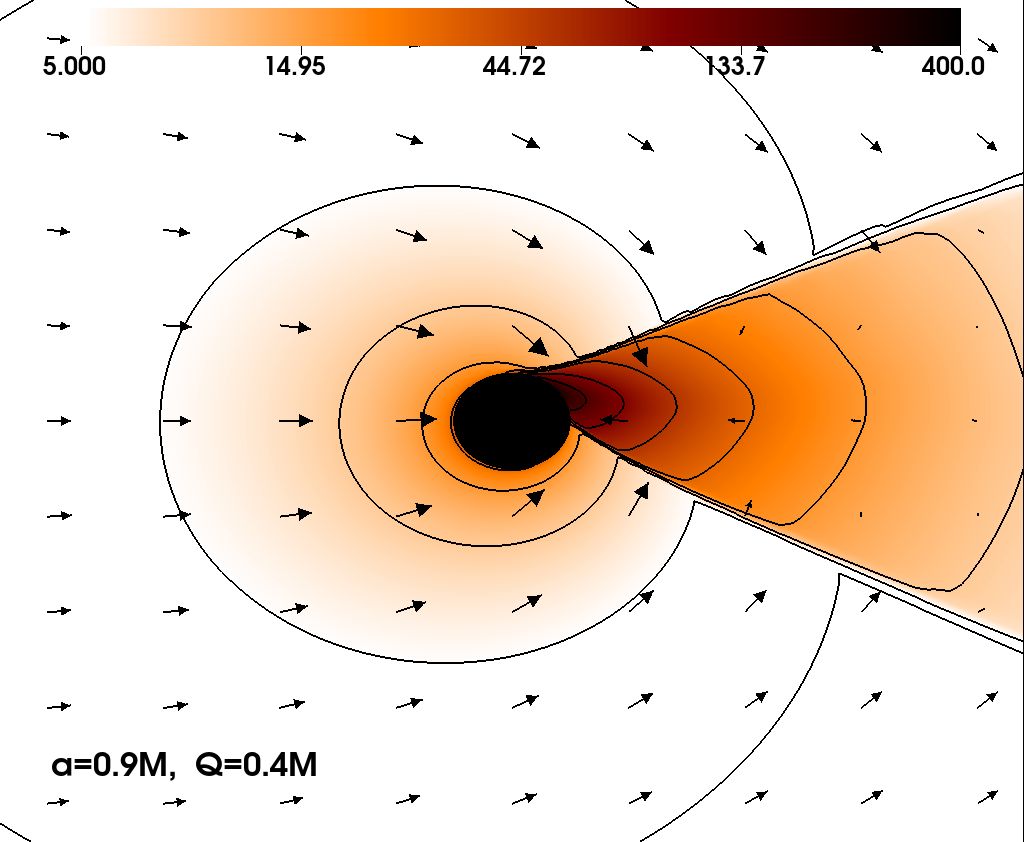}          
     \caption{Plasma dynamics in the equatorial plane and the corresponding evolution of the shock cone structure are examined as functions of the BH spin parameter $a$ and the EEH charge parameter $Q$. The left column shows results for Schwarzschild and Kerr BHs, while the right column provides a detailed representation of the EEH case. The rest-mass density distributions are presented as both color maps and contour plots, accompanied by velocity vector fields that illustrate the BHL accretion flow, the formation of the shock cone, and the inward motion of matter trapped within the cone. For enhanced clarity, zoomed-in views are provided with spatial boundaries defined by $[x_{\min},y_{\min}] = [-20M,-20M]$ and $[x_{\max},y_{\max}] = [20M,20M]$\label{fig13}.
     }
\vspace{1cm}
\label{color_den}
\end{figure*}

The plasma and shock cone structures formed around Schwarzschild, Kerr, and EEH BHs with different values of the $Q$ parameters are shown in Fig.\ref{color_den}. To better analyze the resulting configurations, the rest-mass density in the equatorial plane is plotted as contour maps together with velocity vector fields. In each model, matter falls supersonically toward the BH from the upstream region via the BHL mechanism, while in the downstream region a shock cone forms. The matter trapped inside the cone, influenced by the parameters $a$ and $Q$, generates turbulence and promotes the growth of instabilities. As seen in the color bar of Fig.\ref{color_den}, a strong density gradient develops inside the cone, and the matter compressed between the stagnation point and the shock cone accretes to the BH.

For $a = 0$, the shock cone forms symmetrically and, in the absence of frame dragging, appears less deformed. The density distribution is more stable, and turbulence is primarily confined to the downstream side inside the cone. In contrast, when $a > 0$, the dragging of the frame distorts the cone and twists the plasma streams by the rotation of the BH. This effect becomes stronger with increasing $a$, producing more pronounced shear layers inside the cone and fueling oscillatory behavior.

In the EEH case, NLED corrections significantly modify the plasma structure. Near the horizon, higher values of $Q$ alter the spacetime curvature and enhance the effective gravitational pull. This leads to a denser plasma accumulation, with the shock cone becoming more compact and more unstable than in GR. With increasing $Q$, the cone develops sharper density gradients and stronger velocity shear, intensifying turbulence. The cavity inside the cone traps oscillation modes more effectively, there by boosting the amplitude of HFQPOs. However, at larger distances from the horizon, the effect of $Q$ is to suppress the flow compared to Kerr, stabilizing the outer cone and reducing large-scale turbulence.

Thus, the coexistence of instability near the horizon and relative stability farther away emerges as a distinctive signature of EEH gravity compared to GR, offering a potentially observable difference.

\begin{figure*}[!htp]
  \vspace{1cm}
  \center
    \includegraphics[width=10.0cm,height=5.8cm]{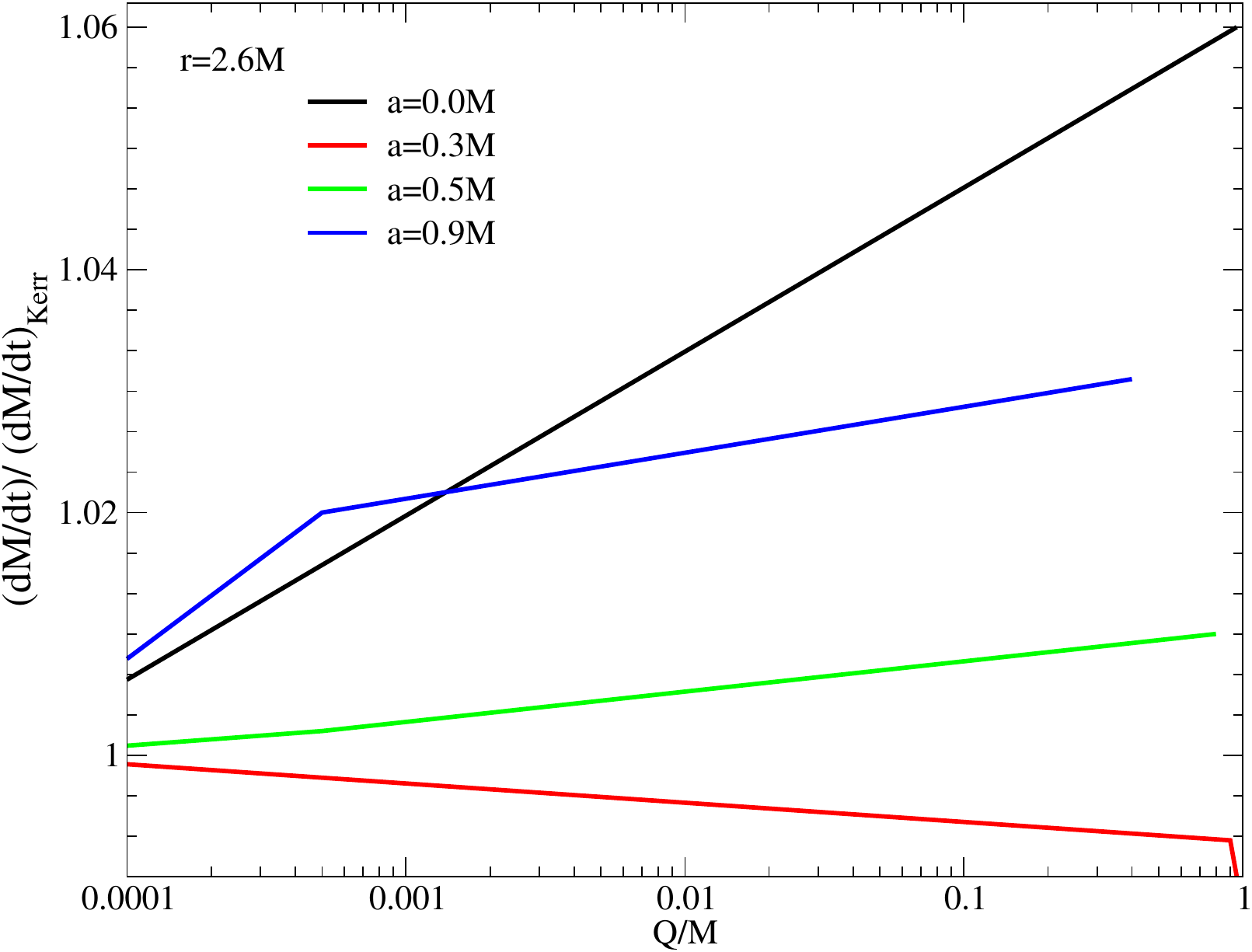} \\
    \vspace{0.3cm}
     \includegraphics[width=10.0cm,height=5.8cm]{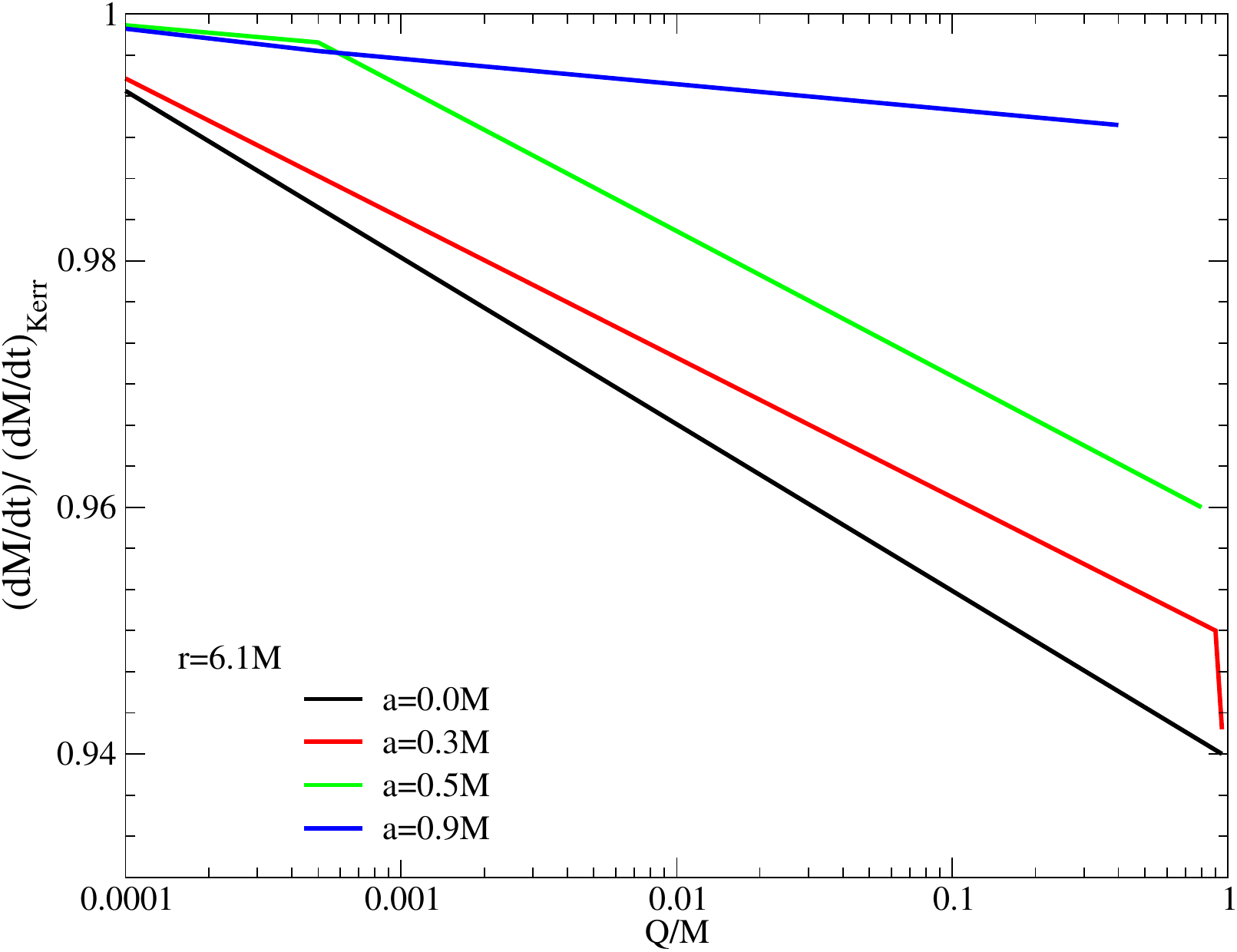} \\
     \vspace{0.3cm} 
      \includegraphics[width=10.0cm,height=5.8cm]{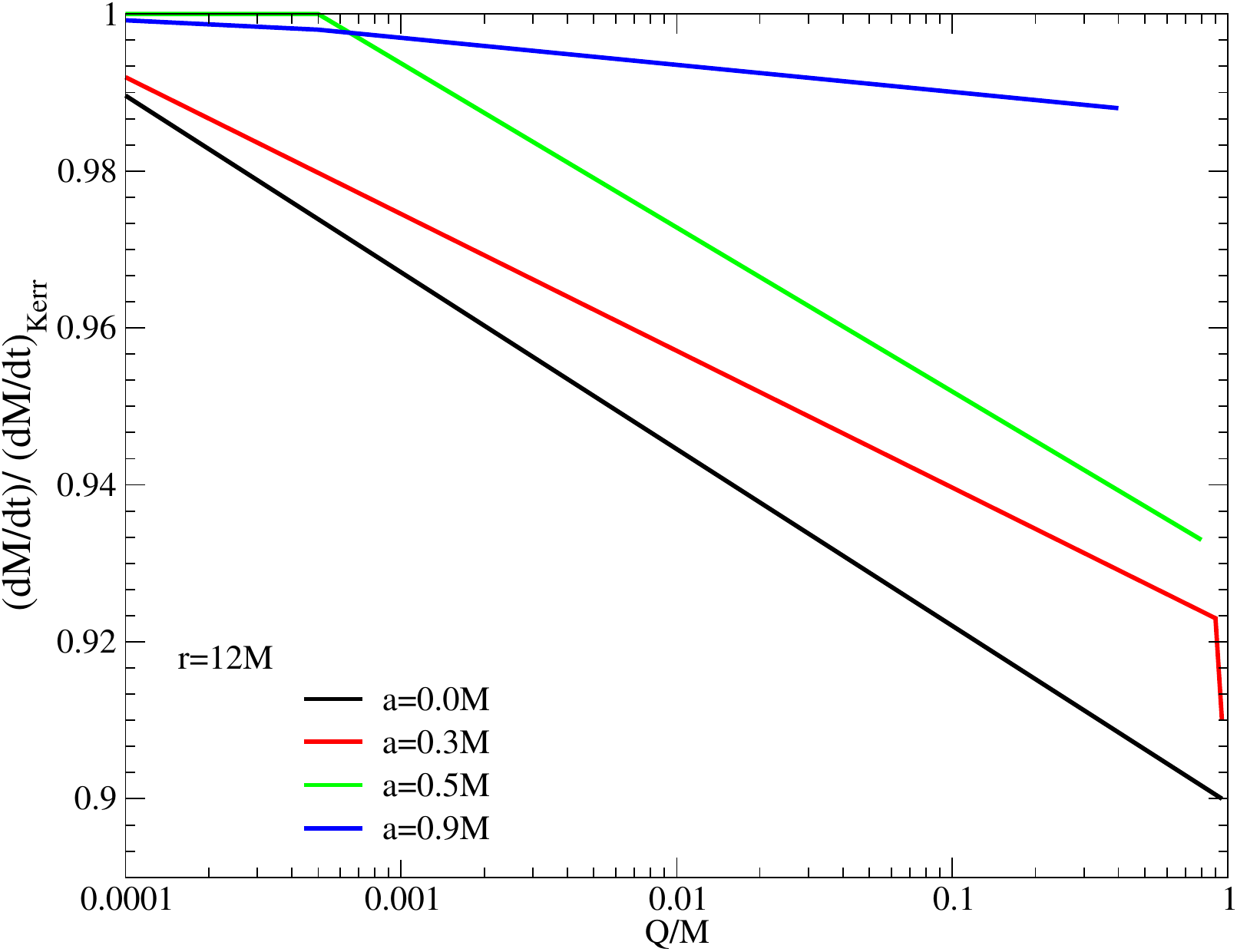} 
     \caption{Summary of the dependence of the mass accretion rate on the EEH charge-to-mass ratio ($Q$) and BH spin across different radial positions. The results show that EEH corrections enhance accretion efficiency in the strong-field regime ($r = 2.3M$), while at larger radii the accretion rate is progressively suppressed with increasing $Q$ compared to the Schwarzschild and Kerr cases\label{fig14}.
    }
\vspace{1cm}
\label{Norm_mass_acc}
\end{figure*}

 After the plasma and shock cone structures around the EEH BH shown in Fig.\ref{color_den} reached a steady state, the average values of the computed mass accretion rates were normalized by the corresponding GR solutions with the same spin parameter, and the results are presented in Fig.\ref{Norm_mass_acc}. In this figure, the behavior of the normalized accretion rate as a function of $Q$ is calculated at different radial points for various spin values, allowing a clearer understanding of how $a$ and $Q$ influence the accretion mechanism, the plasma structure, and even the emergence of QPOs. The top panel of Fig.\ref{Norm_mass_acc}, which shows the behavior near the horizon, supports the formation of the high-density plasma region discussed in Fig.\ref{color_den}. The stronger density gradients and shear layers in these regions make the shock cone more unstable, amplifying turbulence and fostering the conditions for strong QPOs activity.

As seen in the middle and bottom panels of Fig.\ref{Norm_mass_acc}, however, as one moves farther from the horizon, the inflow of matter toward the BH is progressively suppressed with increasing $Q$. This results in the outer regions becoming more stable and less turbulent. In terms of the accretion mechanism, this suppression demonstrates how EEH corrections act to stabilize the shock cone, diminishing large-scale turbulence, and lowering the amplitude of oscillations at larger distances. This dual behavior, enhancement of instability near the BH and suppression at larger radii, represents a distinctive feature of EEH gravity compared to Schwarzschild and Kerr spacetimes, where accretion trends are smoother and less sensitive to charge effects.

Hence, the combined analysis of Figs.\ref{color_den}  and \ref{Norm_mass_acc} shows that the EEH BH produces a more compact and more unstable structure near the horizon, distinguishing it clearly from the GR cases. This behavior is of observational relevance, since it provides a potential signature that could be identified in real data, making EEH gravity testable through accretion flow and QPOs measurements.

\subsection{Shock Cone Evolution}\label{S3-3}

\begin{figure*}[!htp]
  \vspace{1cm}
  \center
  \includegraphics[width=8.0cm,height=6.8cm]{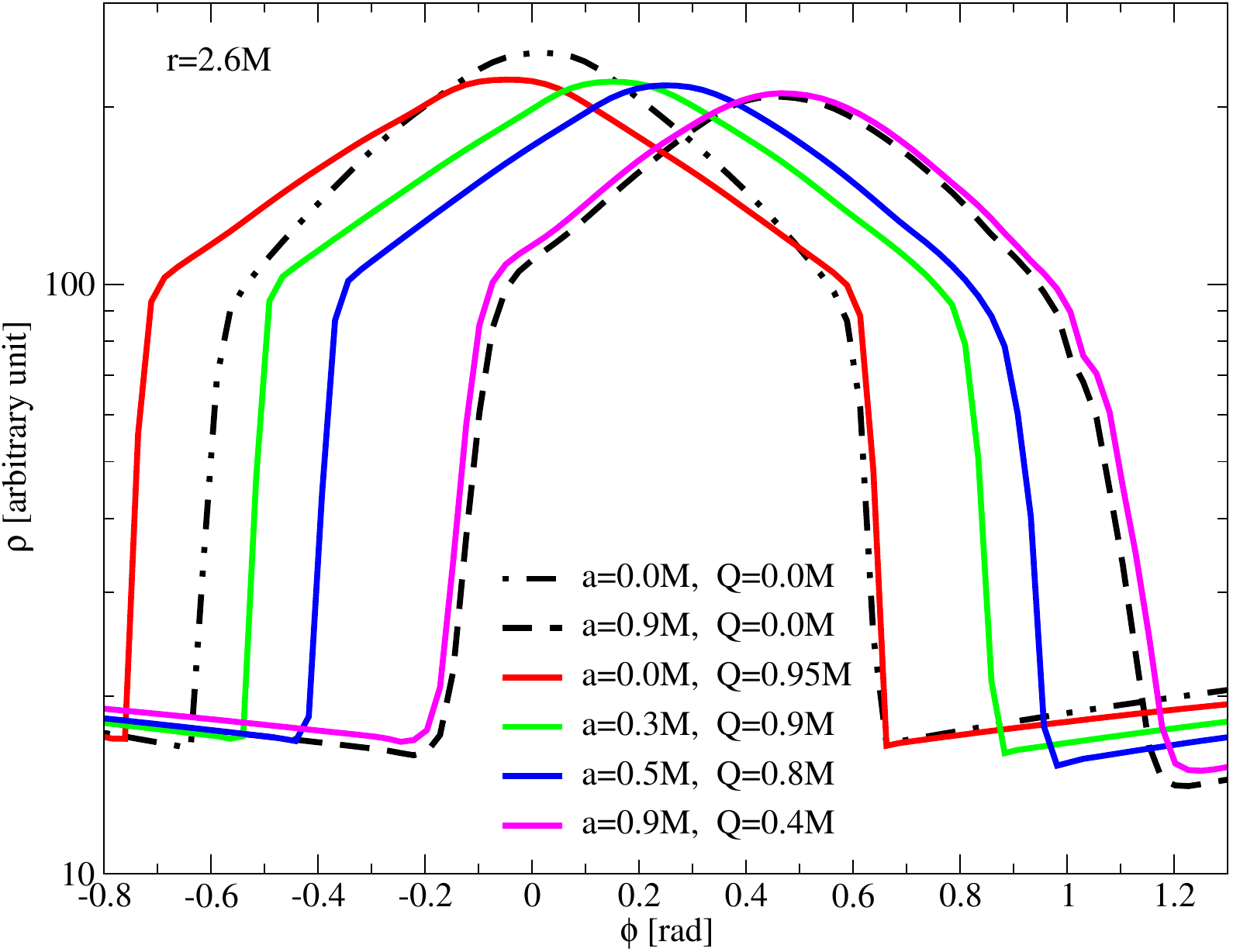} 
       \vspace{0.3cm} 
   \includegraphics[width=8.0cm,height=6.8cm]{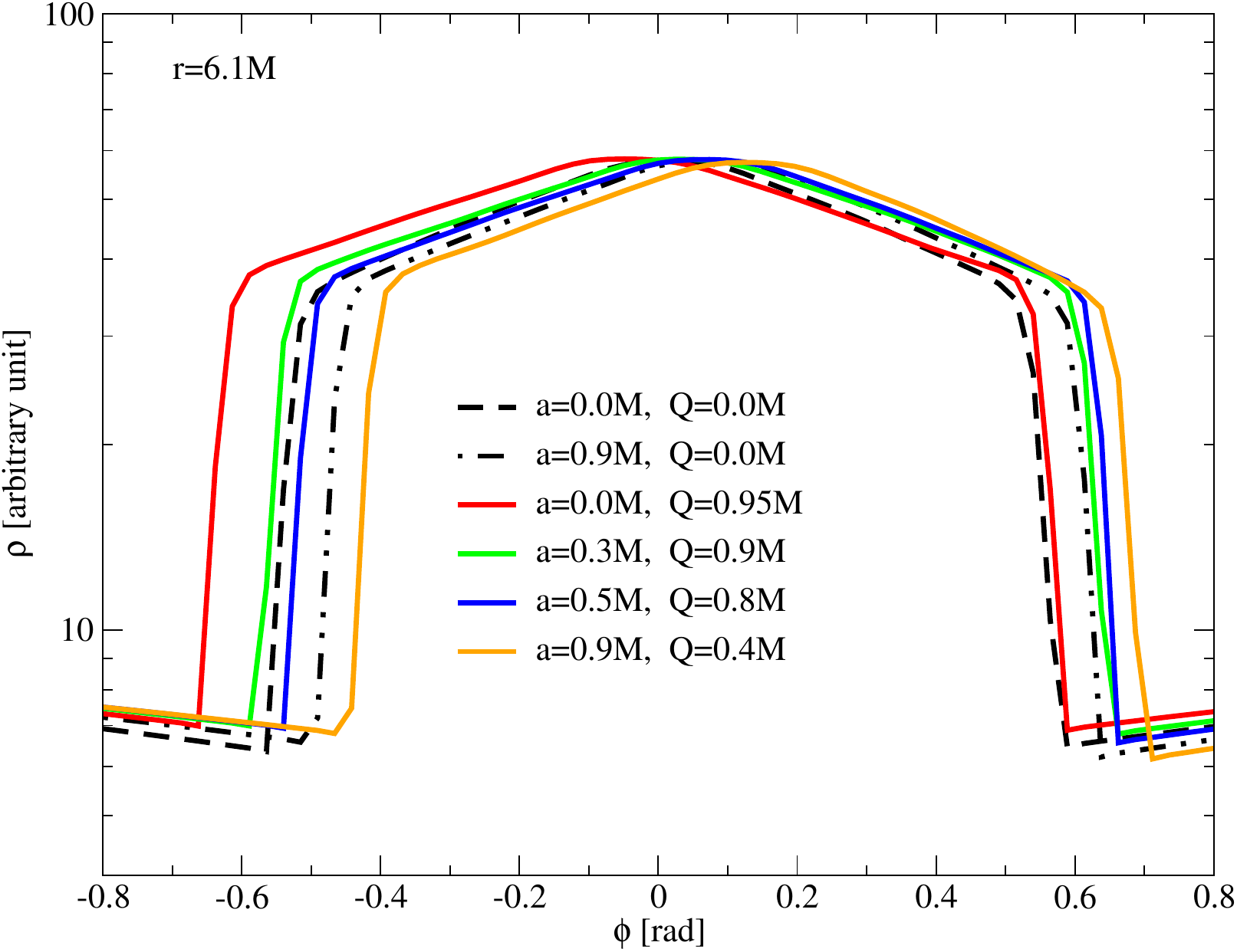}
     \caption{Azimuthal profiles of the rest-mass density at $r = 2.6M$ and $r = 6.1M$, illustrating the morphology and evolution of the shock cone formed during accretion. The structural variations are presented as functions of the BH spin parameter $a$ and the EEH charge parameter $Q$, highlighting the dependence of shock dynamics on both rotational and electromagnetic effects in strong and weak field regimes\label{fig15}.
    }
\vspace{1cm}
\label{dens_azim_diff_r_a_Q}
\end{figure*}

\begin{figure*}[!htp]
  \vspace{1cm}
  \center
   \includegraphics[width=8.0cm,height=5.8cm]{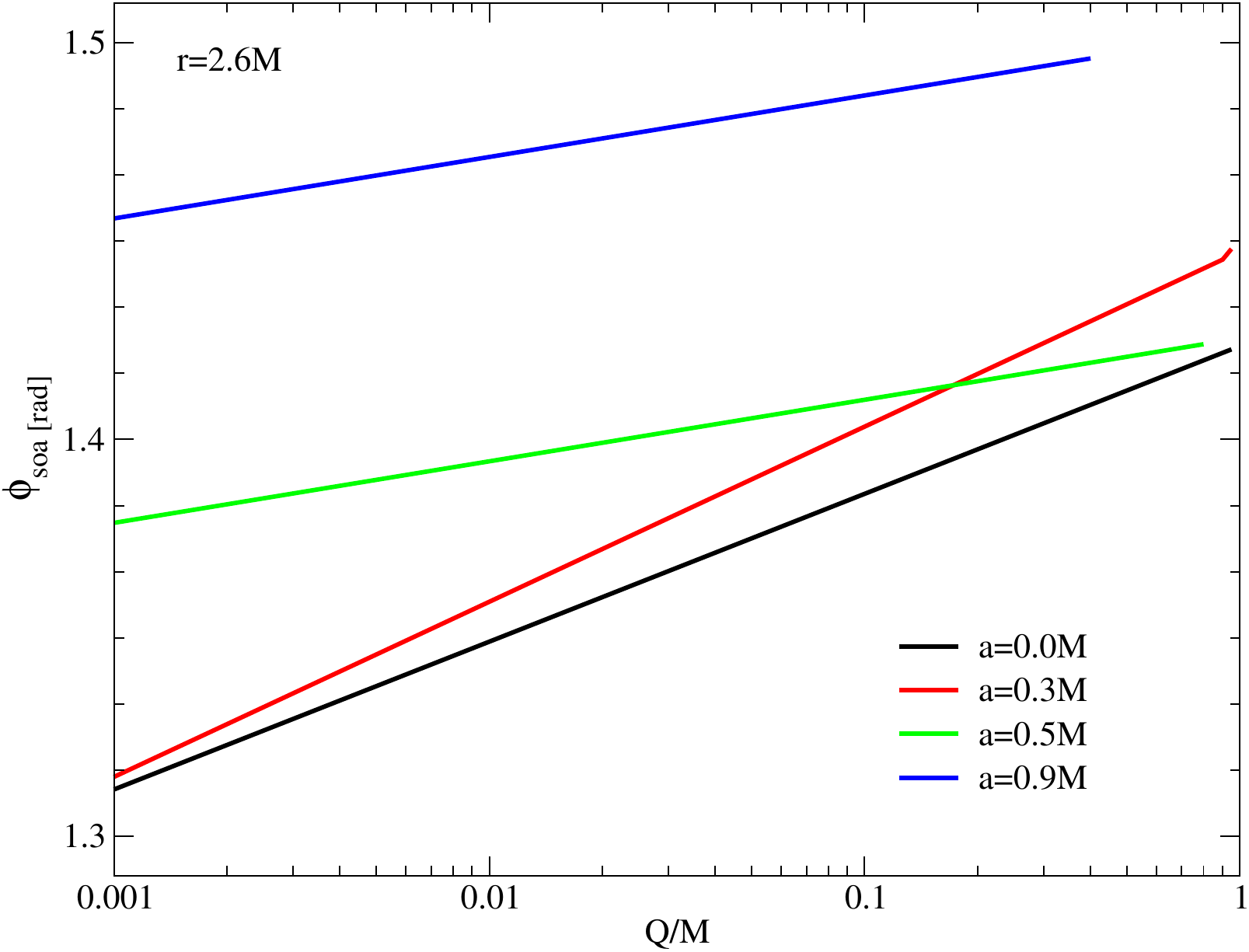} 
       \vspace{0.3cm} 
      \includegraphics[width=8.0cm,height=5.8cm]{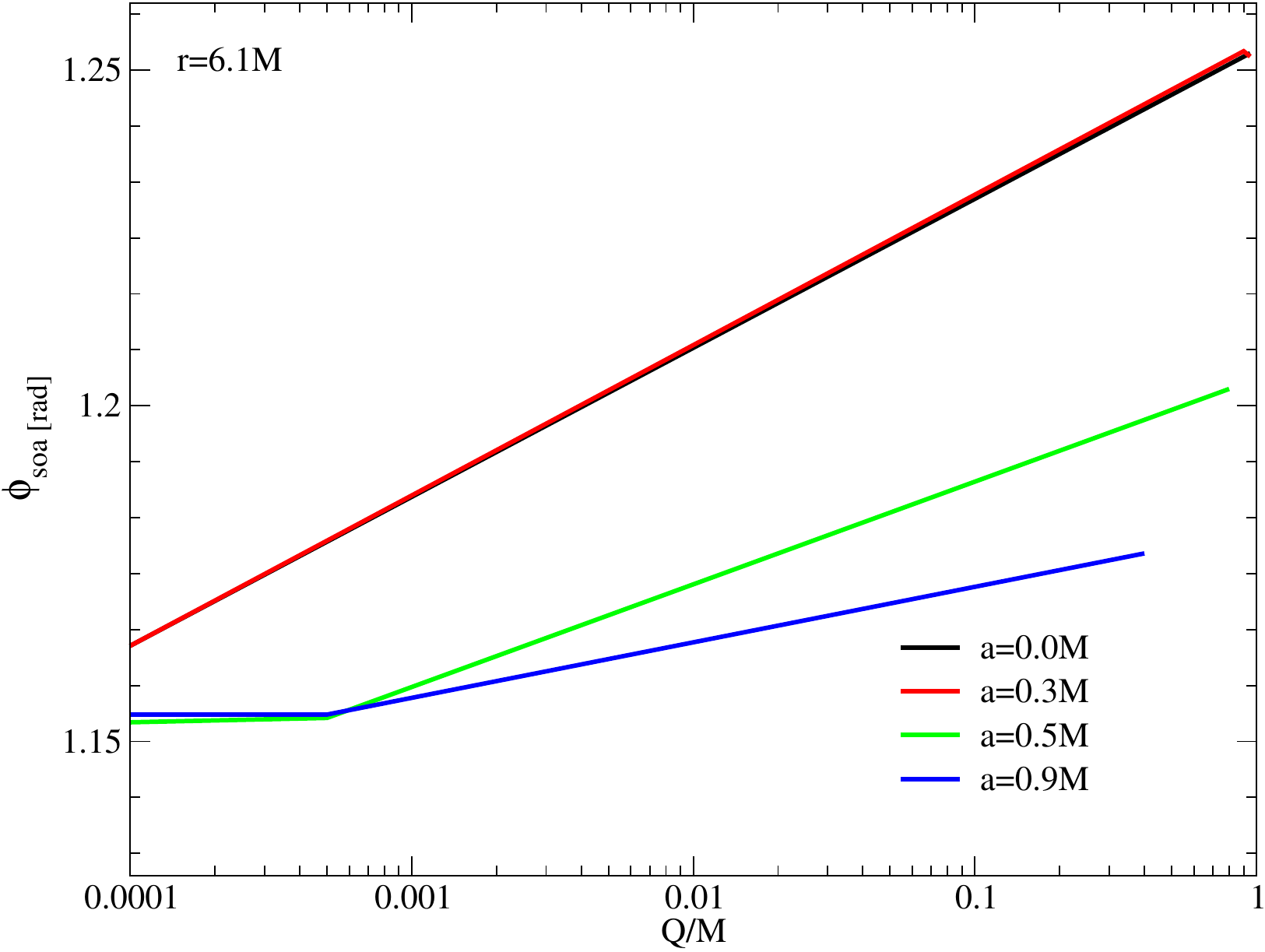} 
     \caption{Comparative analysis of the shock cone opening angle, $\phi_{soa}$, as a function of the EEH charge parameter $Q$ for different BH spin values $a$. The upper panel displays the dependence of $\phi_{soa}$ at $r = 2.6M$, whereas the lower panel presents the corresponding variation at $r = 6.1M$, highlighting the influence of spin on the shock cone geometry in both strong and weak field regions\label{fig16}.
    }
\vspace{1cm}
\label{Shock opening angle}
\end{figure*}

The formation of a shock cone through the BHL mechanism is a common phenomenon in both GR and alternative theories of gravity. However, properties such as the cone opening angle, the displacement of the stagnation point, and the rest-mass density of matter inside the cone vary depending on the parameters that characterize the underlying gravitational theory. Figs.\ref{dens_azim_diff_r_a_Q} and \ref{Shock opening angle} reveal the structures of such cones formed around Schwarzschild, Kerr, and EEH BHs with different values of the charge parameter $Q$. In Fig.\ref{dens_azim_diff_r_a_Q}, the azimuthal variation of the rest-mass density at $r=2.6M$ is plotted, thus illustrating the physical structure of the cone. In this strong-field region, the cone morphology depends on both spin $a$ and the charge parameter $Q$. As shown in our earlier studies, increasing values of $a$ bend the shock cone due to the curvature of spacetime, and consequently the locations that define the boundary of the cone are shifted \cite{Donmez2024PDU,Donmez2025JHEA}. In addition, with larger values of $Q$, the opening angle of the cone increases significantly, as is clearly seen in both Figs.\ref{dens_azim_diff_r_a_Q} and \ref{Shock opening angle}, while the maximum peak density remains nearly unchanged. The physical reason is that, at high $Q$, the NLED corrections associated with EEH gravity modify the effective gravitational attraction and simultaneously give rise to stronger turbulence. As expected, at $r=6.1M$, the influence of both $a$ and $Q$ is weaker than in the strong-field region, and therefore the variation in the cone opening angle is less pronounced, as can be seen in the bottom panels of Figs.\ref{dens_azim_diff_r_a_Q} and \ref{Shock opening angle}. In general, variations in $Q$ produce cones that are wider, slightly denser, and highly unstable. The dependence of the shock cone structure on the EEH BH parameters directly affects the modification and excitation of QPOs, as discussed in Section \ref{S4}. Furthermore, in the upper panel of Fig.\ref{Shock opening angle}, at $r=2.6M$, both $a$ and $Q$ contribute significantly to altering the cone geometry in the strong gravitational field, while in the lower panel ($r=6.1M$), farther away from the BH, the combined effect of spin and charge produces only modest changes in the opening angle of the cone compared to the cases without rotation or slowly rotating. This clearly shows that the combined influence of $a$ and $Q$ modifies the QPOs that may arise in such instabilities.

The dependence of shock cone morphology on the parameters $a$ and $Q$ in EEH gravity may yield signatures that can be traced in astrophysical observations, thus providing a means of testing the theory. In particular, in the strong-field region, the widening of the shock cone and the enhanced instabilities driven by turbulence directly affect both the amplitude and excitation of QPOs. The opening angle of the cone and the distribution of density within it produce observable modifications in the timing and spectral properties of accretion systems. From such data, information on the NLED corrections inherent in EEH gravity can be extracted. Thus, by identifying systematic deviations from the Schwarzschild and Kerr predictions, EEH gravity may be effectively tested in the strong-gravity regime.

\section{QPOs from Numerical Data}\label{S4}

In sections \ref{S3-1}, \ref{S3-2}, and \ref{S3-3}, the physical processes discussed in detail demonstrate that parameters $a$ and $Q$ play a crucial role in shaping the dynamics of matter trapped inside the shock cone. Specifically, the matter confined in the cavity undergoes QPOs, where the resulting modes are either amplified or suppressed depending on the BH parameters. Consequently, both classical Schwarzschild and EEH BH generate shock cones and associated cavities that give rise to distinct QPOs frequencies, which are directly related to the parameter $Q$ of the EEH solution.

\begin{figure*}[!htp]
  \vspace{1cm}
  \center
    \includegraphics[width=8.0cm,height=8.0cm]{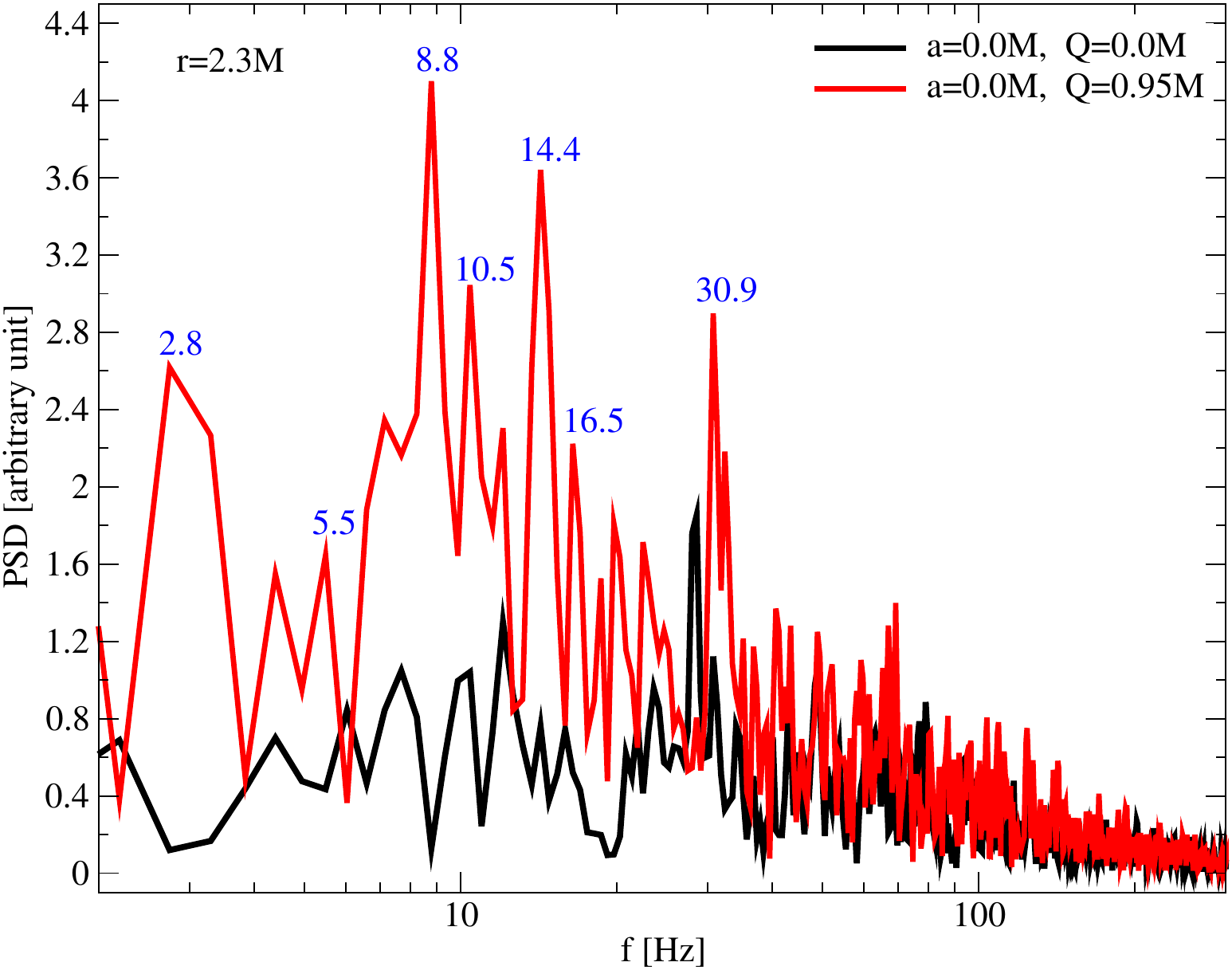} \;\;\;
    \includegraphics[width=8.0cm,height=8.0cm]{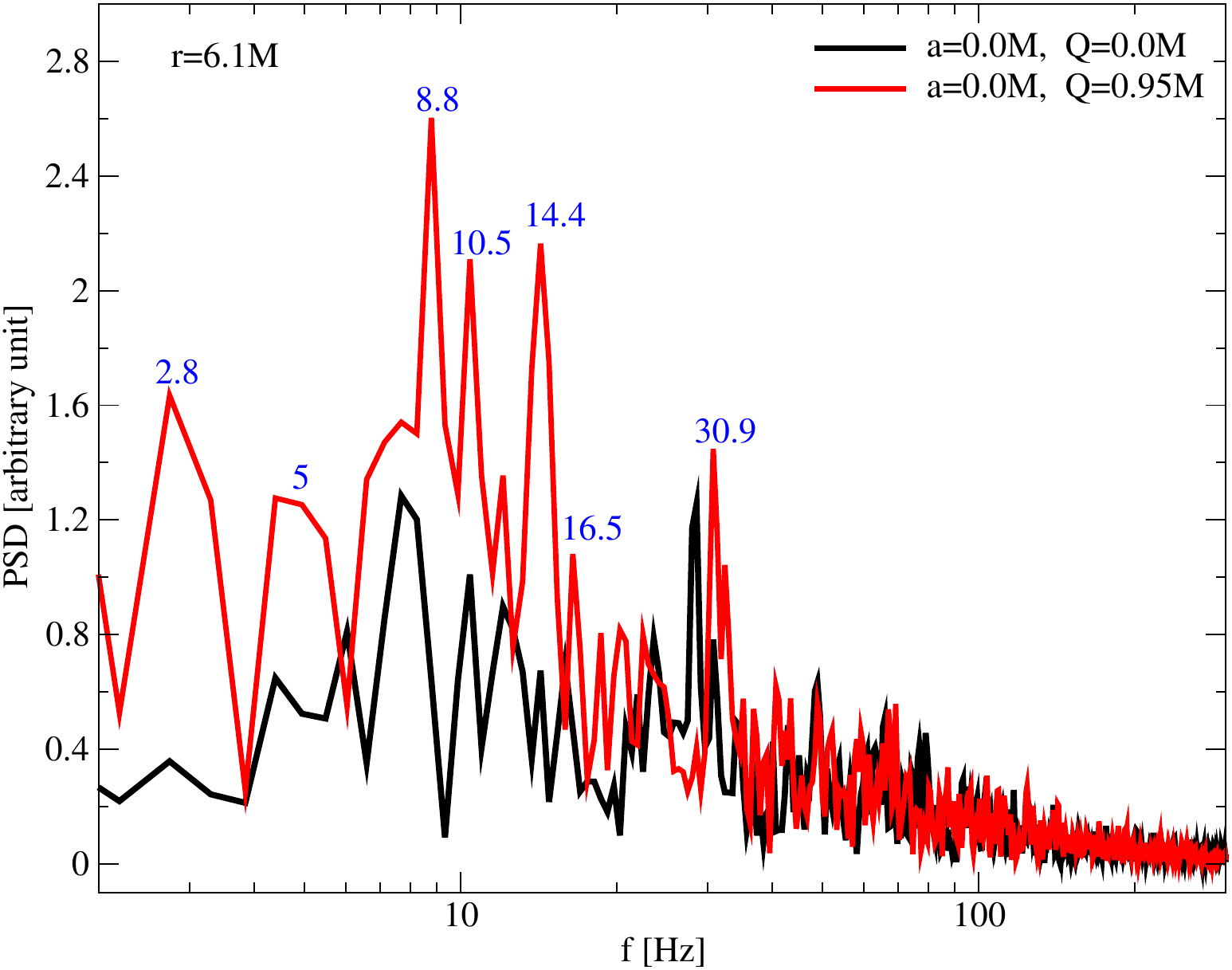}   
     \caption{The PSD analysis illustrating the fundamental QPOs modes and their nonlinear couplings at two distinct radial positions around the Schwarzschild and non-rotating EEH BHs with $M=10M_{\odot}$. The left panel corresponds to $r = 2.3M$, while the right panel corresponds to $r = 6.1M$. The presence of coincident peaks at identical frequencies in both cases indicates that these QPOs features represent a global oscillatory behavior of the accretion flow rather than localized phenomena\label{fig17}.
    }
\vspace{1cm}
\label{QPO_a00}
\end{figure*}

In Fig.\ref{QPO_a00}, we present a comparative analysis between a Schwarzschild BH and a non-rotating EEH BH with $Q=0.95M$, using the power spectral density (PSD) at two characteristic radii, which are a strong gravity region ($r=2.3M$) and a relatively weaker gravity region ($r=6.1M$). The numerical results reveal that, for both radii, the same set of centroid frequencies (e.g. $2.8$, $5.5$, $8.8$, $10.5$, $14.4$, $16.5$, $30.9$ Hz) appear in the PSD analysis. This demonstrates that the observed modes are not transient or artificial features, but rather global oscillation modes trapped within the shock cone cavity. Compared with the Schwarzschild case, the QPOs peaks observed around the EEH BH at $r=2.3M$ show a significant enhancement in power, making the peaks more pronounced and revealing a richer overtone structure. However, at $r=6.1M$, the amplitudes of the peaks are comparable to or slightly lower than those of Schwarzschild. This indicates that the NLED corrections associated with the EEH charge are particularly effective in the strong-field regime close to the BH.

The redistribution of QPOs amplitudes in the PSD carries important observational implications as given in Fig.\ref{QPO_a00}. For EEH BHs, the inner disk region generates stronger QPOs, which likely contribute to the hard $X-$ray variability, whereas the outer disk region mainly contributes to the soft $X-$ray band. Moreover, the stronger inner peaks emphasize the harmonic relations. Specifically, the low-frequency ladder formed by $2.8$, $5.5$, and $8.8$ Hz displays near-integer harmonics, while the higher-frequency set of $10.5$, $14.4$, $16.5$, and $30.9$ Hz yields commensurate ratios such as $10.5:16.5 \approx 3:2$, $10.5:14.4 \approx 4:3$, $5.5:10.5 \approx 2:1$, and $10.5:30.9 \approx 3:1$. These near-integer ratios are the hallmark of nonlinear coupling and parametric resonance, mechanisms frequently invoked to explain twin-peak QPOs in X-ray binaries.

In contrast, the Schwarzschild case exhibits a more balanced distribution of peak power between the inner and outer disk regions, with less prominent harmonic structures. Thus, although both geometries produce the same set of centroid frequencies, the EEH charge enhances the amplitudes of the observable modes, increasing their detectability. This distinction suggests that with future high-sensitivity telescopes, observational evidence of stronger inner-disk harmonics and relatively weaker or Schwarzschild like outer disk harmonics could serve as a test of EEH gravity, potentially revealing deviations from classical GR.

\begin{figure*}[!htp]
  \vspace{1cm}
  \center
     \includegraphics[width=8.0cm,height=8.0cm]{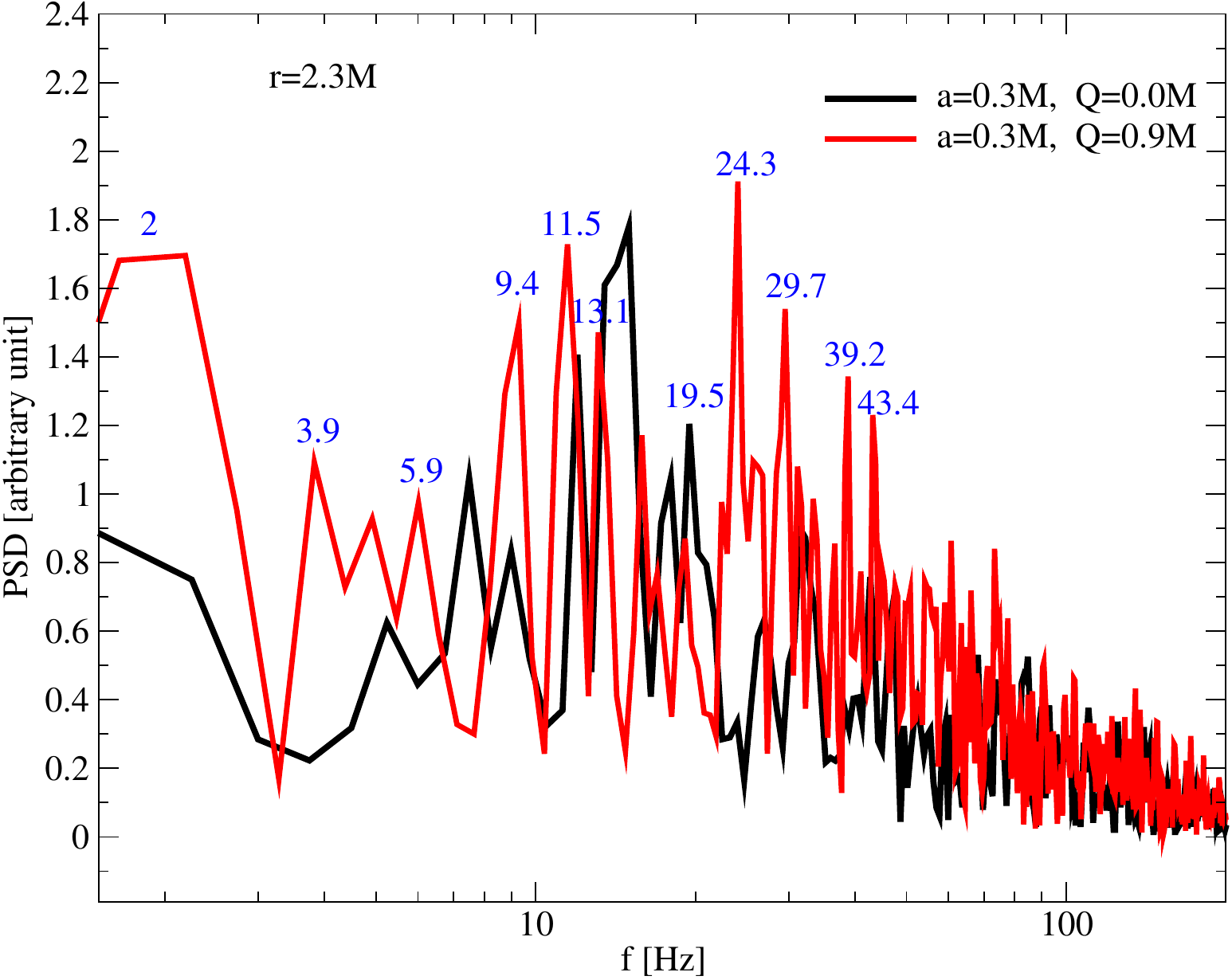} \;\;  
      /;\includegraphics[width=8.0cm,height=8.0cm]{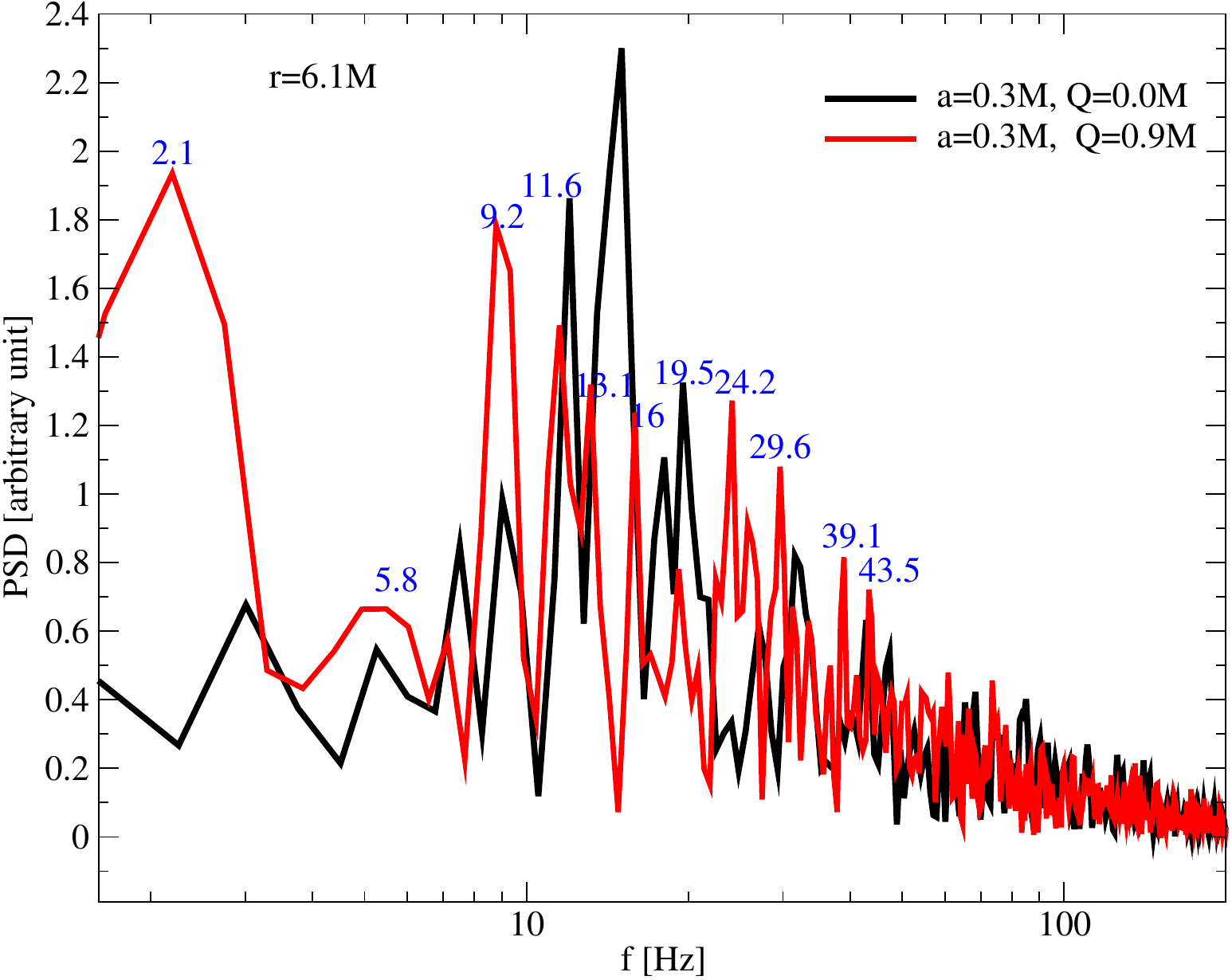}  \\ 
  \vspace{0.2cm}
      \includegraphics[width=8.0cm,height=8.0cm]{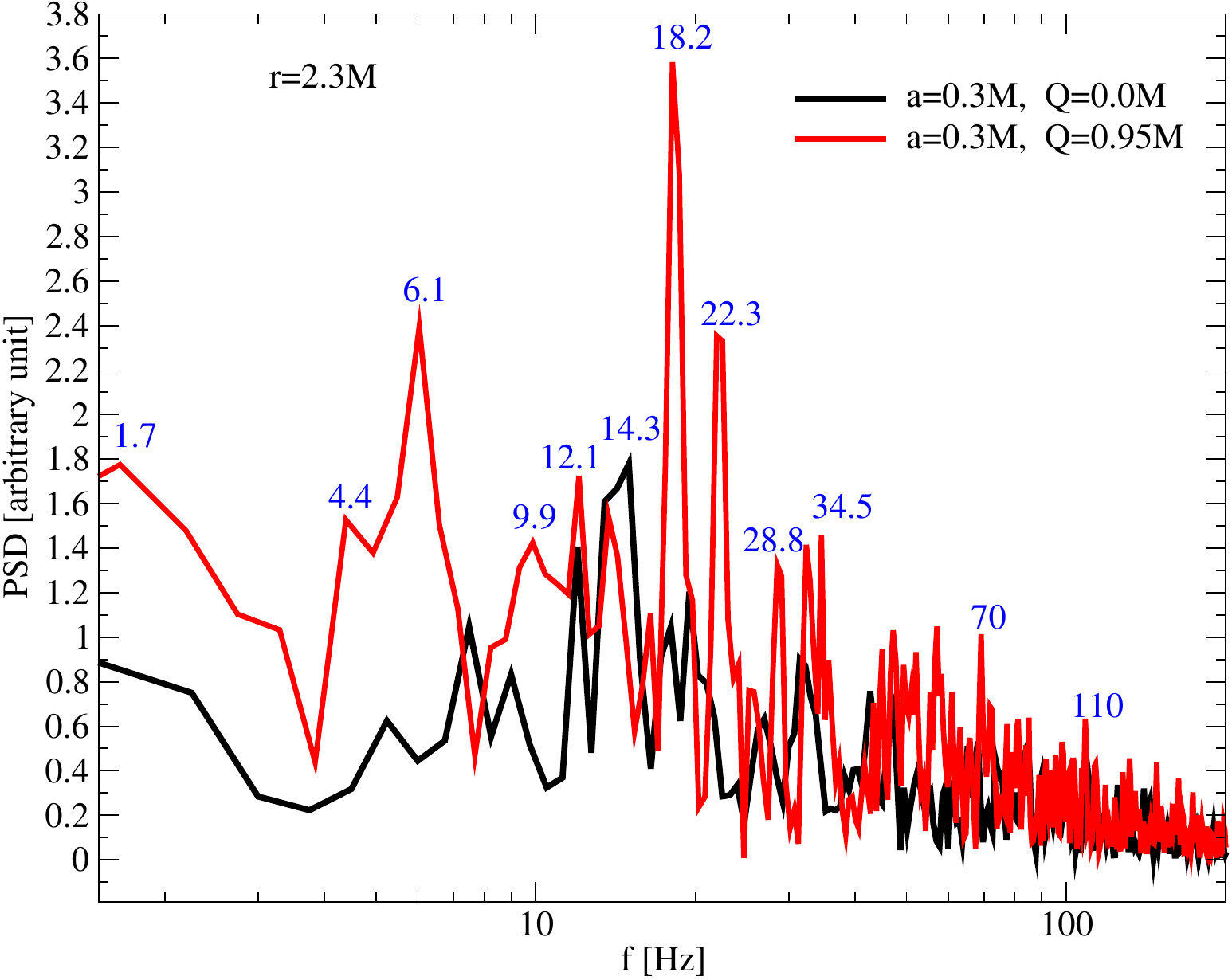} \;\;  
      /;\includegraphics[width=8.0cm,height=8.0cm]{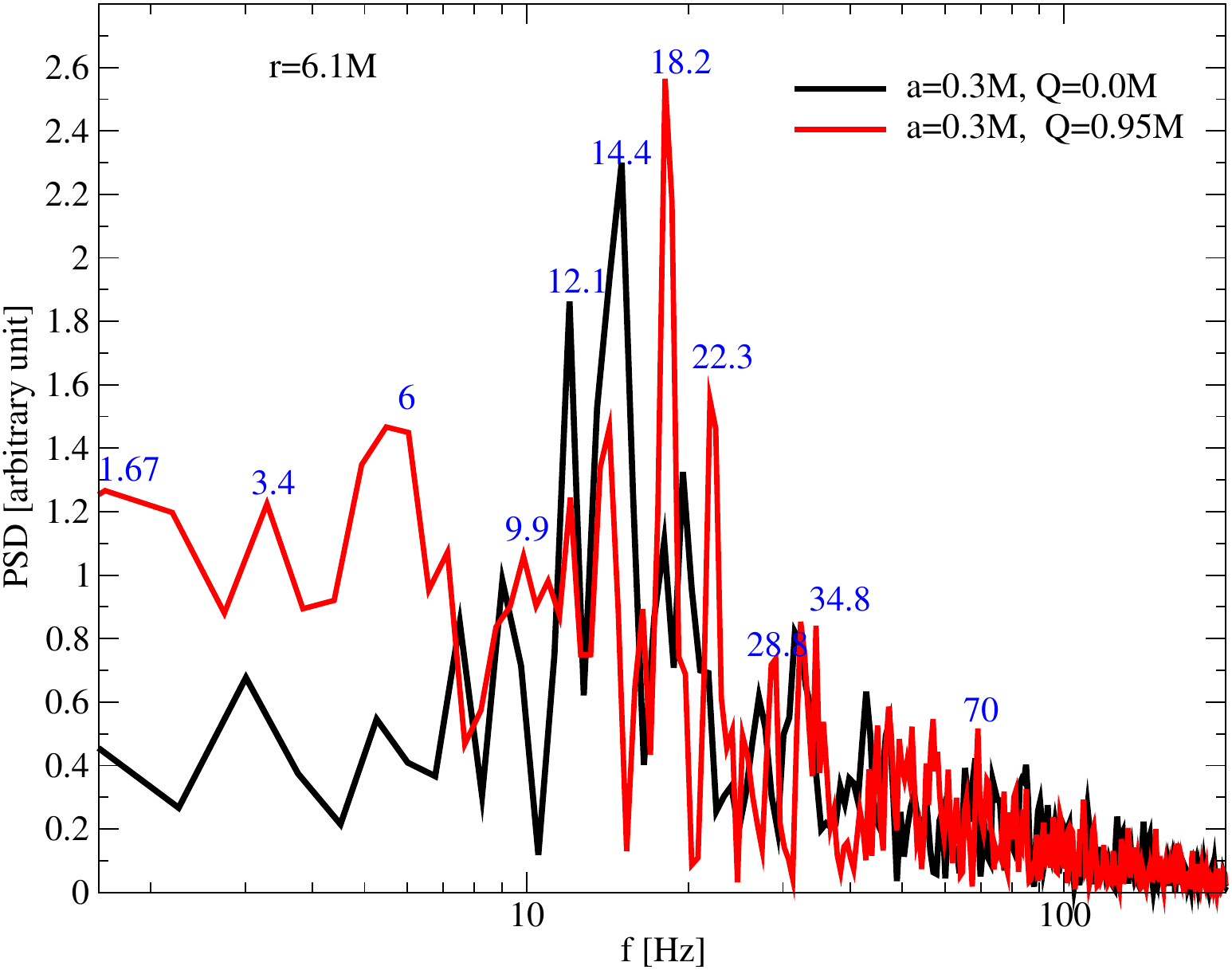}   
     \caption{As in Fig.\ref{QPO_a00}, a PSD analysis is performed, but this time we compare the QPOs arising at different $Q$ values around the EEH BH with spin parameter $a = 0.3M$ to the corresponding results for the Kerr BH. The top row presents the QPOs PSD analysis for $Q = 0.9M$, while the bottom row shows the calculations for $Q = 0.95M$, both contrasted with the Kerr case\label{fig18}.
    }
\vspace{1cm}
\label{QPO_a03}
\end{figure*}

In Fig.\ref{QPO_a03}, we present the PSD analysis of QPOs around a slowly rotating Kerr BH ($a=0.3M$) and EEH BHs with charge parameters $Q=0.9M$ and $Q=0.95M$, allowing a direct comparison between the EEH results and the Kerr baseline. The PSD spectra computed at different radial locations show that the centroid frequencies remain nearly identical across all cases. This demonstrates that the modes are global in nature and not artifacts of localized numerical effects. At the same time, when compared with the Kerr solution, the amplitudes of QPOs in the EEH case are significantly enhanced in the strong-gravity region because of the presence of the charge parameter. As a result, QPOs extracted from the EEH BH exhibit higher observability than those produced in the slowly rotating Kerr model. However, in the weaker-gravity region near the ISCO ($r=6.1M$), the amplitudes of the QPOs frequencies are suppressed, indicating that the plasma flow in this region becomes more stable. Increasing the charge further to $Q=0.95M$ accentuates this behavior. In the inner disk, the PSD reveals a richer overtone structure, with nonlinear coupling among harmonics generating HFQPOs that extend up to $\sim 70{-}110$ Hz. However, in the outer disk, these frequencies are strongly suppressed, reducing their relative detectability.

Systematic analysis also reveals the presence of quasi-comparative frequency relationships, similar to those observed in Fig.\ref{QPO_a00}. These are $18.2:12.1 \approx 3:2$, $34.5:22.3 \approx 3:2$, $12.1:6.1 \approx 2:1$,  $28.8:14.4 \approx 2:1$, $14.4:9.9 \approx 4:3$ .
These ratios are clear signatures of harmonic resonances and nonlinear couplings, which are well-known physical mechanisms underlying QPOs phenomenology. 

Interestingly, the slowly rotating Kerr case shows a more balanced power distribution between the inner and outer disk regions. In contrast, the EEH BH shifts the dominant QPOs power inward, strengthening HFQPOs and making them more observable. This redistribution has direct observational implications. In the EEH scenario, the inner harmonics are expected to dominate the hard $X-$ray band, while the diminished outer peaks reduce variability in the soft $X-$ray band.

\begin{figure*}[!htp]
  \vspace{1cm}
  \center
 \includegraphics[width=8.0cm,height=8.0cm]{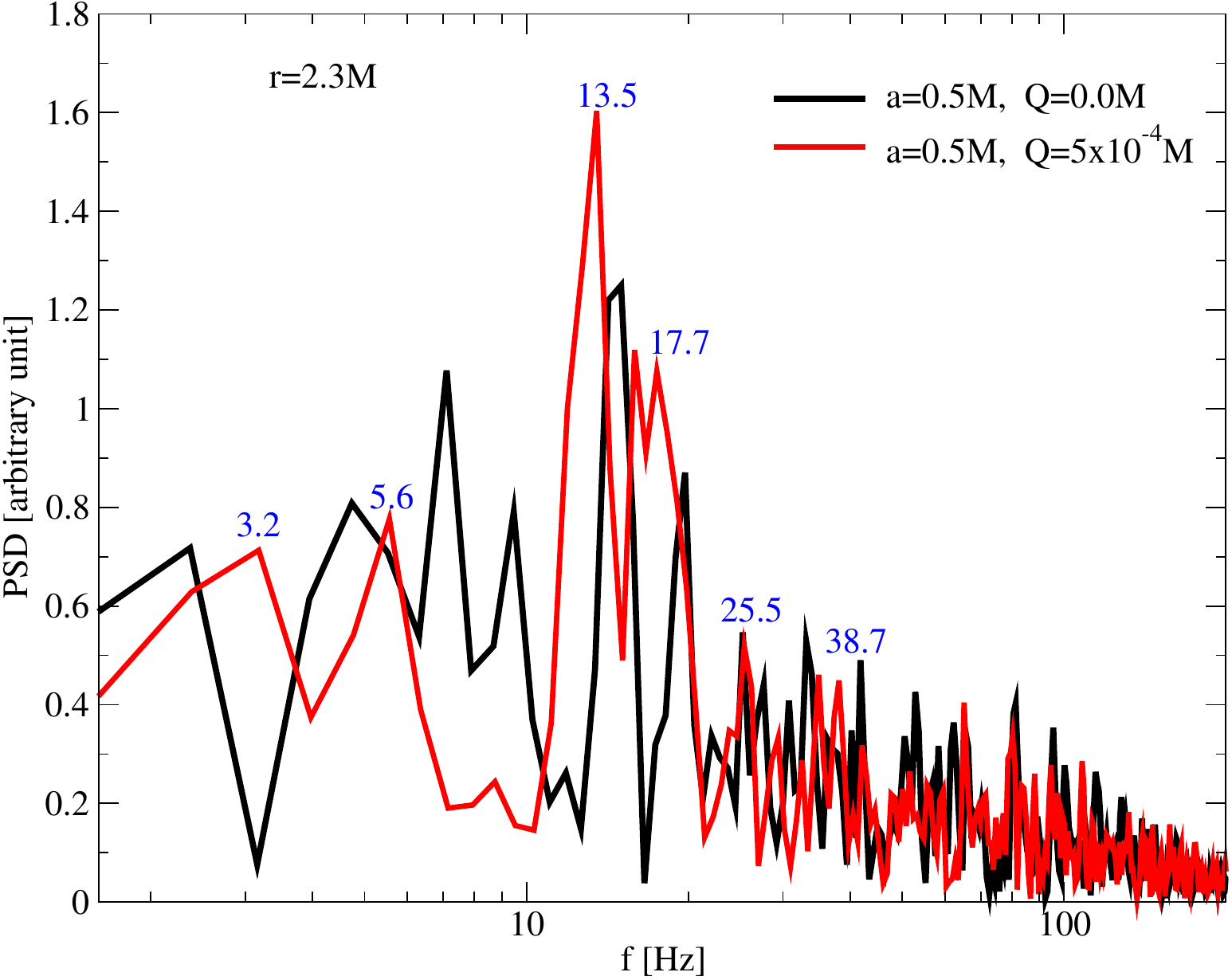} \;\; 
   /;\includegraphics[width=8.0cm,height=8.0cm]{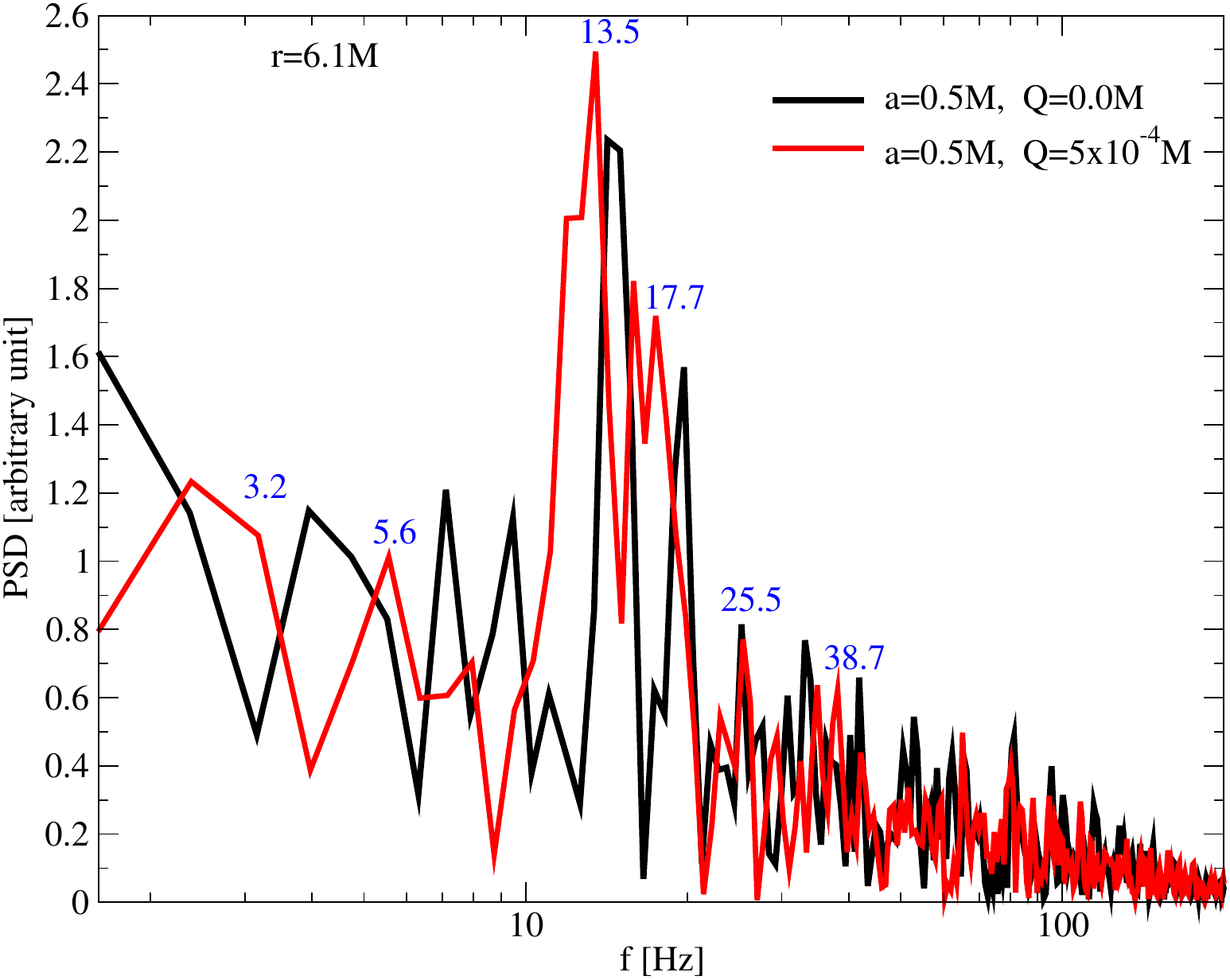} \\
  \vspace{0.2cm}
  \includegraphics[width=8.0cm,height=8.0cm]{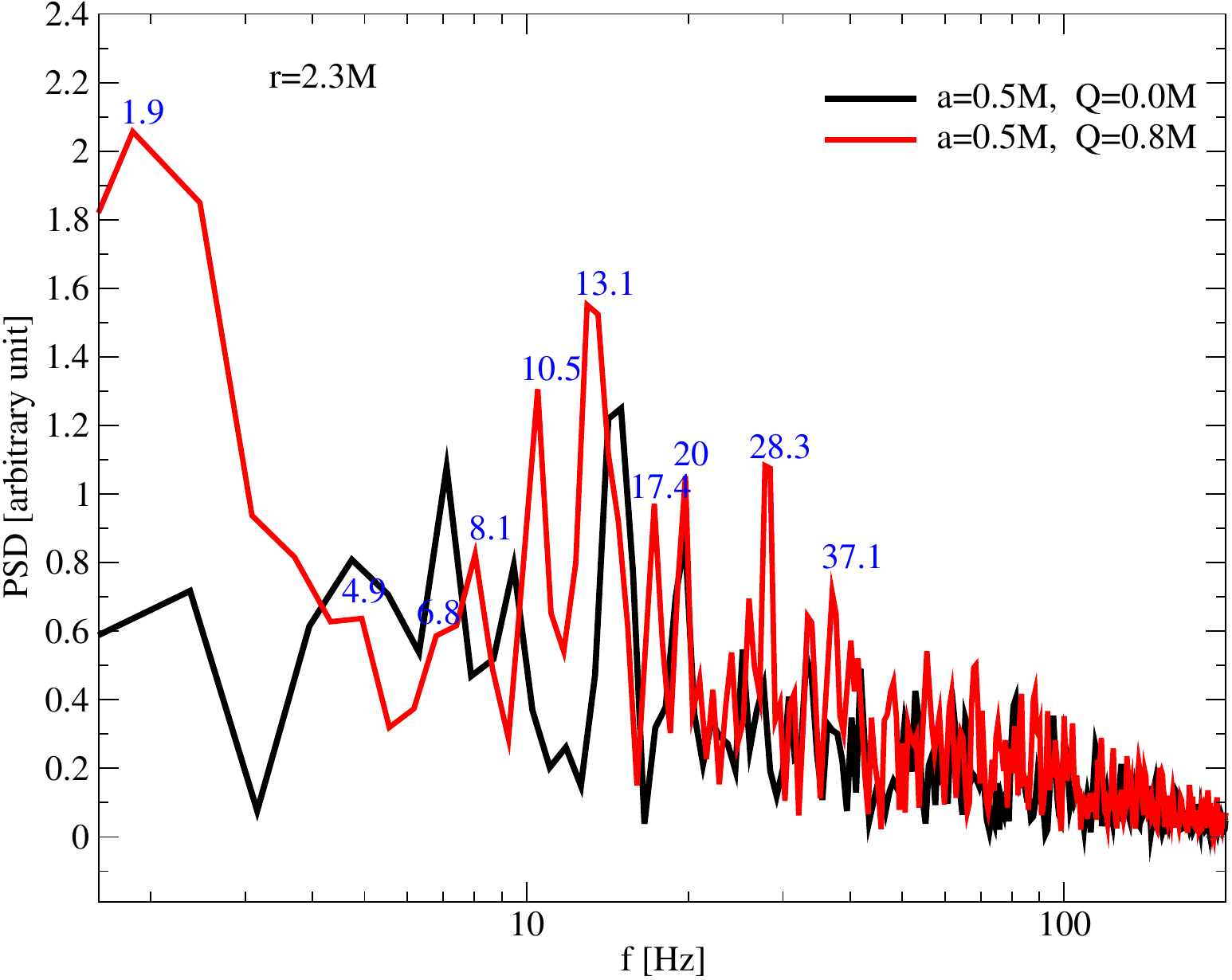} \;\;
    /;\includegraphics[width=8.0cm,height=8.0cm]{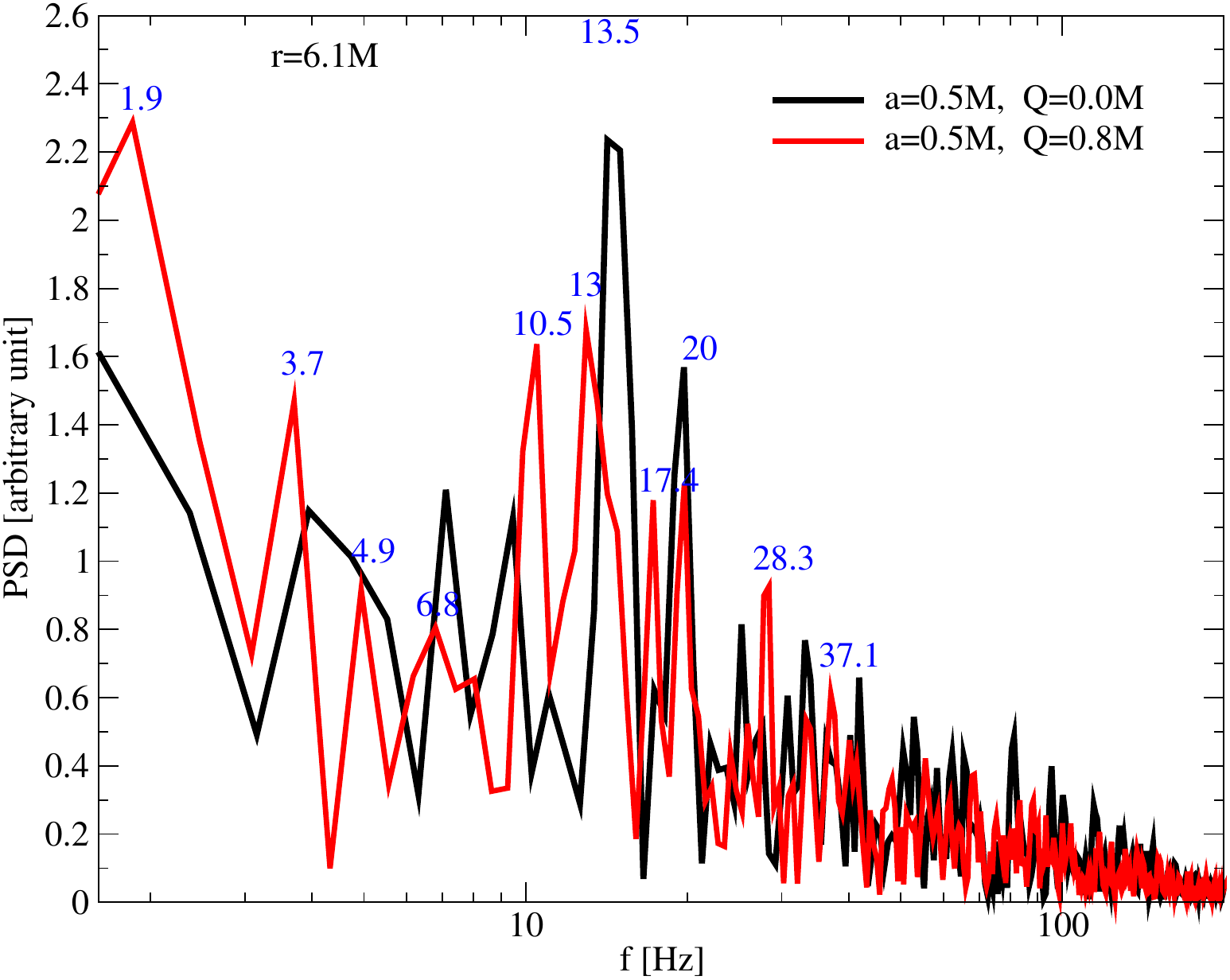} 
     \caption{Same analysis as in Fig. \ref{QPO_a03}, but for the EEH BH with spin parameter $a = 0.5M$ and varying charge $Q$. The resulting QPOs frequencies are computed and compared with those of the Kerr BH at the same spin, highlighting the impact of EEH corrections on oscillatory behavior\label{fig19}.
    }
\vspace{1cm}
\label{QPO_a05}
\end{figure*}

In Fig.\ref{QPO_a05}, the shock cone around Kerr and EEH BHs with spin parameter $a=0.5M$ is analyzed using PSD. Comparisons are made for the EEH charge parameters $Q=5\times10^{-4}M$ and $Q=0.8M$. For a very small charge ($Q=5\times10^{-4}M$), the centroid frequencies remain essentially unchanged. This indicates that the oscillation modes are natural and do not strongly depend on such a small charge parameter. At $r=2.3M$, PSD analysis reveals peaks at $3.2$, $5.6$, $13.5$, $17.7$, $25.5$, and $38.7$ Hz. Almost the same peaks also appear at $r=6.1M$ at very similar frequencies. In this regime, the EEH BH produces a PSD nearly identical to the Kerr case, with only slight shifts in the peak frequencies in both the inner and outer disk regions. The harmonic relationships identified in this case are $38.7:25.5 \approx 3:2$, $25.5:17.7 \approx 4:3$, $17.7:13.5 \approx 4:3$, $5.6:3.2 \approx 7:4$. These ratios correspond to well-known resonant patterns frequently associated with twin-peak QPOs phenomenology, providing evidence of nonlinear coupling even at a very small charge.  

When the charge parameter increases to $Q=0.8M$, the influence of the EEH correction becomes very pronounced. In the inner disk region ($r=2.3M$), the QPOs peaks are strongly amplified, and the presence of overlapping modes produces a much richer overtone structure. In contrast, in the outer disk region ($r=6.1M$), LFQPOs remain visible with strong peaks, but HFQPOs are more strongly suppressed. Although the centroid frequencies remain broadly consistent with the Kerr mode set, the redistribution of amplitudes clearly emphasizes inner-disk dominance.  In this regime, the outer disk still exhibits commensurate ratios such as $3:2$, $2:1$, and $4:3$, demonstrating that parametric resonance and nonlinear couplings also occur in these regions. Observationally, this creates a strong contrast with Schwarzschild or Kerr BHs. For EEH BHs with large $Q$, the most significant changes occur in the inner disk, where strong QPOs drive the variability that contributes to hard X-ray emission. Meanwhile, the outer disk continues to produce variability in the soft X-ray band, similar to earlier models.  

Due to the effects of the $Q$ parameter, QPOs around EEH BHs display modifications that increase the detectability of HFQPOs and simultaneously introduce measurable changes in time dependent spectral data. These signatures may therefore provide a potential observational test of EEH gravity.

\begin{figure*}[!htp]
  \vspace{1cm}
  \center
   \includegraphics[width=8.0cm,height=8.0cm]{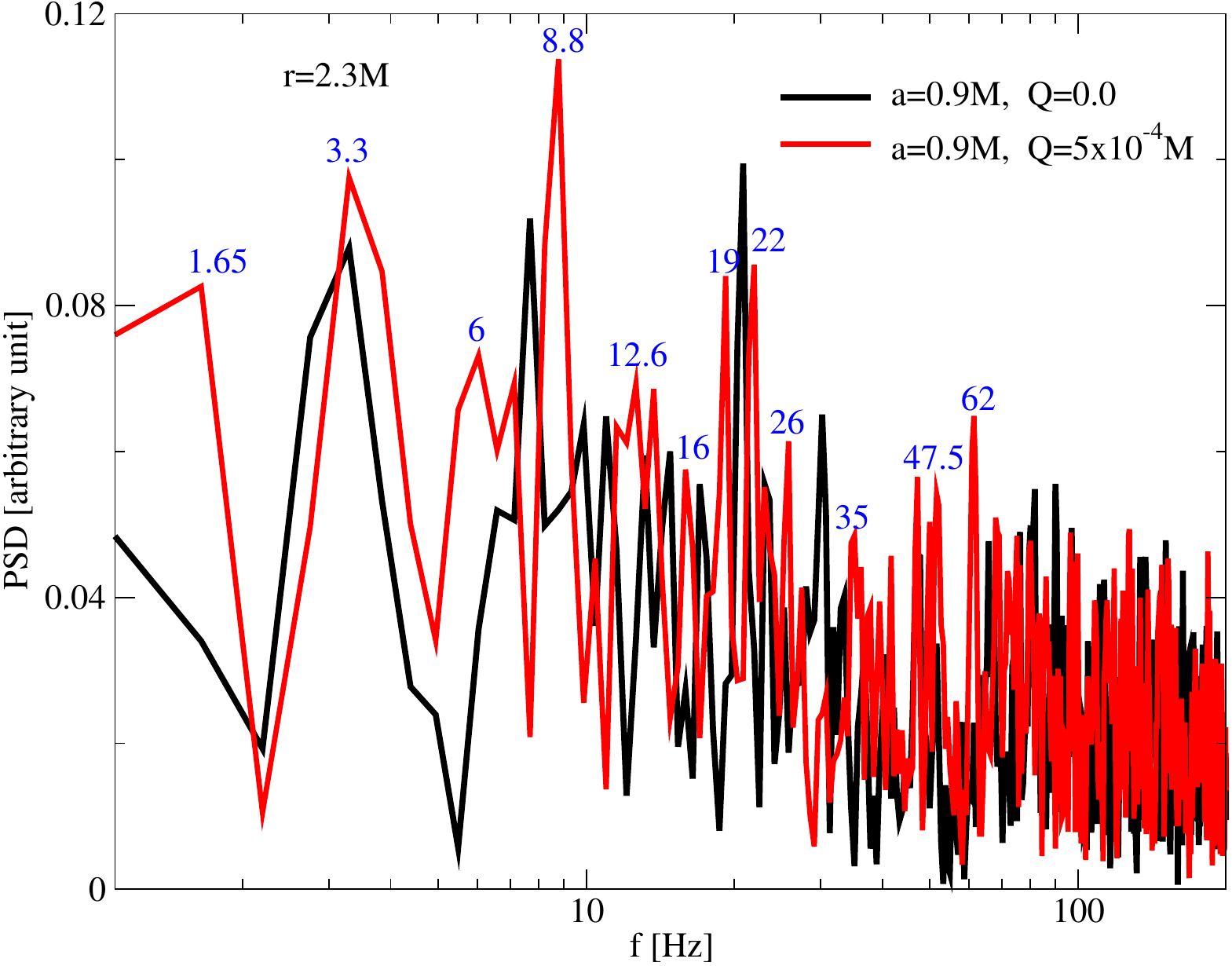}\;\;\;\; \includegraphics[width=8.0cm,height=8.0cm]{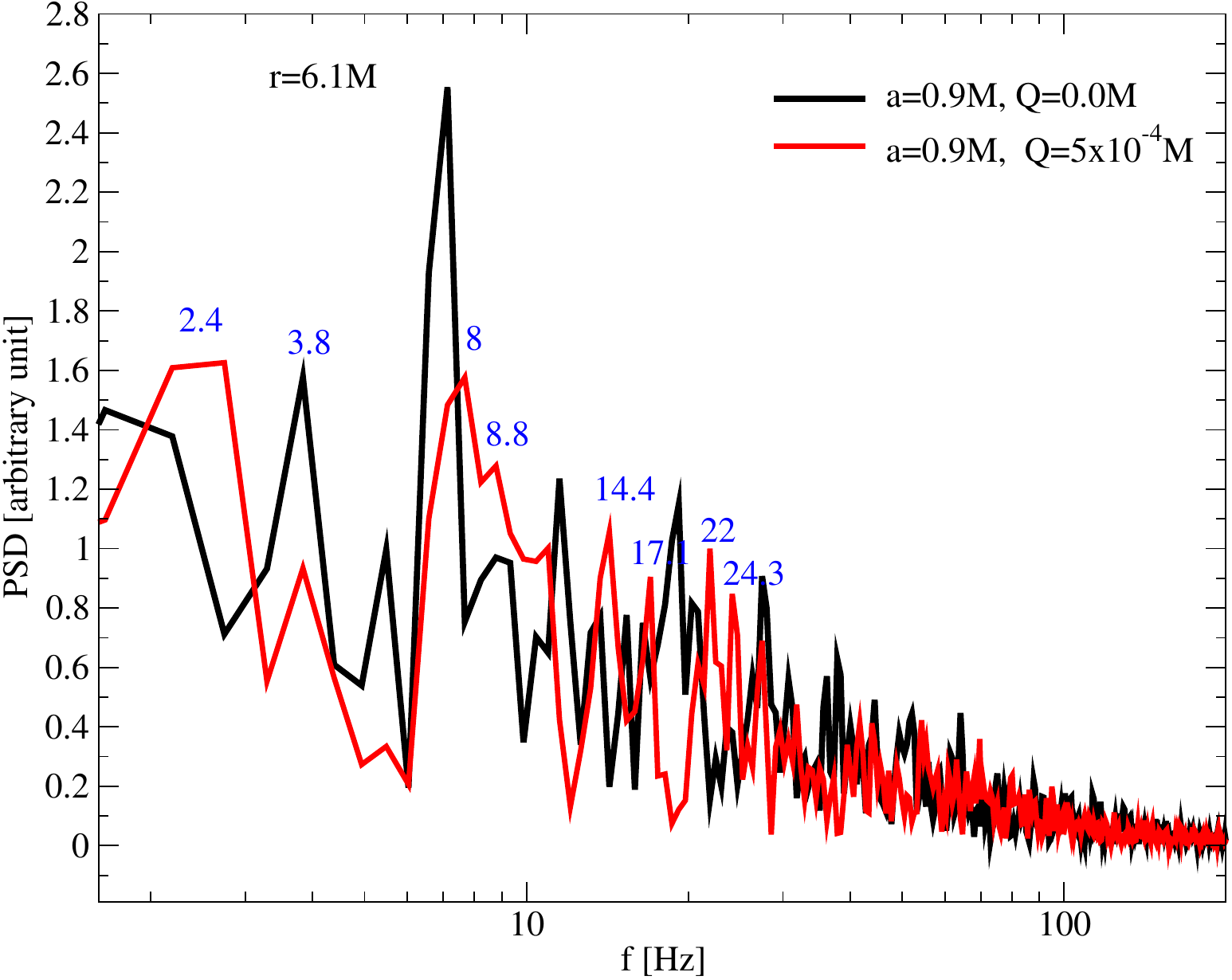} \\
  \vspace{0.2cm}
   \includegraphics[width=8.0cm,height=8.0cm]{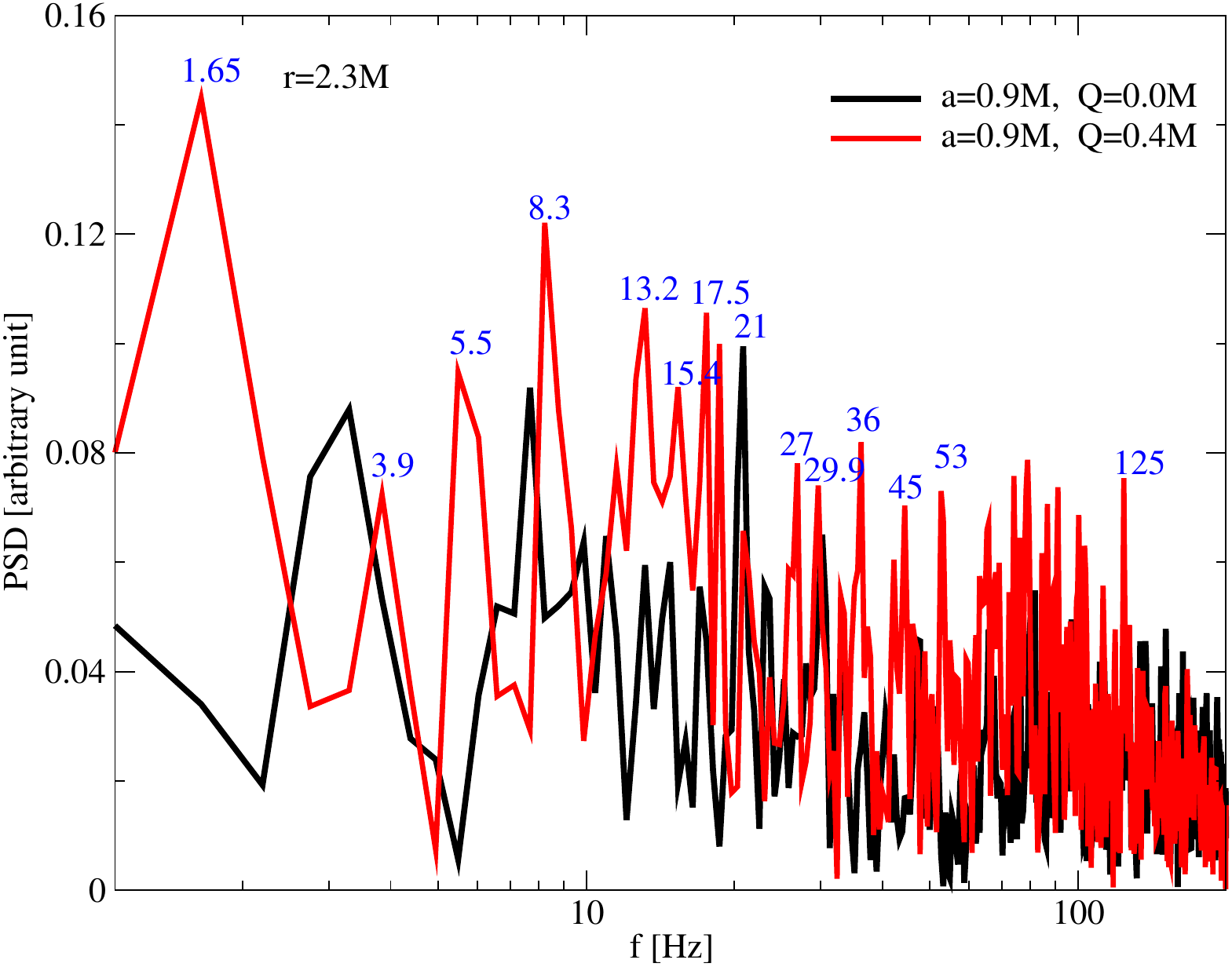}\;\;\;\; \includegraphics[width=8.0cm,height=8.0cm]{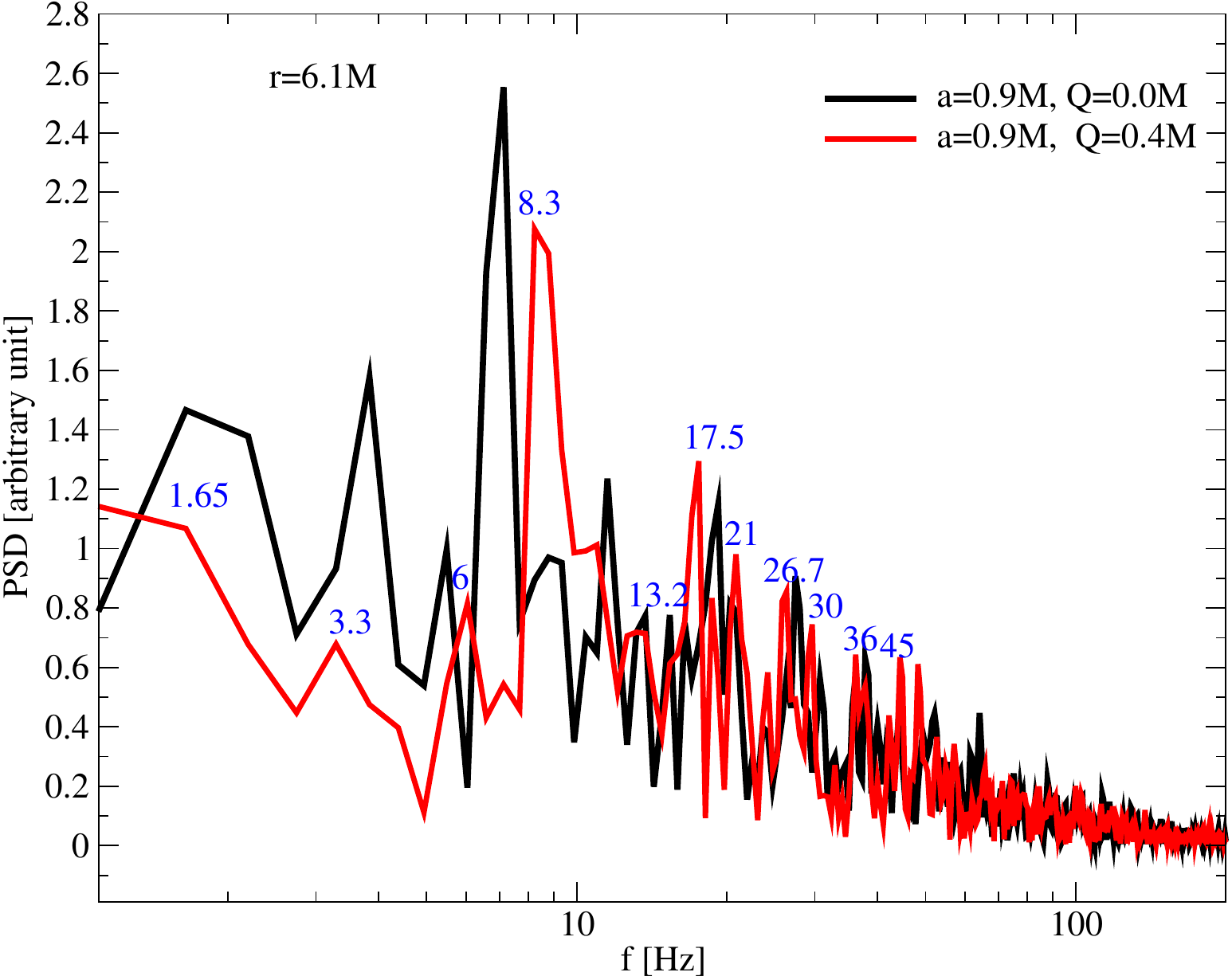} 
     \caption{Same analysis as in Fig. \ref{QPO_a05}, now extended to the EEH BH with spin parameter $a = 0.9M$ and varying charge $Q$, where the resulting QPOs frequencies are computed to assess the influence of strong spin-charge coupling on the oscillatory behavior\label{fig20}.
    }
\vspace{1cm}
\label{QPO_a09}
\end{figure*}

In Fig.\ref{QPO_a09}, the PSD analysis of QPOs is presented for a rapidly rotating BH ($a=0.9M$), comparing the Kerr solution with EEH BHs for charge parameters $Q=5\times10^{-4}M$ and $Q=0.4M$.  For the case of $Q=5\times10^{-4}M$, the centroid frequencies are virtually identical to those of the Kerr BH. At $r=2.3M$, peaks appear at $1.65$, $3.3$, $6$, $8.8$, $12.6$, $16$, $19$, $22$, $26$, $35$, $47.5$,  and $62$ Hz. By contrast, at $r=6.1M$, peaks are observed at $2.4$, $3.8$, $8$, $8.8$, $14.4$, $17.1$, $22,$ and $24.3$ Hz. The close similarity of frequencies at these two radial points, even when considering numerical uncertainties, shows that the oscillation modes are global rather than local in origin. This trend was also confirmed in Figs.\ref{QPO_a00}, \ref{QPO_a03}, and \ref{QPO_a05}. As in previous BH models, resonant ratios such as $8.8:6 \approx 3:2$, $26:12.6 \approx 2:1$, $22:14.4 \approx 3:2$ are observed. Interestingly, although the charge is very small, the QPOs peaks around the EEH BH at $r=2.3M$ are somewhat stronger than in the Kerr case, while at $r=6.1M$ the opposite occurs. This inversion may be due to the strong spacetime curvature in the inner region of the rapidly rotating BH. The combined effects of spin and charge result in stronger EEH peaks in the inner disk region, whereas Kerr dominates in the outer disk.  

When the charge parameter is increased to $Q=0.4M$, the influence of EEH corrections on the PSD becomes much clearer. In the inner disk at $r=2.3M$, a very rich overtone structure emerges, with peaks at $1.65$, $3.9$, $5.5$, $8.3$, $13.2$, $15.4$, $17.5$, $21$, $27$, $29.9$, $36$, $45$, $53$, and $125$ Hz. This dense frequency distribution creates strong harmonic families, including $8.3:5.5 \approx 3:2$, $27:13.2 \approx 2:1$, $29.9:21 \approx 4:3$, $53:36 \approx 3:2$. At the same time, the outer disk region at $r=6.1M$ also exhibits commensurate pairs, and in some cases the outer-disk amplitudes are even stronger than those in the inner disk. This result is different from Fig.\ref{QPO_a09} for lower-spin cases, because in the $a=0.9M$ scenario, the severe spacetime curvature combines with the EEH charge parameter to strongly influence the PSD. The large rotational angular momentum of the high-spin BH suppresses turbulence in the inner cavity, redistributing the oscillatory power outward. Consequently, unlike the Schwarzschild or low-spin EEH picture, where QPOs power is concentrated in the inner disk, the high-spin EEH BH with moderate charge can produce relatively stronger outer-disk oscillations.  

In addition, the strong Lense-Thirring precession in the rapidly rotating spacetime overlaps with cavity-trapped frequencies and other resonances, generating a very rich set of peaks as seen in the left column of the Fig.\ref{QPO_a09}. Although the amplitudes of the QPOs frequencies in the outer disk ($r=6.1M$) may exceed those of the inner disk ($r=2.3M$), testing EEH gravity from the outer region appears less feasible, because the Kerr solution still produces stronger oscillations there. In contrast, in the inner disk region, advances in observational technology may enable the detection of EEH corrections. The QPOs around EEH BHs in the inner regions are stronger and display a richer overtone structure than those around Kerr, providing a promising means of testing the effects of the EEH charge parameter.

\section{Parameter Space Analysis}\label{S5}

Having established the dynamical features of accretion flows and the resulting QPOs described in the previous sections, we now investigate in greater detail their dependence on the physical parameters of the BH. In particular, we explore how the EEH charge parameter $Q$ and the spin parameter $a$, individually or in combination, affect the strength, frequency, and observability of QPOs modes at different radial positions. This systematic analysis enables us to distinguish universal trends from local effects, demonstrating how strong-field amplification, ISCO-scale modulations, and hydrodynamical interactions collectively shape the QPOs phenomenology. Revealing these details provides the possibility of testing EEH gravity using QPOs data obtained from astrophysical observations.

\begin{figure*}[!htp]
  \vspace{1cm}
  \center
     \includegraphics[width=8.0cm,height=8.0cm]{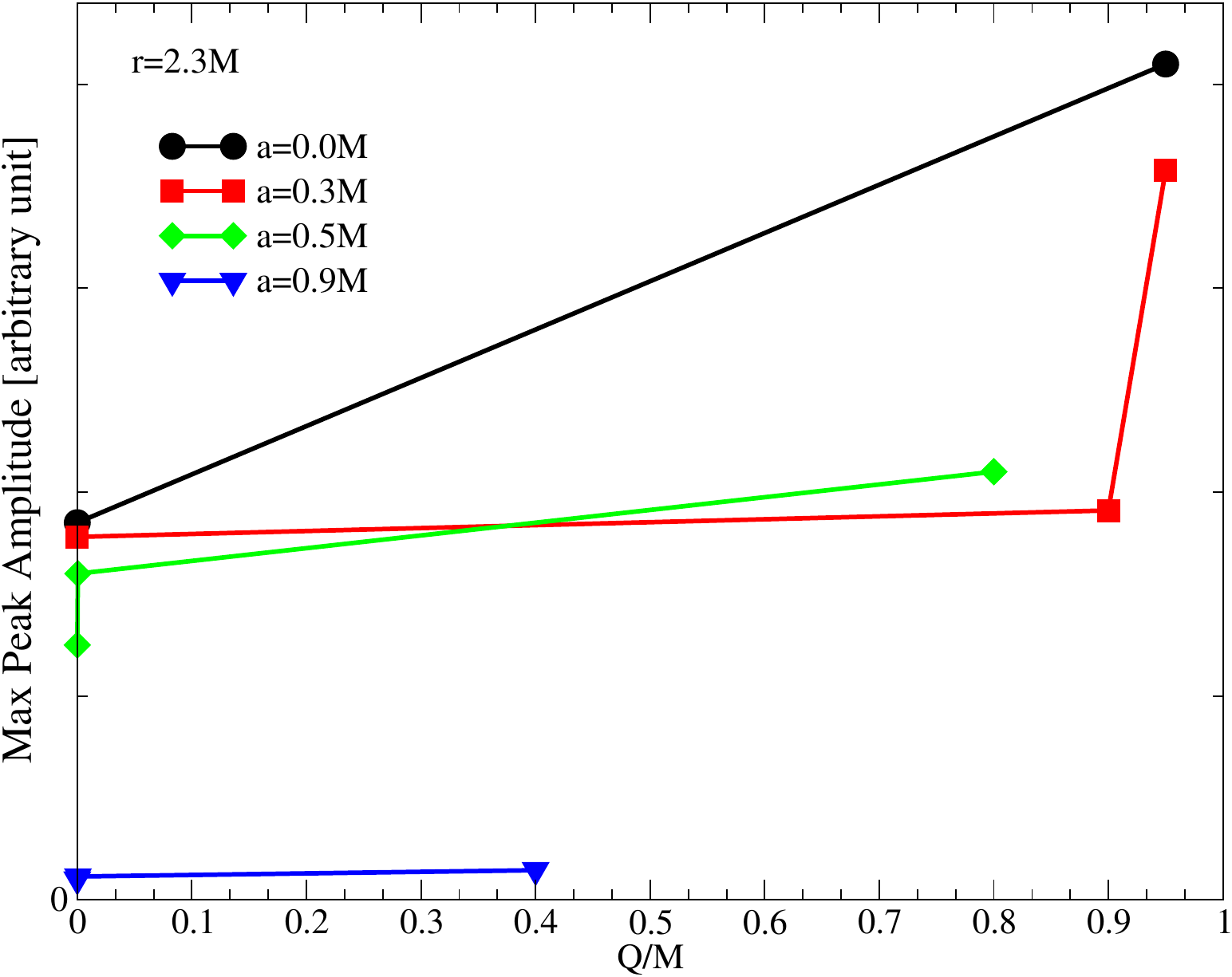} \;\;\;
     \includegraphics[width=8.0cm,height=8.0cm]{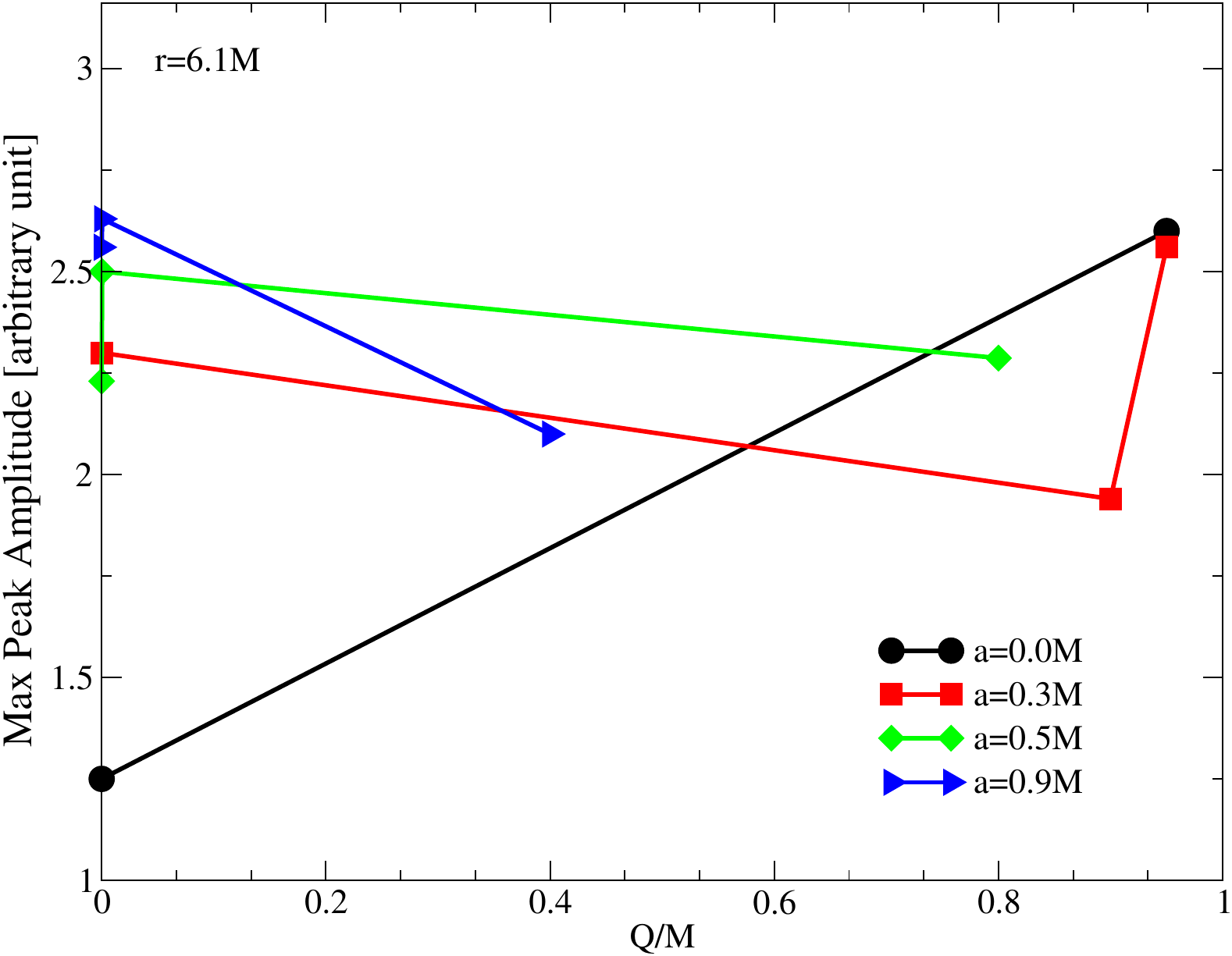}
     \caption{Variation of the maximum QPOs peak amplitudes with respect to $Q$ for different spin parameters at $r = 2.3M$ (left) and $r = 6.1M$ (right). In the strong-field region ($r = 2.3M$), amplitudes increase monotonically with $Q$, becoming exponential for $Q > 0.9M$ at low spins. At larger radii ($r = 6.1M$), amplitudes generally decrease with $Q$, though for $Q > 0.9M$ they rise again, highlighting the strong influence of EEH corrections on QPOs observability\label{fig21}.
    }
\vspace{1cm}
\label{QPO_max_peak_ampt}
\end{figure*}

Fig.\ref{QPO_max_peak_ampt} shows the dependence of the maximum QPOs peak amplitudes on the EEH charge-to-mass ratio ($Q$) for different BH spin parameters at two representative radii. The left panel corresponds to $r = 2.3M$, a region deep in the strong-field regime where the gravitational potential is maximal and QPOs are generated from modes trapped inside the cavity. In this case, the peak amplitudes grow monotonically with increasing $Q$, with an exponential rise observed for $Q > 0.9M$ in the non-rotating ($a=0M$) and slowly rotating ($a=0.3M$) cases. The right panel corresponds to $r = 6.1M$, near the ISCO where the gravitational field is weaker. Here, the amplitudes generally decrease as $Q$ increases, except at high charge ($Q > 0.9M$), where the peaks again exhibit a growth trend. For $a = 0M$, only a single value of $Q$ is allowed, so the global trend cannot be established.

In general, these results indicate that the EEH correction parameter $Q$ plays a decisive role in modulating the strength of QPOs. In the strong-field region, the EEH effects significantly amplify the oscillatory modes, while at larger radii the amplitudes are suppressed before rising again at high $Q$. Compared with Schwarzschild and Kerr BHs, EEH corrections enhance the excitation of modes in the strong-gravity regime, thereby improving the potential observability of QPOs in realistic astrophysical systems.

Figure \ref{QPO_max_peak_ampt} presents the variation of the QPOs frequencies associated with the maximum amplitude peaks (given in Fig.\ref{QPO_max_peak_ampt} as functions of the EEH charge-to-mass ratio ($Q$) for different BH spin parameters. The left panel corresponds to $r = 2.3M$, a region deep in the strong-field regime where the gravitational potential dominates the plasma dynamics. Here, the QPOs frequencies exhibit a clear transition from HFQPOs to LFQPOs as $Q$ increases. The lowest LFQPOs are observed for the rapidly rotating case ($a = 0.9M$), consistent with theoretical expectations of the Lense-Thirring effect, which predicts that LFQPOs are preferentially generated around rapidly rotating BHs.

In contrast, the right panel shows the behavior at $r = 6.1M$, near the ISCO and in a relatively weaker gravitational field. Unlike the monotonic trends seen closer to the horizon, the frequencies here display a nonlinear evolution: with increasing $Q$, they initially shift toward the LFQPOs regime but subsequently return toward the HFQPO side. This non-monotonic behavior appears only for intermediate and high spin values ($a = 0.3M$ and $a = 0.9M$). The results suggest that while strong gravity near the horizon enforces a robust transition toward LFQPOs, farther out the interplay between gravitational and hydrodynamical forces introduces irregularities in the frequency evolution. In general, the EEH charge parameter $Q$ emerges as a critical factor in the regulation of QPOs modes, its effects being most pronounced in the strong-gravity regime and increasingly complex in weaker-field regions around the ISCO.

\begin{figure*}[!htp]
  \vspace{1cm}
  \center
   \includegraphics[width=8.0cm,height=8.0cm]{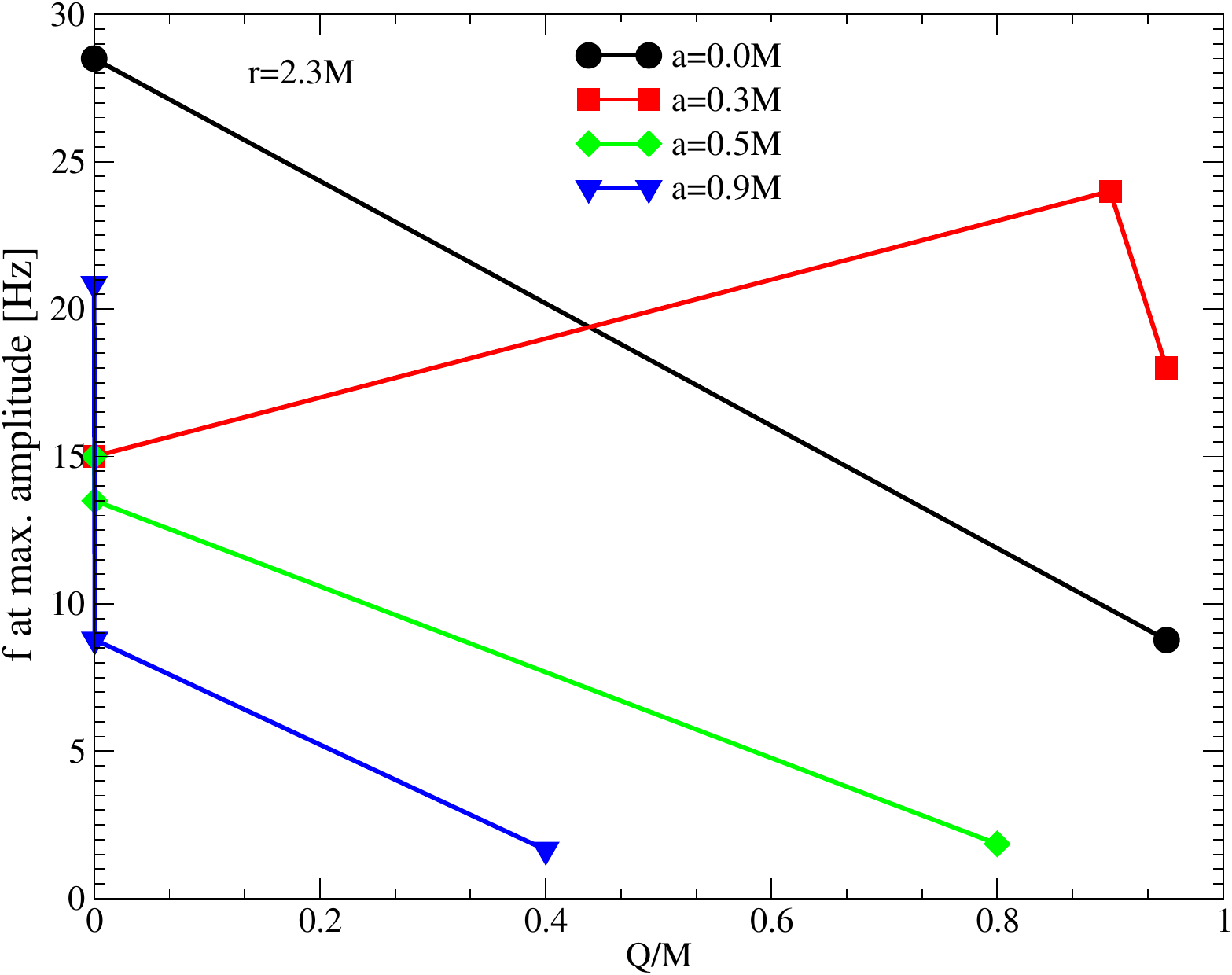} \;\;\;\includegraphics[width=8.0cm,height=8.0cm]{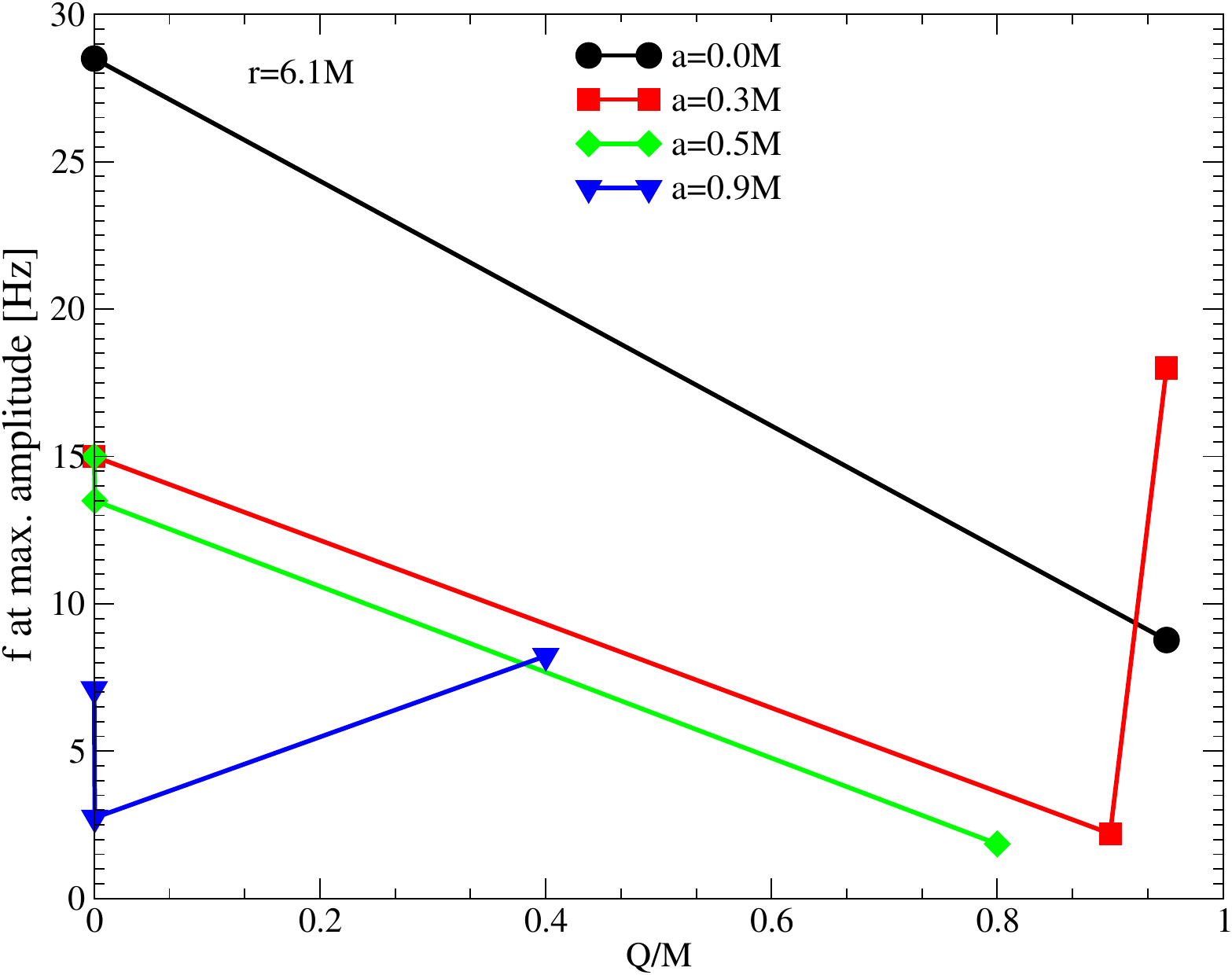} 
     \caption{Variation of QPOs frequencies corresponding to the maximum-amplitude peaks given in  Fig. \ref{QPO_max_peak_ampt}  with respect to $Q$ for different spin parameters. At $r=2.3M$ (left), the frequencies transition from HFQPOs to LFQPOs, with the lowest LFQPOs occurring for $a=0.9M$ due to the Lense–Thirring effect. At $r=6.1M$ (right), the behavior is less regular: frequencies initially shift toward LFQPOs and then back toward HFQPOs for $a=0.3M$ and $a=0.9M$, reflecting the competition between gravitational and hydrodynamical effects near the ISCO\label{fig22}.
    }
\vspace{1cm}
\label{QPO_max_peak_ampt_f}
\end{figure*}

\section{Astrophysical Implications and Observational Signatures}\label{S6}

The phenomena occurring around an EEH BH, together with the plasma and the dynamical structure of the shock cone, can be explained in terms of the numerical results obtained in this work. The spin parameter $a$ and the charge parameter $Q$ of the EEH BH not only influence the accretion mechanism around the BH, but also affect the long-term stability of the shock cone and the formation of QPOs trapped and excited within it. Variations in these parameters directly regulate the amount of matter accreted onto the BH, the intensity of turbulence generated inside the shock cone, and the strength of QPOs mode excitation and their nonlinear couplings. All of these processes have a significant impact on the spectral timing properties of the observational data. In particular, our results demonstrate that the EEH correction parameter $Q$ alters the power of oscillation frequencies in the innermost disk region and reorganizes the distribution of modes, allowing transitions between hard and soft $X-$ray states (or vice versa). Such transitions are not predicted within classical Schwarzschild or Kerr spacetimes and could therefore serve as a diagnostic of NLED effects in strong gravity. When analyzed together with observational data, these features provide a pathway for testing EEH gravity using $X-$ray binaries, ultraluminous $X-$ray sources, and active galactic nuclei. Since both LFQPOs and HFQPOs are commonly observed in X-ray binaries \cite{Ingram2019} and AGNs \cite{Gierlinski2008,Pan2016}, the QPOs signatures identified in our simulations can be directly compared with these systems to search for possible imprints of EEH gravity.

One of the most significant outcomes of our simulations is the emergence of a clear inner outer disk dichotomy in the accretion flow. In the inner disk region at $r=2.3M$, the EEH parameter regulates the inflow of matter toward the BH, steepens the density gradient, and enables the development of strong turbulence inside the shock cone. This environment excites a rich spectrum of LFQPOs and HFQPOs, producing resonance patterns such as $3:2$, $2:1$, and $4:3$. The oscillations amplified by the $Q$ parameter of the EEH BH contribute directly to the $X-$ray emission from the innermost disk region. In contrast, as one moves outward toward $r=6.1M$ and beyond, accretion is suppressed, turbulence is damped, and the oscillation modes become weaker. As a result, LFQPOs become more dominant and play a key role in shaping $X-$ray variability. This redistribution of variability power across radial zones with enhanced HFQPOs in the inner disk and weakened LFQPOs in the outer disk is not predicted by Schwarzschild or Kerr BHs and therefore represents a distinctive observational signature of EEH gravity.

It has been shown that the fundamental modes trapped inside the cavity of the shock cone formed by BHL accretion undergo parametric resonance and nonlinear couplings. According to the numerical results, these modes are not local artificial fluctuations, but rather global oscillations whose properties are modified by the spin and charge parameters of EEH gravity. The simulations reveal that resonance ratios such as $3:2$, $2:1$, and $4:3$ naturally arise for different values of $a$ and $Q$, making their occurrence a robust feature of the model. These numerically obtained ratios are in close agreement with the twin-peak QPOs observed in microquasars such as GRO J1655-40 and XTE J1550-564 \cite{Kluzniak2005,Remillard2006,Wang2008,Belloni2014}. The persistence of these commensurate ratios indicates that resonant interactions among the radial, azimuthal, and Lense-Thirring \cite{Ingram2009} frequencies are naturally supported within the EEH framework. Moreover, the amplification of inner-disk harmonics by the charge parameter $Q$ suggests that EEH gravity provides a viable mechanism for generating unusually strong HFQPOs that remain difficult to explain in pure Kerr spacetimes. Therefore, the EEH model emerges as an alternative gravity framework capable of reproducing the harmonic oscillations and overtone structures observed in various $X-$ray binary systems. These results demonstrate that the systematic resonance patterns present in observational data can be explained by considering the combined influence of the charge parameter and the spin parameter of the BH in the strong-gravity regime of EEH gravity.

For a rapidly rotating BH with $a=0.9M$, the combined effects of frame dragging and NLED corrections produce a complex and enriched frequency spectrum. This behavior differs significantly from the spectrum predicted by the Kerr geometry. Our simulations show that while the inner disk generally amplifies the HFQPOs, the presence of the charge parameter $Q$ redistributes the oscillatory power outward. As a result, in the case $a=0.9M$, the frequency power of the outer disk can become stronger than that of the inner-disk modes. Numerical calculations also reveal a correlation between LFQPO frequencies and radial location, which is in good agreement with the predictions of the Lense-Thriring precession model \cite{Remillard2006,Ingram2009}. Thus,  $Q$ emerges as a parameter that modifies the behavior of QPOs in a measurable way. This provides a potential test through spectral analyses of LFQPOs observed in X-ray binaries and AGN. The combined influence of $Q$ and $a$ generates a highly complex spectrum in the strong-gravity regime, a feature absent in Kerr spacetimes, and therefore offers a distinctive behavior through which the NLED corrections of EEH gravity may be probed observationally.

The numerical findings of strengthened inner-disk harmonics, altered PSD structures, and suppressed QPOs amplitudes in the outer disk provide observationally testable predictions. In $X-$ray binaries, EEH gravity anticipates HFQPOs with strong inner-disk amplitudes and pronounced harmonic ladders, which can be directly compared with NICER \cite{Rani2025} and RXTE data \cite{Ingram2019} on Galactic microquasars. This feature may help explain the unusually strong HFQPOs observed in sources such as $GRO J1655-40$ and $XTE J1550-564$, phenomena that remain only partially understood within Kerr geometry. For a stellar-mass BH of $M=10M_{\odot}$, the QPOs frequencies obtained in our simulations can be rescaled to supermassive BHs using the relation $f(Hz)=f(Hz) \left(10M_{\odot}/M_{SMBH}\right)$. In this way, the same resonance patterns extend naturally to AGNs and ULXs \cite{Agrawal2015} that host massive central BHs. The numerical frequencies derived for stellar-mass systems are consistent with QPOs reported in Seyfert galaxies and ULXs \cite{Gierlinski2008,Pasham2014}. In particular, energy-dependent amplitude changes, enhanced variability in the hard $X-$ ray band, and suppressed power in the soft band associated with the outer disk may represent characteristic signatures of EEH gravity. Furthermore, forthcoming high-resolution $X-$ray missions such as Athena \cite{Nandra2013}, eXTP \cite{Zhang2019}, and STROBE-X \cite{Say2019} will provide the spectral and temporal precision necessary to observe these rich QPOs structures in the innermost disk regions. If confirmed, such observations would offer the opportunity to test NLED corrections in the strong-gravity regime and extend EEH-based diagnostics across the full mass spectrum, from stellar BHs to supermassive systems.

\section{Conclusions}\label{S7}

The EEH framework extends the conventional Einstein-Maxwell theory by illustrating QED corrections that account for nonlinear effects in the electromagnetic field. Also, these corrections, originally defined by Euler and Heisenberg and later developed within the Plebański approach models, effectively describe vacuum polarization phenomena where virtual electron-positron pairs modify the electromagnetic field structure in regions of extreme intensity. Also, in this formalism, the standard electromagnetic invariants $X$ and $Y$, constructed from the Faraday tensor and its dual, enter the Lagrangian through higher-order terms proportional to the fine-structure constant $\alpha$, which quantifies the strength of electromagnetic interactions. The presence of $\alpha$ in the EH Lagrangian encodes the nonlinearity of the vacuum. It governs the departure from Maxwell's electrodynamics, particularly when the field strengths approach or exceed the Schwinger critical field $E_c \approx 1.32\times10^{18}\,\mathrm{V/m}$. As a result, the QED vacuum acts as a polarizable medium, screening the effective electric charge and thereby altering the geometry of spacetime only through the re-normalized charge parameter $\Tilde{Q}$. The resulting metric, obtained from the Einstein equations coupled to this NLED, describes a rotating EEH BH whose line element generalizes the Kerr-Newman solution by introducing $\Tilde{Q}$ as a field-dependent effective charge that varies with both radial and angular coordinates. This correction depends on a dimensionless parameter $\beta$, which scales inversely with the square of the BH mass, implying that nonlinear effects are most pronounced for compact objects of smaller mass. Physically, the EEH geometry manifests as a Kerr-Newman like spacetime in which the vacuum polarization suppresses the electromagnetic contribution to the curvature, thereby modifying the location of horizons and the permissible charge-to-mass ratio. Indeed, for a fixed rotation parameter $a$, the maximum admissible charge decreases as the spin increases, indicating that higher rotational energy constrains the existence of classical BH solutions and favors the emergence of naked singularities for larger $Q$. The dynamics of test particles in this background, analyzed through the Hamiltonian formalism, reveal substantial deviations from the Kerr model due to the nonlinear electromagnetic corrections. The conserved quantities, namely the specific energy $\mathcal{E}$ and angular momentum $\mathcal{L}$, exhibit distinct dependencies on the magnetic parameter $B$ and the rotation parameter $a$: increasing $B$ enhances both the energy and the angular momentum required for circular orbits, while increasing $a$ has the opposite effect, lowering these quantities and stabilizing the orbit at smaller radii. Comparative analysis between Kerr and rotating EEH BHs shows that particles orbiting the latter possess higher energies and angular momenta, reflecting the influence of nonlinear vacuum polarization in deepening the effective potential wells. The effective potential $V_{\mathrm{eff}}(r)$, obtained from the normalization condition of the four-velocity, shows these effects: its minima correspond to stable circular orbits, while its maxima denote unstable configurations. In addition, the potential profile reveals that as the magnetic parameter $B$ increases, the depth of the potential also increases, thereby improving the trapping of particles. In contrast, a stronger rotation parameter $a$ tends to flatten the potential, indicating a reduced confinement efficiency. These results illustrate the relation between electromagnetic nonlinearity and rotation in shaping the orbital dynamics near the BH. The associated effective force, derived from the gradient of $V_{\mathrm{eff}}(r)$, further confirms that higher magnetic fields amplify the attractive gravitational interaction, while rotation mitigates it. In this case, this intricate balance translates into measurable astrophysical consequences, such as shifts in the position of the ISCOs, alterations in the energy extraction mechanisms, and modifications of precession phenomena. In particular, the analysis of fundamental frequencies, periastron precession, and Lense-Thirring precession around rotating EEH BHs demonstrates that NLED corrections yield higher precession rates compared to the Kerr BH models and illustrate an observational signature of QED-induced modifications in strong gravity environments. Also, the EEH model encapsulates a consistent semi-classical description of gravity coupled to quantum-corrected electromagnetism, where the vacuum polarization parameter $\alpha$ quantifies the strength of the nonlinear coupling and governs the geometric and dynamical features of the resulting BH spacetime. This theoretical framework thus bridges quantum electrodynamics and GR in an astrophysical relevant regime, providing a pathway to probe vacuum polarization effects through the orbital characteristics and precessional dynamics of particles in the vicinity of rotating BHs. 

The analysis of harmonic oscillations as perturbations of circular orbits provides deep insight into the underlying geometry and dynamical properties of neutral test particles in the vicinity of a rotating EEH BH. By introducing small perturbations around stable circular trajectories, one uncovers a rich structure of epicyclic motions, in which the particle executes QPOs characterized by three fundamental frequencies, namely the radial ($\omega_{r}$), the vertical ($\omega_{\theta}$) and azimuthal ($\omega_{\phi}$) components. These frequencies, defined through the second derivatives of the effective potential with respect to $r$ and $\theta$, encode the local stability and response of the system to small displacements from equilibrium. In the Newtonian case, these frequencies coincide, leading to closed elliptical orbits around spherically symmetric configurations. However, in a relativistic spacetime such as that of the Schwarzschild or rotating EEH BH models, this degeneracy is lifted due to curvature effects and frame dragging, resulting in $\omega_{r} < \omega_{\theta} = \omega_{\phi}$, a condition that manifests itself as orbital precession and the characteristic deviation from Newtonian orbits. In addition, the transformation of these locally measured frequencies to those perceived by distant static observers involves the inclusion of the gravitational redshift factor $\mathrm{d}\tau/\mathrm{d}t$, linking proper-time oscillations to asymptotic measurements. Once rescaled in physical units using the factor $c^{3}/(GM)$, the resulting dimensionless frequencies $\Omega_{j}$ $(j \in \{r, \theta, \phi\})$ reveal the intricate dependence of oscillatory motion on the BH parameters, including the spin $a$, charge $Q$, and the non-linear electrodynamics coupling $\beta$. In this case, the detailed expressions for $\Omega_{r}$, $\Omega_{\theta}$, and $\Omega_{\phi}$ exhibit complex polynomial structures in $r$ that encapsulate the influence of higher-order electromagnetic corrections inherent in the EEH model. Physically, the behavior of these frequencies underscores the competition between gravitational attraction, rotation, and magnetic interactions. The numerical analysis demonstrates that an increase in the magnetic parameter $B$ shifts the frequency profiles closer to the event horizon, indicating that magnetic effects enhance the binding of neutral particles, thereby reducing the effective radius of stable oscillatory motion. In contrast, increasing the rotation parameter $a$ counteracts this tendency, pushing the frequency peaks outwards and indicating that frame-dragging weakens the confining potential around the BH. Furthermore, the distinctions among $\Omega_{r}$, $\Omega_{\theta}$, and $\Omega_{\phi}$ give rise to two notable relativistic phenomena: the periapsis precession and the Lense-Thirring precession. In this case, defined as $\Omega_{P}$, quantifies the azimuthal advance of the orbit within the equatorial plane, while the latter, $\Omega_{L}$, captures the nodal precession induced by spacetime rotation. Furthermore, the analysis reveals that the frequency of periapsis precession decreases with increasing magnetic parameters $B$, but increases with the rotation parameter $a$, implying that magnetic fields act to suppress orbital precession, while rotation enhances it through stronger frame-dragging effects. In contrast, both $B$ and $a$ contribute to a decrease in the Lense-Thirring precession frequency, suggesting that the combined influence of magnetism and rotation tends to attenuate the vertical precession of the orbital plane. Remarkably, the magnitude of the periapsis precession frequency in the rotating EEH spacetime exceeds that of the Kerr geometry, showing the signature of nonlinear electromagnetic corrections on orbital dynamics. These findings establish a coherent physical picture in which the harmonic oscillations and associated precessions serve as precise probes of the spacetime structure and the interplay between gravitational, rotational, and electromagnetic effects in NLED BHs. 

Our numerical simulations test how the EEH charge parameter $Q$ and the BH spin parameter $a$ jointly govern the morphology and dynamics of BHL accretion flows. In this case, by solving the GRH equations within the fixed spacetime metric of the EEH BH, we reveal a distinct two-regime behavior in the accretion process that emerges as a direct consequence of the nonlinear coupling between gravity, charge, and rotation. In addition, in the strong-field region close to the event horizon ($r = 2.3M$), the EEH charge parameter substantially enhances the mass inflow rate compared to the Schwarzschild and Kerr cases, amplifying the density and velocity gradients within the shock cone. This amplification triggers powerful hydrodynamical instabilities that act as an efficient mechanism for the excitation of fundamental QPOs, whose amplitudes increase markedly with $Q$ and $a$. In contrast, at larger radial distances ($r = 6.1M$ and $12M$), where the gravitational field is weaker, the accretion rate is systematically suppressed, leading to a smoother and more stable plasma morphology with reduced oscillatory activity. The introduction of BH rotation further enriches this picture: in the low-spin regime ($a = 0.3M$), frame-dragging effects distort the shock cone and promote the emergence of LFQPOs, while moderate rotation ($a = 0.5M$) induces strong inner-region instabilities that couple nonlinearly to outer-region oscillations, giving rise to mixed low- and high-frequency QPOs spectra. In the rapidly rotating configuration ($a = 0.9M$), frame dragging dominates the accretion dynamics, producing violent oscillatory behavior and high-amplitude HFQPOs that persist in multiple radial zones. In addition, these results illustrate that QPOs are not random fluctuations but organized resonant phenomena arising from the synergistic influence of $Q$ and $a$ on the excitation of trapped modes within the shock cone. The EEH charge parameter modulates the inflow of matter and regulates the growth of instabilities, whereas spin governs the angular momentum transfer, precession effects, and nonlinear resonance conditions. In this case, this interplay provides a robust theoretical framework linking the microscopic dynamics of accreting plasma to macroscopic observables, thereby offering a promising avenue for testing EEH gravity through future high-precision astrophysical observations. 

The numerical analysis testing of plasma and shock cone structures around the Schwarzschild, Kerr, and EEH BH models shows how the interplay between spin $a$ and the NLED parameter $Q$ fundamentally alters the morphology and stability of the accretion. Also, in the case of the Schwarzschild BH model, the plasma infall proceeds symmetrically, and the resulting shock cone remains relatively steady, with turbulence largely confined within the downstream region. Also, when rotation is introduced ($a > 0$), frame-dragging effects twist the plasma flow lines, producing deformed shock cones with pronounced shear layers that seed oscillations and turbulent mixing. The inclusion of EEH corrections leads to even more striking behavior: near the horizon, the NLED terms associated with $Q$ modify the spacetime curvature and enhance the effective gravitational attraction, giving rise to denser, more compact plasma distributions. This enhancement steepens density gradients and amplifies velocity shear, thus strengthening turbulence and promoting the growth of oscillatory modes that can manifest as HFQPOs. In contrast, at larger radii, the EEH corrections act to suppress the accretion flow, stabilizing the outer regions of the shock cone, and reducing large-scale turbulence. This coexistence of inner instability and outer stability constitutes a distinct feature of EEH gravity compared to GR, imprinting a characteristic two-zone structure in the accretion dynamics. Another important result, the dependence of the shock cone opening angle and stagnation geometry on both $a$ and $Q$, illustrates the nonlinear coupling between rotation and electromagnetic self-interaction in the strong-field regime. Increasing $Q$ systematically widens the cone and enhances turbulence without significantly altering the peak density, testing that EEH-induced modifications primarily reshape the dynamical morphology rather than the total mass load. In the strong-field region ($r=2.6M$), both parameters act coherently to distort the cone and modify its oscillatory response, whereas farther out ($r=6.1M$), their influence diminishes, producing only mild geometric variations. In this case, these results suggest that the EEH framework naturally predicts localized zones of enhanced instability near the horizon, surrounded by relatively quiescent outer regions, a pattern that may leave observable imprints on the timing and spectral properties of accreting systems. Hence, systematic deviations in shock cone geometry, accretion rate modulation, and QPOs excitation relative to Schwarzschild or Kerr predictions could serve as empirical signatures of EEH gravity in the strong-field regime, offering a promising route to test NLED effects through high-resolution astrophysical observations. 

The numerical analysis illustrates that QPOs in the accretion flow around EEH BHs are global oscillation modes intrinsically linked to the nonlinear dynamics of the shock cone cavity. Also, across all configurations, ranging from non-rotating to rapidly spinning cases, the PSD reveals that the same set of centroid frequencies appears at distinct radial positions, confirming that these oscillations are not localized numerical artifacts but coherent, large-scale modes of the accreting plasma. In this case, the results show that the EEH parameters $a$ and $Q$ exert a pronounced influence on both the amplitude and spatial distribution of the oscillation power. In particular, the charge parameter $Q$ enhances the oscillatory amplitudes in the strong-field regime, leading to pronounced inner-disk harmonics and a richer overtone structure compared to the Schwarzschild or Kerr baselines. This enhancement is especially evident at $r=2.3M$, where NLED corrections amplify HFQPOs, while at larger radii, such as $r=6.1M$, the amplitudes become comparable to or weaker than those in the classical solutions, indicating that EEH corrections are most effective in regions of intense curvature. Also, the PSD spectra consistently exhibit near-integer frequency ratios such as $3:2$, $4:3$, and $2:1$, reflecting the presence of nonlinear coupling and parametric resonance, mechanisms that underpin the observed twin-peak QPOs phenomenology in X-ray binaries. Increasing either the spin parameter $a$ or the charge $Q$ modulates the power redistribution within the disk: for slowly rotating BHs, the inner disk dominates the QPOs activity, whereas for high-spin cases ($a=0.9M$) with moderate charge, the combination of strong curvature, frame dragging, and spin-charge coupling shifts part of the oscillatory power outward, yielding a more complex interplay between inner and outer disk variability. Also, the persistence of the same centroid frequencies across these spatial zones confirms the global nature of the trapped modes, while the redistribution of QPOs amplitudes delineates a clear observational distinction between EEH and Kerr geometries. In physical terms, the enhancement of HFQPOs in the EEH BH spacetime implies stronger variability in the hard X-ray band originating from the inner disk, whereas the suppressed oscillations in the outer region correspond to the soft X-ray band. These trends suggest that forthcoming high-sensitivity X-ray missions could potentially distinguish EEH gravity signatures by detecting stronger inner-disk harmonics, richer overtone structures, and characteristic frequency ratios that deviate from classical GR predictions. Thus, the interplay between NLED and rotation in EEH gravity not only modifies the oscillatory behavior of accreting matter but also opens a potential observational window into the strong-field regime beyond the Kerr paradigm. 

The numerical investigation testing the dynamical behavior of QPOs in the vicinity of an EEH BH model is strongly governed by the interplay between the spin parameter $a$ and the NLED charge parameter $Q$. In this case, a systematic exploration of the parameter space reveals that $Q$ acts as a powerful regulator of both the amplitude and frequency of oscillatory modes, producing a distinctive radial dependence that differentiates EEH BHs from their Kerr and Schwarzschild counterparts. Also, in the strong-field region ($r=2.3M$), the maximum QPOs amplitudes increase monotonically with $Q$ and exhibit an exponential enhancement for $Q>0.9M$ in low-spin configurations, indicating that nonlinear electromagnetic corrections significantly amplify the oscillatory energy trapped within the inner accretion cavity. Simultaneously, the associated frequencies undergo a transition from HFQPOs to LFQPOs as $Q$ increases, revealing that the EEH corrections shift the characteristic variability from the rapid inner-disk oscillations to slower global modes dominated by Lense-Thirring precession. In contrast, at larger radii near the ISCOs ($r=6.1M$), the QPOs amplitudes generally decrease with $Q$ before rising again for $Q>0.9M$, while the frequencies display a non-monotonic evolution that reflects the competition between gravitational, hydrodynamical, and NLED effects. These behaviors imply that the EEH parameter $Q$ governs a complex redistribution of oscillatory power across the accretion flow, steepening the density gradient and triggering turbulence in the inner disk while damping oscillations farther out. In this case, the combined influence of $a$ and $Q$ thus produces a multi-scale QPOs spectrum characterized by resonance ratios such as $3:2$, $2:1$, and $4:3$, naturally emerging from parametric couplings among the radial, azimuthal, and Lense-Thirring modes. Such resonance patterns, commonly observed in X-ray binaries and active galactic nuclei, suggest that EEH gravity offers a compelling mechanism for the amplification of HFQPOs and the modulation of LFQPOs, yielding observable spectral-timing signatures that deviate from Kerr predictions. In this context, the amplification of inner-disk harmonics, the emergence of nonlinear frequency couplings, and the charge-induced reorganization of variability power across the disk represent distinctive imprints of NLED in the strong-gravity regime, providing a promising avenue for testing EEH gravity through high-resolution X-ray observations with missions such as NICER, Athena, and eXTP.

As shown in Fig. (\ref{fig1}), the variation of the horizon radii $(r_{\pm}/M)$ in the EEH spacetime is plotted as a function of the charge-to-mass ratio $(Q/M)$ for different rotation parameters $a$. In this case, the plot illustrates that increasing $a$ reduces the permissible range of $Q$ that maintains a regular BH, while exceeding these limits results in a naked singularity. Also, Figure (\ref{fig2}) presents the particle energy $\mathcal{E}$ around a rotating EEH BH, where the first column displays the influence of the magnetic parameter $B$ and the second shows the variation with $a$. It is observed that $\mathcal{E}$ increases with $B$ but decreases with $a$, implying that stronger magnetic effects raise orbital energy, whereas rotation reduces it. Also, Figure (\ref{fig3}) depicts the angular momentum $\mathcal{L}$ as a function of $r$, the left column shows its dependence on $B$, and the right column on $a$. The results show that $\mathcal{L}$ grows with both $r$ and $B$, but decreases with increasing $a$. Another important result, Figure (\ref{fig4}) illustrates the effective potential $V_{\text{eff}}(r)$ for various $Q/M$ and $a/M$ values at fixed $\mathcal{L}=3.5$ and $\beta/M=0.01$, showing that larger $B$ lowers the potential minima, while higher $a$ raises them. The potential minimum for the Kerr BH lies above that of the EEH BH, indicating stronger binding in the EEH case. Also, Figure (\ref{fig5}) presents the effective force $F$ as a function of $r/M$ for several $Q/M$ and $a/M$ values. The results show that for small $B$, the force is mainly attractive, and its magnitude increases with $B$ but decreases with $a$. Figure (\ref{fig6}) illustrates the radial profiles of the oscillation frequencies $\nu_{r}$, $\nu_{\theta}$, and $\nu_{\phi}$ for different combinations of $a/M$ and $Q/M$, showing that increasing $B$ shifts the frequencies inward toward the horizon, while higher $a$ moves them outward. Figures (\ref{fig7}) and (\ref{fig8}) illustrate the periastron precession frequency $\Omega_{P}$ and Lense–Thirring frequency $\Omega_{L}$ versus $r/M$ for varying $a/M$ and $Q/M$ at fixed $\beta/M=0.01$. It is shown that $\Omega_{P}$ decreases with $B$ but increases with $a$, while $\Omega_{L}$ decreases with both parameters, remaining smaller in the Kerr limit. Figures (\ref{fig9})–(\ref{fig12}) show the time evolution of the mass accretion rate $(dM/dt)$ at $r=2.3M$, $6.1M$, and $12M$ for different $a/M$ and $Q/M$. Also, the EEH corrections enhance accretion near the horizon and suppress it farther away, with the strongest effect at higher spin. Also, Figure (\ref{fig13}) shows contour maps of the rest-mass density and velocity fields in the equatorial plane for Schwarzschild, Kerr, and EEH BHs, showing that EEH corrections produce denser, less stable plasma near the horizon and stronger turbulence within the shock cone. Figure (\ref{fig14}) shows the normalized accretion rate $(dM/dt)/(dM/dt)_{\text{Kerr}}$ versus $(Q/M)$ for various $a/M$ at three radii, illustrating enhancement in the strong-field region $(r=2.6M)$ and suppression at larger distances. Figures (\ref{fig15}) and (\ref{fig16}) present the azimuthal variation of rest-mass density at $r=2.6M$ and $r=6.1M$, showing that increasing $a$ and $Q$ widens the shock cone and shifts the stagnation point, with higher $Q$ producing greater instability. In this context, Fig. (\ref{fig17}) shows the PSD comparison between Schwarzschild and EEH BHs at $r=2.3M$ and $r=6.1M$. The EEH BH exhibits stronger and illustrated QPOs peaks, including harmonic ratios like $3:2$ and $2:1$, while at larger radii, amplitudes decrease, indicating that NLED corrections amplify QPOs activity in the strong-field region. 

Also, Figures (\ref{fig18})-(\ref{fig22}) present the numerical results of QPOs for rotating EEH BHs, illustrating the effects of the spin parameter $(a/M)$ and charge-to-mass ratio $(Q/M)$ on the PSD and oscillation frequencies at a fixed coupling $\beta/M=0.01$. Figure (\ref{fig18}) displays the PSD for the non-rotating configuration $(a=0)$ at $r=2.3M$ and $r=6.1M$ for $Q/M=0$ and $Q/M=0.95$. The results illustrate that near the horizon ($r=2.3M$), higher charge values generate stronger QPOs peaks with dominant frequencies around 2.8-30.9 Hz, while the amplitudes reduce significantly at $r=6.1M$, confirming that EEH corrections mainly enhance inner-disk oscillations. Figure (\ref{fig19}) shows the PSD for a moderately rotating case $(a/M=0.5)$, comparing $Q/M=5\times10^{-4}$ and $Q/M=0.8$. Also, the analysis indicates that for $r=2.3M$, increasing $Q/M$ intensifies the PSD and yields stronger HFQPOs, whereas at $r=6.1M$ the oscillation power decreases and the spectral peaks shift slightly toward lower frequencies, reflecting weaker outer-disk activity. Also, Figure (\ref{fig20}) corresponds to the rapidly rotating case $(a/M=0.9)$, where $Q/M=5\times10^{-4}$ and $Q/M=0.4$ are analyzed at the same radial points. The plots reveal that for small charge values, the PSD behaves similarly to the Kerr limit, while larger $Q/M$ introduces additional high-frequency peaks in the range $1.6$-$62$ Hz near the horizon. In this case, at $r=6.1M$, low-frequency modes dominate with reduced amplitude, showing that nonlinear electromagnetic effects amplify the inner-disk oscillations. Figure (\ref{fig21}) illustrates the normalized maximum amplitude $(A_{\text{max}}/A_{\text{Kerr}})$ versus $Q/M$ for spin parameters $a/M=0.0$, $0.3$, $0.5$, and $0.9$, evaluated at $r=2.3M$ and $r=6.1M$. The plots show that at $r=2.3M$, the amplitude increases steadily with $Q/M$ and becomes exponential beyond $Q/M>0.9$, especially for lower $a/M$, while at $r=6.1M$ the amplitude initially decreases before rising at higher charge values, confirming that the strong-field region dominates the amplification. In this context, Figure (\ref{fig22}) depicts the evolution of dominant QPOs frequencies corresponding to the maximum PSD amplitudes as a function of $Q/M$ for the same spin values and radii. It is found that at $r=2.3M$, the frequencies shift from HFQPOs to LFQPOs as $Q/M$ increases, with the lowest values associated with the fastest rotation $(a/M=0.9)$, consistent with the Lense-Thirring precession. At $r=6.1M$, the frequency variation becomes irregular, showing alternating transitions between HFQPOs and LFQPOs as $Q/M$ increases. In this context, Figures (\ref{fig18})-(\ref{fig22}) show that the EEH charge parameter $(Q/M)$ plays a crucial role in modulating both the amplitude and frequency structure of QPOs, leading to harmonic ratios such as $3:2$, $2:1$, and $4:3$, with the strongest enhancements observed in the inner region for small $\beta/M$ and high spin values.\\ 

\section*{Acknowledgments}
All simulations were performed using the Phoenix High
Performance Computing facility at the American University of the Middle East
(AUM), Kuwait.\\

\section*{Data Availability Statement}
The data generated in this study were produced using high-performance computing resources and are available from the corresponding author upon reasonable request.

\section*{Appendix}

\begin{widetext}
\begin{eqnarray*}
    \hat{p}_{1}(r)&=&-3 p^{2}_{r} \beta ^3 Q^{18}+7 p^{2}_{r} r^4 \beta ^2 Q^{14}+6 p^{2}_{r} r^6 \beta ^2 Q^{12}-13 p^{2}_{r} r^5 \beta ^2 Q^{12}+\mathcal{L}^2 r^4 \beta ^2 Q^{12}-5 p^{2}_{r} r^8 \beta  Q^{10}-8 p^{2}_{r} r^{10} \beta  Q^8\nonumber\\&+&18 p^{2}_{r} r^9 \beta  Q^8-2 \mathcal{L}^2 r^8 \beta  Q^8+p^{2}_{r} r^{12} Q^6-3 \mathcal{E}^2 r^{12} \beta  Q^6-3 p^{2}_{r} r^{12} \beta  Q^6+14 p^{2}_{r} r^{11} \beta  Q^6-2 \mathcal{L}^2 r^{10} \beta  Q^6\nonumber\\&-&16 p^{2}_{r} r^{10} \beta  Q^6+4 \mathcal{L}^2 r^9 \beta  Q^6+2 p^{2}_{r} r^{14} Q^4-5 p^{2}_{r} r^{13} Q^4+\mathcal{L}^2 r^{12} Q^4+\mathcal{E}^2 r^{16} Q^2+p^{2}_{r} r^{16} Q^2-6 p^{2}_{r} r^{15} Q^2\nonumber\\&+&2 \mathcal{L}^2 r^{14} Q^2+8 p^{2}_{r} r^{14} Q^2-4 \mathcal{L}^2 r^{13} Q^2-\mathcal{E}^2 r^{17}-p^{2}_{r} r^{17}+\mathcal{L}^2 r^{16}+4 p^{2}_{r} r^{16}-4 \mathcal{L}^2 r^{15}-4 p^{2}_{r} r^{15}\nonumber\\&+&4 \mathcal{L}^2 r^{14}+a^6 p^{2}_{r} r^{12}+r^4 \left(-\beta  Q^6+r^4 Q^2+(r-2) r^5+a^2 r^4\right)^2 \text{p$\theta $}(\tau )^2+2 a^3 \mathcal{E} \mathcal{L} r^8 \left(3 \beta  Q^6+r^5-Q^2 r^4\right)\nonumber\\&-&2 a \mathcal{E} \mathcal{L} r^4 \left(\beta ^2 Q^{12}-2 r^4 \beta  Q^8-4 (r-1) r^5 \beta  Q^6+r^8 Q^4+2 (r-2) r^9 Q^2+r^{10} (4-3 r)\right)+a^4 r^8 \Big[p^{2}_{r} \big[-5 \beta  Q^6\nonumber\\&+&3 r^4 Q^2+r^5 (2 r-5)\Big]-\mathcal{E}^2 \left(3 \beta  Q^6+r^5-Q^2 r^4\right)\Big]+a^2 r^4 \Big[\Big[\beta ^2 Q^{12}-2 r^4 \beta  Q^8+2 r^5 (2-3 r) \beta  Q^6\nonumber\\&+&r^8 Q^4+2 (r-2) r^9 Q^2-2 (r-2) r^{10}\Big] \mathcal{E}^2-\mathcal{L}^2 \left(r^9-Q^2 r^8+3 Q^6 \beta  r^4\right)+p^{2}_{r} \Big[7 \beta ^2 Q^{12}-10 r^4 \beta  Q^8\nonumber\\&+&2 r^5 (9-4 r) \beta  Q^6+3 r^8 Q^4+2 r^9 (2 r-5) Q^2+r^{10} \left(r^2-6 r+8\right)\Big]\Big],\\
    \hat{p}_{2}(r)&=&\sqrt{\frac{a^2 \left(\beta  Q^6-Q^2 r^4+r^5 (r+2)\right)+r^6 \left(\mathcal{L}^2+r^2\right)}{a^2 r^6-\beta  Q^6 r^2+Q^2 r^6+(r-2) r^7}},\\
    \hat{p}_{3}(r)&=&\frac{1}{\left(r^8-a^2 \left(3 \beta  Q^6-Q^2 r^4+r^5\right)\right)^2}\Big[2 a^3 r^2 \left(3 \beta  Q^6-Q^2 r^4+r^5\right)^{3/2}-a^2 \Big[12 \beta ^2 Q^{12}-10 \beta  Q^8 r^4+\beta  Q^6 r^5 (9 r+13)\nonumber\\&+&2 Q^4 r^8-Q^2 r^9 (3 r+5)+3 r^{10} (r+1)\Big]+2 a r^4 \left(4 \beta  Q^6-2 Q^2 r^4+3 r^5\right) \sqrt{3 \beta  Q^6-Q^2 r^4+r^5}\nonumber\\&+&r^8 \left(-4 \beta  Q^6+2 Q^2 r^4+(r-3) r^5\right)\Big],\\
    \hat{p}_{4}(r)&=&r^6 \left(a^2 \mathcal{E} \left(\beta  Q^6-Q^2 r^4+r^5 (r+2)\right)+a \mathcal{L} \left(-\beta  Q^6+Q^2 r^4-2 r^5\right)+\mathcal{E} r^8\right)^2,\\
    \hat{p}_{5}(r)&=&\mathcal{E}^2 \left(8 \beta ^2 Q^{12}-30 \beta  Q^8 r^4-9 \beta  Q^6 r^5 (7 r-8)+6 Q^4 r^8+Q^2 r^9 (9 r-20)-6 (r-2) r^{10}\right)+\mathcal{L}^2 \nonumber\\&\times&\left(-21 \beta  Q^6 r^4+3 Q^2 r^8-2 r^9\right),\\
    \hat{p}_{6}(r)&=&-\mathcal{E}^2 \Big[3 \beta ^3 Q^{18}-9 \beta ^2 Q^{14} r^4+3 \beta ^2 Q^{12} r^5 (r+6)+9 \beta  Q^{10} r^8+2 \beta  Q^8 r^9 (19 r-18)+3 Q^6 r^{10} \Big[12 \beta +(21 \beta -1) r^2\nonumber\\&-&36 \beta  r\Big]-9 Q^4 (r-2) r^{13}-9 Q^2 (r-2)^2 r^{14}+2 r^{15} \left(3 r^2-14 r+12\right)\Big]-\mathcal{L}^2 r^4 \Big[-8 \beta ^2 Q^{12}+30 \beta  Q^8 r^4\nonumber\\&+&\beta  Q^6 r^5 (43 r-72)-6 Q^4 r^8+Q^2 r^9 (20-9 r)+r^{10} \left(r^2+6 r-12\right)\Big],\\
    \hat{p}_{7}(r)&=&3 \beta ^3 Q^{18}-9 \beta ^2 Q^{14} r^4-3 \beta ^2 Q^{12} (r-6) r^5+9 \beta  Q^{10} r^8+4 \beta  Q^8 r^9 (7 r-9)+3 Q^6 r^{10} \left(12 \beta +(12 \beta -1) r^2-24 \beta  r\right)\nonumber\\&-&9 Q^4 (r-2) r^{13}-2 Q^2 r^{14} \left(5 r^2-18 r+18\right)+4 r^{15} \left(3 r^2-8 r+6\right),\\
    \hat{p}_{8}(r)&=&\left(15 \beta ^2 Q^{12}+\beta  Q^6 r^5 (21 r-20)+Q^4 r^8-3 Q^2 r^{10}+2 r^{11}\right),\\
    \hat{p}_{9}(r)&=&r^2 \left(a^2 \mathcal{E} \left(\beta  Q^6-Q^2 r^4+r^5 (r+2)\right)+a \mathcal{L} \left(-\beta  Q^6+Q^2 r^4-2 r^5\right)+\mathcal{E} r^8\right)^2,\\
    \hat{p}_{10}(r)&=&\mathcal{L}^2 \left(31 \beta  Q^6 r^4-Q^2 r^8+2 r^9\right)-\mathcal{E}^2 \Big[\beta ^2 Q^{12}-2 \beta  Q^8 r^4+2 \beta  Q^6 r^5 (2-31 r)+Q^4 r^8+2 Q^2 (r-2) r^9\nonumber\\&-&4 (r-1) r^{10}\Big],\\
        \hat{p}_{11}(r)&=&\left(\beta ^2 Q^{12}-2 \beta  Q^8 r^4+\beta  Q^6 r^5 (4-31 r)+Q^4 r^8+Q^2 (r-4) r^9-2 (r-2) r^{10}\right),\\
\hat{p}_{12}(r)&=&\mathcal{E}^2 r^2 \left(\beta ^2 Q^{12}-2 \beta  Q^8 r^4+\beta  Q^6 r^5 (4-31 r)+Q^4 r^8+Q^2 (r-4) r^9-2 (r-2) r^{10}\right)\nonumber\\&+&\mathcal{L}^2 \left(\beta ^2 Q^{12}-2 \beta  Q^8 r^4+4 \beta  Q^6 r^5+Q^4 r^8-4 Q^2 r^9-r^{10} \left(r^2-4\right)\right).
    \end{eqnarray*}

\end{widetext}


\bibliography{mybib}
\bibliographystyle{ieeetr}

\end{document}